\documentclass[twocolumn]{aastex63}
\usepackage{amsmath}
\usepackage{lineno}
\usepackage{longtable}
\usepackage{ulem} % provides \sout

\newcommand{\msun}{${\rm M_{\sun}}$}

\def\ltsima{$\; \buildrel < \over \sim \;$}
\def\simlt{\lower.5ex\hbox{\ltsima}}
\def\gtsima{$\; \buildrel > \over \sim \;$}
\def\simgt{\lower.5ex\hbox{\gtsima}}
\def\kms{{\rm\,km\,s^{-1}}}
\def\mas{{\rm\,mas}}
\def\masyr{{\rm\,mas/yr}}
\def\kpc{{\rm\,kpc}}
\def\mpc{{\rm\,Mpc}}
\def\msun{{\rm\,M_\odot}}
\def\lsun{{\rm\,L_\odot}}

\def\pc{{\rm\,pc}}

\def\g{{\rm\,g}}

\def \r {r$^{1/4}$ }

\newcommand{\fmmm}[1]{\mbox{$#1$}}
\newcommand{\scnd}{\mbox{\fmmm{''}\hskip-0.3em .}}


\def\AA{$\; \buildrel \circ \over {\rm A}$}

\def\deg{^\circ}
\def\degg{\hbox{$\null^\circ$\hskip-3pt .}}
\def\s{\ifmmode \widetilde \else \~\fi}
\def\={\overline}

\def\spose#1{\hbox to 0pt{#1\hss}}

\def\lta{\mathrel{\spose{\lower 3pt\hbox{$\mathchar"218$}}
     \raise 2.0pt\hbox{$\mathchar"13C$}}}
\def\gta{\mathrel{\spose{\lower 3pt\hbox{$\mathchar"218$}}
     \raise 2.0pt\hbox{$\mathchar"13E$}}}
\def\Dt{\spose{\raise 1.5ex\hbox{\hskip3pt$\mathchar"201$}}}    % upper case
\def\dt{\spose{\raise 1.0ex\hbox{\hskip2pt$\mathchar"201$}}}    % lower case

\def\r{${\rm r^{1/4}}$}

\def\dotsfill{\leaders\hbox to 1em{\hss.\hss}\hfill}
\def\sun{\odot}
\def\earth{\oplus}

\def\Gyr{{\rm\,Gyr}}

\def\ltsima{$\; \buildrel < \over \sim \;$}
\def\gtsima{$\; \buildrel > \over \sim \;$}
\def\lsim{\lower.5ex\hbox{\ltsima}}
\def\gsim{\lower.5ex\hbox{\gtsima}}
\def\lapp{\ifmmode\stackrel{<}{_{\sim}}\else$\stackrel{<}{_{\sim}}$\fi}
\def\gapp{\ifmmode\stackrel{>}{_{\sim}}\else$\stackrel{<}{_{\sim}}$\fi}

\submitjournal{ApJ}

\shorttitle{A global model of the mass distribution in the Milky Way from stellar streams}
\shortauthors{Ibata et al.}
\graphicspath{{./}{figures/}}
\begin{document}

\title{Charting the Galactic acceleration field\\
II. A global mass model of the Milky Way from the STREAMFINDER Atlas of Stellar Streams detected in \emph{Gaia} DR3}

\correspondingauthor{Rodrigo Ibata}
\email{rodrigo.ibata@astro.unistra.fr}

\author[0000-0002-3292-9709]{Rodrigo Ibata}
\affiliation{Universit\'e de Strasbourg, CNRS, Observatoire astronomique de Strasbourg, UMR 7550, F-67000 Strasbourg, France}
\nocollaboration{1}

\author[0000-0002-8318-433X]{Khyati Malhan}
\affiliation{Max-Planck-Institut f\"ur Astronomie, K\"onigstuhl 17, D-69117, Heidelberg, Germany}
\affiliation{Myrspoven AB, V\"astg\"otagatan 1, 11827 Stockholm, Sweden}
\nocollaboration{1}

\author[0000-0001-8392-3836]{Wassim Tenachi}
\affiliation{Universit\'e de Strasbourg, CNRS, Observatoire astronomique de Strasbourg, UMR 7550, F-67000 Strasbourg, France}
\nocollaboration{1}

\author[0000-0002-0544-2217]{Anke Ardern-Arentsen}
\affiliation{ Institute of Astronomy, University of Cambridge, Madingley Road, Cambridge CB3 0HA, UK}
\nocollaboration{1}

\author[0000-0001-8200-810X]{Michele Bellazzini}
\affiliation{INAF - Osservatorio di Astrofisica e Scienza dello Spazio, via Gobetti 93/3, I-40129 Bologna, Italy}
\nocollaboration{1}

\author[0000-0002-0358-4502]{Paolo Bianchini}
\affiliation{Universit\'e de Strasbourg, CNRS, Observatoire astronomique de Strasbourg, UMR 7550, F-67000 Strasbourg, France}
\nocollaboration{1}

\author[0000-0002-1014-0635]{Piercarlo Bonifacio}
\affiliation{GEPI, Observatoire de Paris, Universit\'e PSL, CNRS, 5 Place Jules Janssen, 92190 Meudon, France}
\nocollaboration{1}

\author[0000-0001-6011-6134]{Elisabetta Caffau}
\affiliation{GEPI, Observatoire de Paris, Universit\'e PSL, CNRS, 5 Place Jules Janssen, 92190 Meudon, France}
\nocollaboration{1}

\author{Foivos Diakogiannis}
\affiliation{Data61, CSIRO, Floreat WA, Australia}
\nocollaboration{1}

\author{Raphael Errani}
\affiliation{Universit\'e de Strasbourg, CNRS, Observatoire astronomique de Strasbourg, UMR 7550, F-67000 Strasbourg, France}
\nocollaboration{1}

\author[0000-0003-3180-9825]{Benoit Famaey}
\affiliation{Universit\'e de Strasbourg, CNRS, Observatoire astronomique de Strasbourg, UMR 7550, F-67000 Strasbourg, France}
\nocollaboration{1}

\author[0000-0003-1623-6643
]{Salvatore Ferrone}
\affiliation{GEPI, Observatoire de Paris, Universit\'e PSL, CNRS, 5 Place Jules Janssen, 92190 Meudon, France}
\nocollaboration{1}

\author[0000-0002-1349-202X]{Nicolas F. Martin}
\affiliation{Universit\'e de Strasbourg, CNRS, Observatoire astronomique de Strasbourg, UMR 7550, F-67000 Strasbourg, France}
\nocollaboration{1}

\author{Paola di Matteo}
\affiliation{GEPI, Observatoire de Paris, Universit\'e PSL, CNRS, 5 Place Jules Janssen, 92190 Meudon, France}
\nocollaboration{1}

\author{Giacomo Monari}
\affiliation{Universit\'e de Strasbourg, CNRS, Observatoire astronomique de Strasbourg, UMR 7550, F-67000 Strasbourg, France}
\nocollaboration{1}

\author[0000-0001-5073-2267]{Florent Renaud}
\affiliation{Universit\'e de Strasbourg, CNRS, Observatoire astronomique de Strasbourg, UMR 7550, F-67000 Strasbourg, France}
\affiliation{University of Strasbourg Institute for Advanced Study, 5 all\'ee du G\'en\'eral Rouvillois, F-67083 Strasbourg, France}
\nocollaboration{1}

\author{Else Starkenburg}
\affiliation{Kapteyn Astronomical Institute, University of Groningen, Landleven 12, 9747 AD Groningen, the Netherlands}
\nocollaboration{1}

\author[0000-0002-2468-5521]{Guillaume Thomas}
\affiliation{Instituto de Astrof\'isica de Canarias, E-38205 La Laguna, Tenerife, Spain}
\affiliation{Universidad de La Laguna, Dpto. Astrofísica, E-38206 La Laguna, Tenerife, Spain}
\nocollaboration{1}

\author{Akshara Viswanathan}
\affiliation{Kapteyn Astronomical Institute, University of Groningen, Landleven 12, 9747 AD Groningen, the Netherlands}
\nocollaboration{1}

\author{Zhen Yuan}
\affiliation{Universit\'e de Strasbourg, CNRS, Observatoire astronomique de Strasbourg, UMR 7550, F-67000 Strasbourg, France}
\nocollaboration{1}

\begin{abstract}
We present an atlas and follow-up spectroscopic observations of 87 thin stream-like structures detected with the STREAMFINDER algorithm in \emph{Gaia} DR3, of which 29 are new discoveries. Here we focus on using these streams to refine mass models of the Galaxy. Fits with a double power law halo with the outer power law slope set to $-\beta_h=3$ yield an inner power law slope $-\gamma_h=0.97^{+0.17}_{-0.21}$, a scale radius of $r_{0, h}=14.7^{+4.7}_{-1.0}\kpc$, a halo density flattening $q_{m, h}=0.75\pm0.03$, and a local dark matter density of $\rho_{h, \odot}=0.0114\pm0.0007 \msun \pc^{-3}$. Freeing $\beta$ yields $\beta=2.53^{+0.42}_{-0.16}$, but this value is heavily influenced by our chosen virial mass limit. The stellar disks are found to have a combined mass of $4.20^{+0.44}_{-0.53}\times10^{10}\msun$, with the thick disk contributing $12.4\pm0.7$\% to the local stellar surface density. The scale length of the thin and thick disks are $2.17^{+0.18}_{-0.08}\kpc$ and $1.62^{+0.72}_{-0.13}\kpc$, respectively, while their scale heights are $0.347^{+0.007}_{-0.010}\kpc$ and $0.86^{+0.03}_{-0.02}\kpc$, respectively. The virial mass of the favored model is $M_{200}=1.09^{+0.19}_{-0.14}\times 10^{12}\msun$, while the mass inside of $50\kpc$ is $M_{R<50}=0.46\pm0.03\times 10^{12}\msun$. We introduce the Large Magellanic Cloud (LMC) into the derived potential models, and fit the ``Orphan'' stream therein, finding a mass for the LMC that is consistent with recent estimates. Some highlights of the atlas include the nearby trailing arm of $\omega$-Cen, and a nearby very metal-poor stream that was once a satellite of the Sagittarius dwarf galaxy. Finally, we unambiguously detect a hot component around the GD-1 stream, consistent with it having been tidally pre-processed within its own DM subhalo.
\end{abstract}


\section{Introduction}
\label{sec:Introduction}

The formation of galaxies involves a complex interplay between the accretion of gas and the merging of previously bound fragments. When these fragments are of low mass compared to their host or their dominant companion, tidal forces can have time to act progressively to slowly unbind material from the lower mass object. If the disruption is gradual, so that the injected energy is only just enough to unbind the most loosely bound stars, the ejected objects will find themselves on similar orbits to their progenitor, and diffuse along a ``tidal stream'' pattern that approximates that orbit \citep{1999A&A...352..149C}. The disrupted stars that by chance acquire higher kinetic energy leave approximately through the L1 Lagrange point, and race ahead of their progenitor, forming the leading tidal arm. Conversely those stars that are perturbed into lower energy orbits emerge roughly through the L2 point, and form the trailing arm (see, e.g. \citealt{2000MNRAS.318..753F}). 

Tidal disruption can take place over hundreds of millions to billions of years, with the dwindling progenitor slowly losing stars that drift away along their slightly different orbits. The requirement that the stream stars originated in their progenitor provides a physical constraint that allows us to attempt to turn back the clock and find the smooth potential and any perturbations that can permit this inversion. This is perhaps the most useful property of stellar streams, giving us a means to peer into the dynamical past of a galaxy without making assumptions about its equilibrium, and so uncover properties of the global and local dark matter distribution that are otherwise invisible. Nevertheless, it should be borne in mind that tidal disruption typically produces ejecta with a spectrum of energies, meaning that stars move along the tidal stream pattern at different speeds, so that position along the pattern is not perfectly correlated with ejection time. Furthermore, epicyclic patterns are also present in streams, which can complicate their analysis \citep{2008AN....329.1061K}.

This promise of being able to map out the acceleration field of our Galaxy motivated us to develop a dedicated stream-detection algorithm, the \texttt{STREAMFINDER}, with the intention to deploy it on the \emph{Gaia} mission catalogs \citep{2016A&A...595A...1G}. The \texttt{STREAMFINDER} is effectively a friend-finding algorithm, with a ``distance'' in the parameter space of observables defined so as to make objects on similar orbits and with similar stellar populations appear close together. The procedure is presented in detail in \citet{2018MNRAS.477.4063M}, \citet[][hereafter Paper~I]{2018MNRAS.481.3442M} and \citet[][hereafter Paper~II]{2021ApJ...914..123I}. Paper~I applied the algorithm to the \emph{Gaia} DR2 catalog, based on 22 months of astrometric observations, while Paper~II extended the search to the \emph{Gaia} EDR3 catalog, with 33 months of observations. In \citet{2022MNRAS.516.5331M} the \texttt{STREAMFINDER} sources were cross-matched with the Pristine survey catalog \citep{2017MNRAS.471.2587S,2023arXiv230801344M}, providing metallicity estimates for the stars and hence better discrimination against contamination, which allowed us to lower the detection threshold and so find further stream candidates. A complete catalog of Galactic stellar streams was compiled by \citet{2023MNRAS.520.5225M}, and a thorough bibliography of previous searches is given therein.

Our aim in this contribution is to present the \texttt{STREAMFINDER} results for the full \emph{Gaia} DR3 dataset \citep{2023A&A...674A...1G}, which supersedes EDR3. The new analysis makes use of the radial velocity measurements for 33 million stars provided by \emph{Gaia} DR3, which we complement with measurements from other large spectroscopic surveys. We will also present measurements from our own dedicated spectroscopic follow-up with VLT/UVES and INT/IDS as well as our earlier CFHT/ESPaDOnS measurements, providing the community the full \emph{Gaia} DR3 stream catalog as uncovered by our algorithm. The main scientific goal of the present contribution is to conduct a detailed analysis of the constraints that these observations impose on the large-scale mass distribution in the Milky Way. 

The layout of the paper is as follows. In Section~\ref{sec:Alterations}, we will begin by explaining the small alterations we made to the algorithm for this new \emph{Gaia} DR3 analysis, while Section~\ref{sec:Observations} will briefly review our follow-up spectroscopic observations, and Section~\ref{sec:DR3_maps} will present the new sky maps and detections. In Section~\ref{sec:Stream_Fitting_isolated}, we will begin the modelling of the streams assuming that the Milky Way is axisymmetric and that its center defines the origin of an inertial frame around which the streams orbit. Section~\ref{sec:Stream_Fitting_test_particles} will generalise the model to allow the Local Group galaxies to move under the influence of each other, and the streams will be modelled as dissolving systems. In Section~\ref{sec:Orbit_Stream_Correction_functions}, we will explain how we use test particle simulations to correct simple orbit integrations, and thus greatly speed up the exploration of the parameters of the model.  The results of the stream-fitting analysis are presented in Section~\ref{sec:Stream_Fitting_results}. In Section~\ref{sec:DR3_Detections}, we briefly summarize some salient streams, and finally draw our conclusions in Section~\ref{sec:Conclusions}.

\begin{figure*}
\begin{center}
\includegraphics[angle=0, viewport= 180 180 1000 1000, clip, width=\hsize]{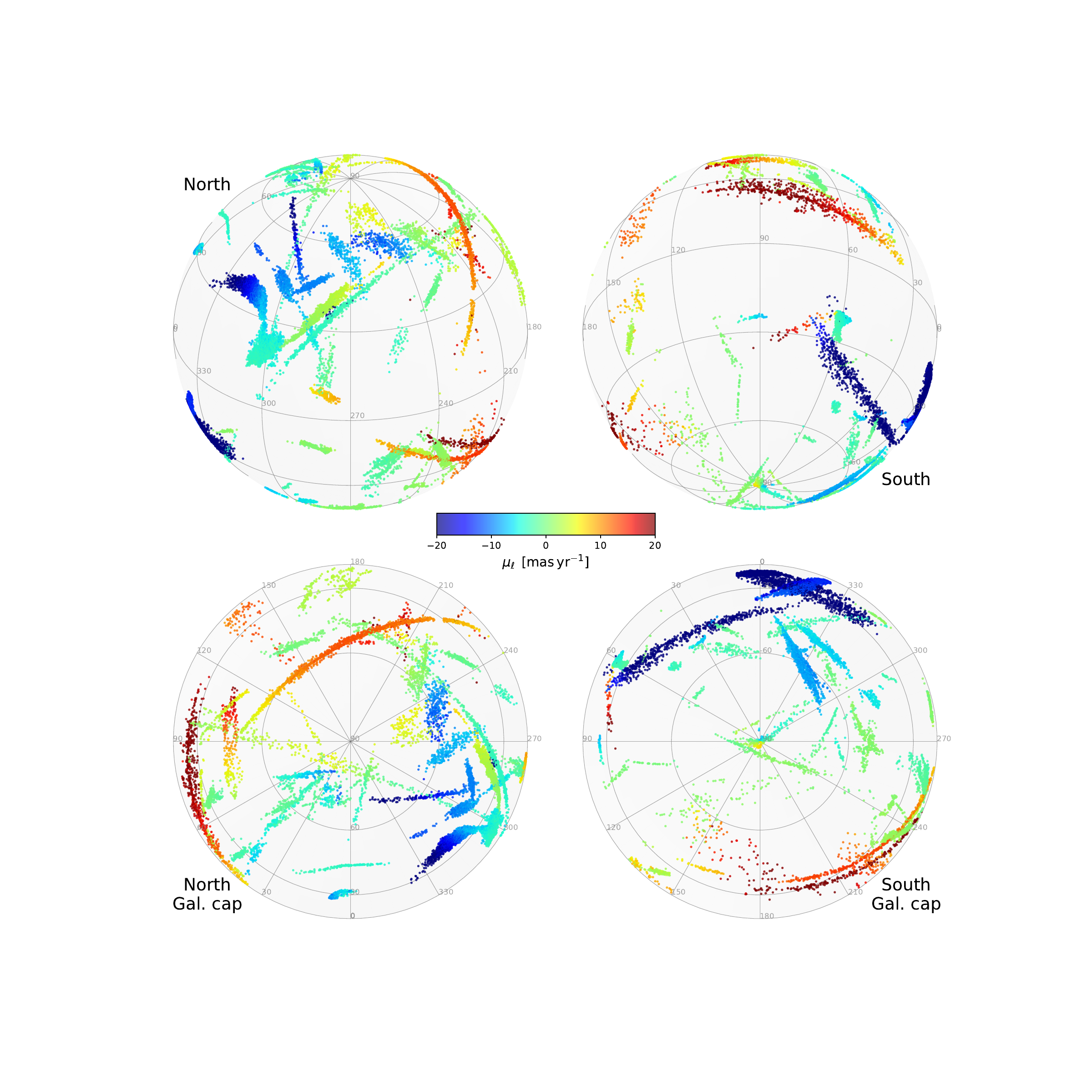}
\end{center}
\caption{Four different projections in Galactic coordinates of the 24,540 \texttt{STREAMFINDER} sources with $\ln{\mathcal{\hat{L}}}>19.8$ (more restrictive than a $6\sigma$ threshold). The color of the points encodes the proper motion $\mu_\ell$ along the Galactic longitude direction, as measured by \emph{Gaia} in the DR3 catalog.}
\label{fig:Ball_mul}
\end{figure*}

\begin{figure*}
\begin{center}
\includegraphics[angle=0, viewport= 180 180 1000 1000, clip, width=\hsize]{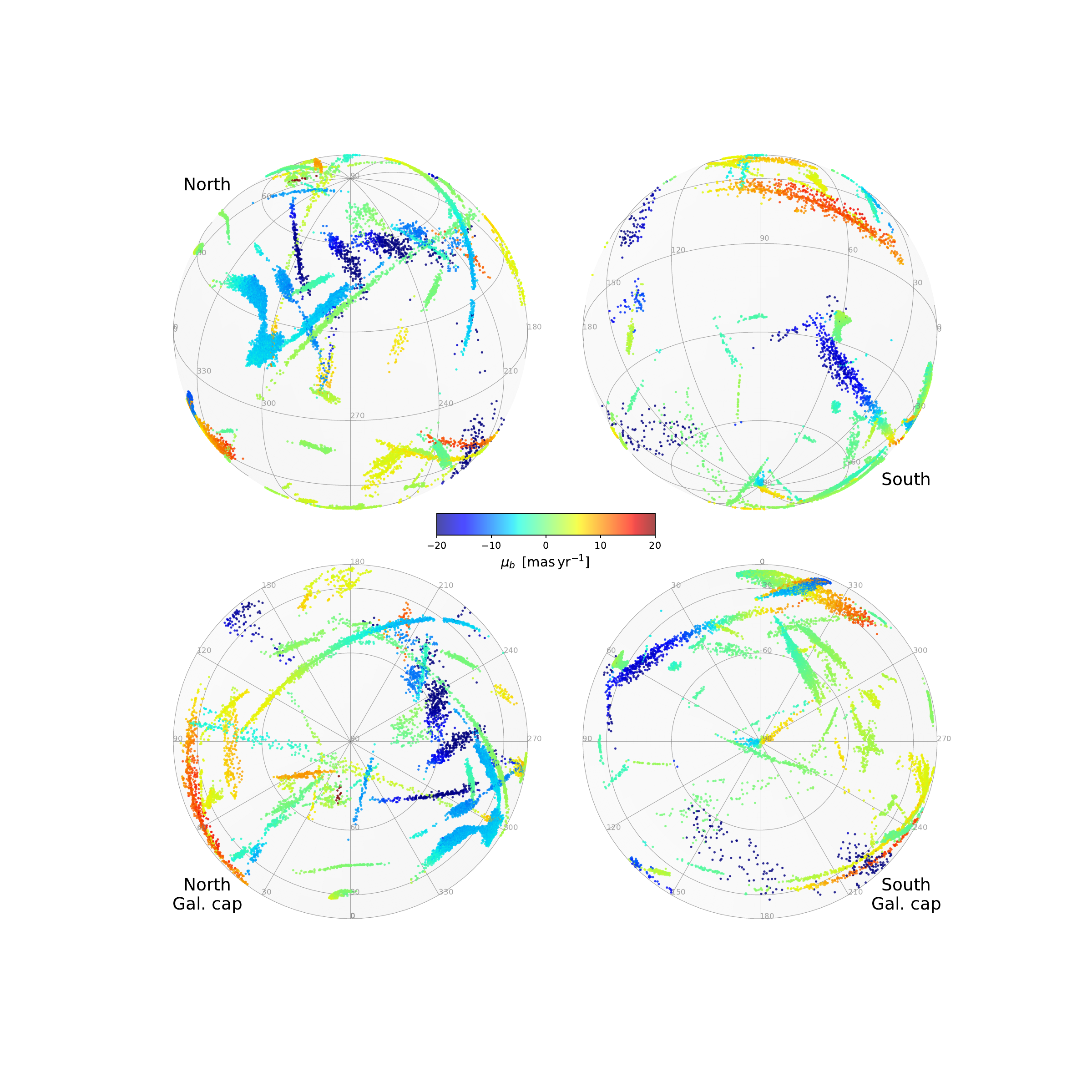}
\end{center}
\caption{As Figure~\ref{fig:Ball_mul}, but showing the measured proper motion $\mu_b$ of the sources along the Galactic longitude direction.}
\label{fig:Ball_mub}
\end{figure*}

\section{STREAMFINDER alterations}
\label{sec:Alterations}

\begin{figure*}
\begin{center}
\includegraphics[angle=0, viewport= 180 180 1000 1000, clip, width=\hsize]{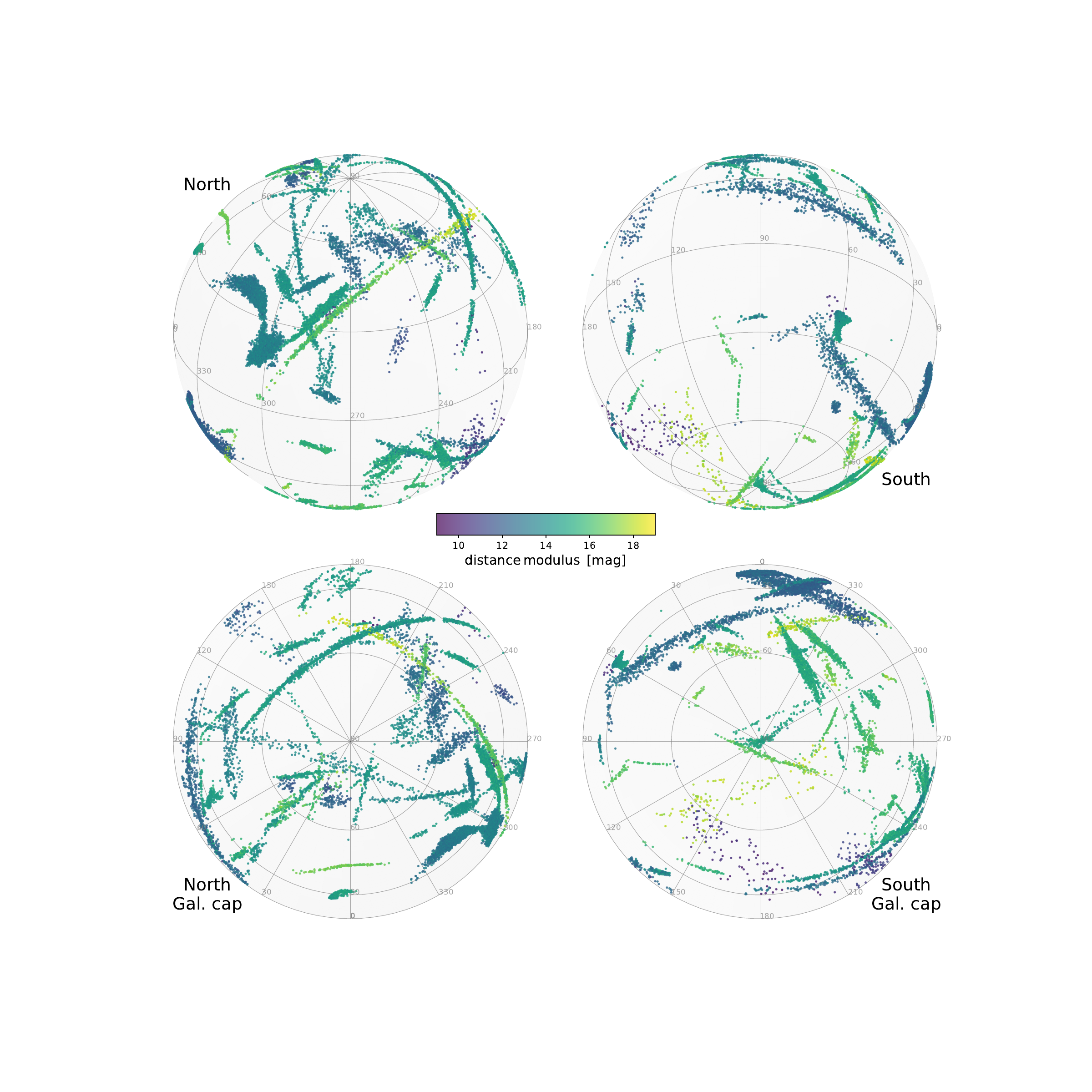}
\end{center}
\caption{As Figure~\ref{fig:Ball_mul}, but showing the distance solutions, displayed as distance modulus.}
\label{fig:Ball_dis}
\end{figure*}

Here we use a very similar version of the search algorithm as that presented in Paper~II. As before, we aim to find a low-contrast stream superimposed on a ``contaminating'' population of normal Milky Way stars, and so we adopt a log-likelihood function
\begin{equation}
{\ln} \mathcal{L} = \rm \sum_{\rm{data}} {\ln} \, [\eta \mathcal{P}_{\rm stream}(\theta) + (1 - \eta) \, \mathcal{P}_{\rm cont} ] \, ,
\label{eqn:likelihood}
\end{equation}
where $\eta$ is the fraction of stars in the stream, $\mathcal{P}_{\rm stream}(\theta)$ is the probability distribution function (PDF) of the stream model with parameters $\theta$, and $\mathcal{P}_{\rm cont}$ is the PDF of the contaminants. The stream model is model of a stellar population along a simple orbit that has been smeared-out with a Gaussian
\begin{align}
\begin{split}
\mathcal{P}_{\rm stream}(\theta)  = & \mathcal{P}_{\rm length} \times \mathcal{P}_{\rm width} \times \mathcal{P}_{\rm LF} \times \mathcal{P}_{\rm color} \times \mathcal{P}_\mu \times \mathcal{P}_{\rm \varpi} \\
&\times \mathcal{P}_{v} \\
= & \mathcal{P}_{\rm stream}^{\rm Paper~II}(\theta) \times \mathcal{P}_{v} \, ,
\label{eqn:prob}
\end{split}
\end{align}
where $\mathcal{P}_{\rm length}$ is a uniform distribution along a $L=20\deg$ stream fragment, $\mathcal{P}_{\rm width}$ is a Gaussian width of $50\pc$ dispersion, $\mathcal{P}_{\rm LF}$ is the probability of being drawn from the luminosity function of the adopted simple stellar population (SSP) model (we adopt the PARSEC isochrone models, \citealt{2012MNRAS.427..127B}), $\mathcal{P}_{\rm color}$ is a probability of the measured $G_{\rm Bp} - G_{\rm Rp}$ color of the star at the observed $G$ magnitude, and $\mathcal{P}_\mu$ and $\mathcal{P}_{\rm \varpi}$ are the Gaussian probabilities of the observed astrometry given an assumed $5\kms$ intrinsic stream velocity dispersion, and a $50\pc$ line of sight dispersion. (The proper motion probability term $\mathcal{P}_\mu$ is a two-dimensional Gaussian PDF incorporating the proper motion uncertainties in right ascension and declination and their correlation cross-terms). Up to here, all is identical to the model described in Paper~II. However, since the DR3 catalog provides many line of sight velocity measurements, we now also include a simple Gaussian probability $\mathcal{P}_{v}$ that the star has a line of sight velocity within $5\kms$ of the orbit. Most stars in the DR3 catalog do not have measured velocities, and for those we set $\mathcal{P}_{v}=1$.

\begin{figure*}
\begin{center}
\includegraphics[angle=0, viewport= 180 180 1000 1000, clip, width=\hsize]{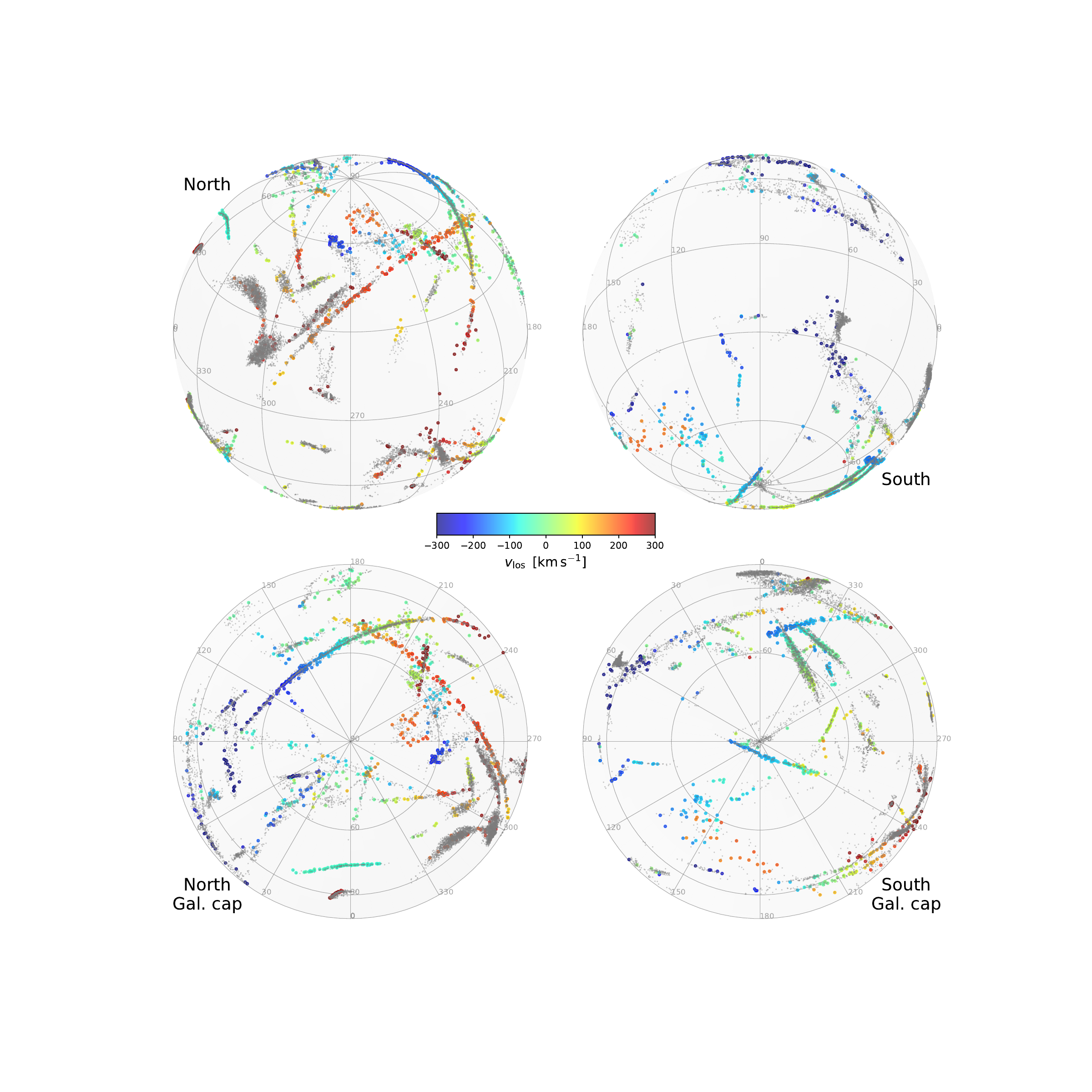}
\end{center}
\caption{As Figure~\ref{fig:Ball_mul}, but showing the line of sight velocity measurements of the stars in the sample. Stars without such measurements are marked with small gray points.}
\label{fig:Ball_vel}
\end{figure*}

Including velocities into the likelihood function of the stream model means that we also have to include them in the contamination model $\mathcal{P}_{\rm cont}$. We decided to retain the previously fitted, completely empirical, Gaussian Mixture Model (GMM) description $\mathcal{P}_{\rm cont}^{\rm Paper~II}(\alpha,\delta,\mu_\alpha,\mu_\delta,G,G_{\rm BP}-G_{\rm RP},\varpi)$ but include a new term $\mathcal{P}_{\rm GMM}$ as follows:
\begin{align}
\begin{split}
\mathcal{P}_{\rm cont}(\alpha,\delta,\mu_\alpha,\mu_\delta,G,G_{\rm BP}-G_{\rm RP},\varpi,v) = & \\
\mathcal{P}_{\rm cont}^{\rm Paper~II}(\alpha,\delta,\mu_\alpha,\mu_\delta,G,G_{\rm BP}-G_{\rm RP},\varpi) & \\
\mathcal{P}_{\rm GMM}(v | \alpha,\delta,G,G_{\rm BP}-G_{\rm RP},\varpi) \, \, . \\
\end{split}
\end{align}
Note that the final term ignores the correlations of line of sight velocity with proper motion, which we felt was a necessary simplifying assumption given the relative scarcity of the velocity information compared to the astrometric information. This $\mathcal{P}_{\rm GMM}$ term is calculated from the available velocities in large $5.6\deg\times5.6\deg$ tiles of a zenithal equal area projection of the sky.

\begin{figure*}
\begin{center}
\includegraphics[angle=0, viewport= 180 180 1000 1000, clip, width=\hsize]{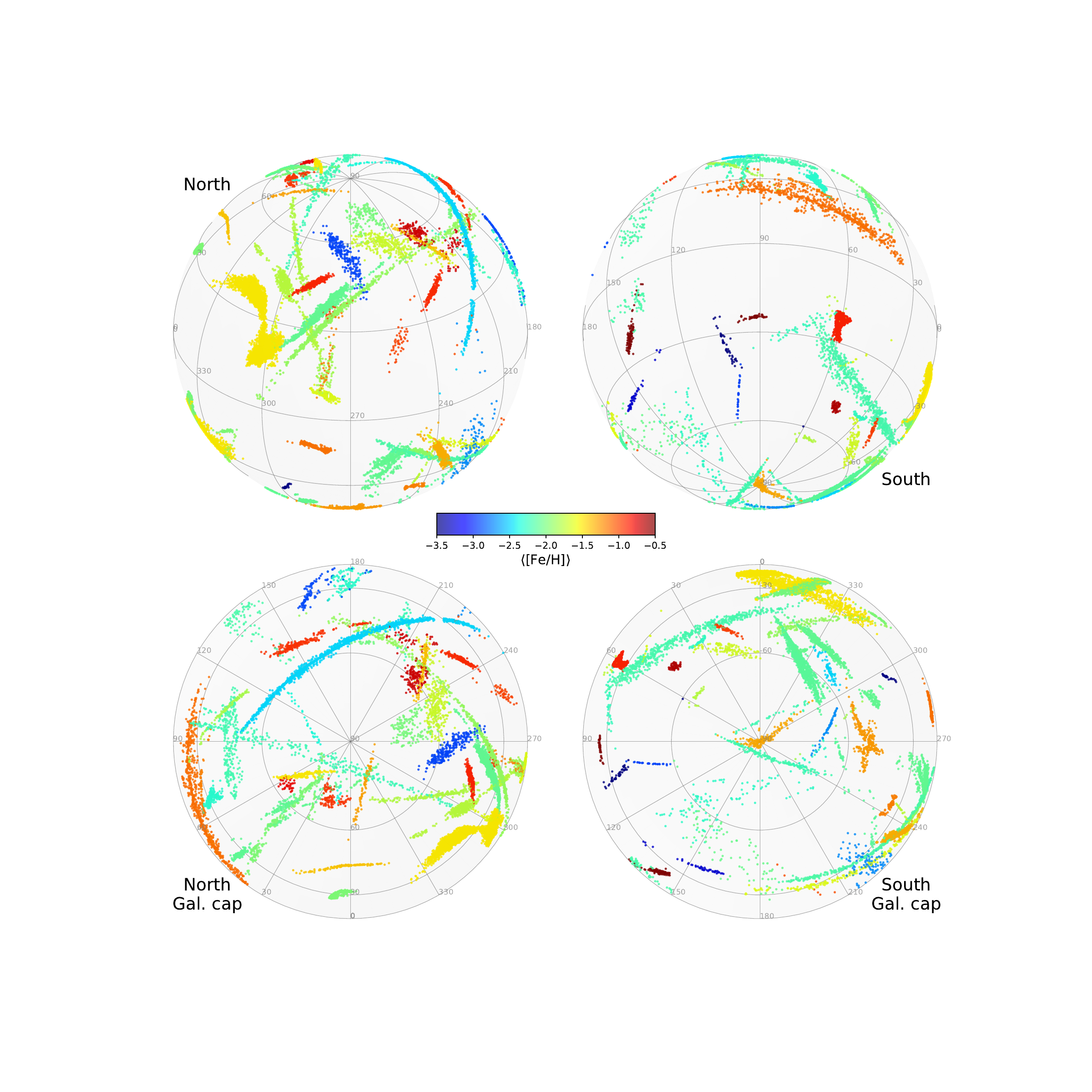}
\end{center}
\caption{As Figure~\ref{fig:Ball_mul}, but showing the mean spectroscopic metallicity ${\rm \langle [Fe/H] \rangle}$  (or photometric metallicity when spectroscopic values are not available) of the streams, as listed in Table~\ref{tab:streams}.}
\label{fig:Ball_FeH}
\end{figure*}

\begin{figure*}
\begin{center}
\includegraphics[angle=0, viewport= 165 1 1000 1000, clip, width=\hsize]{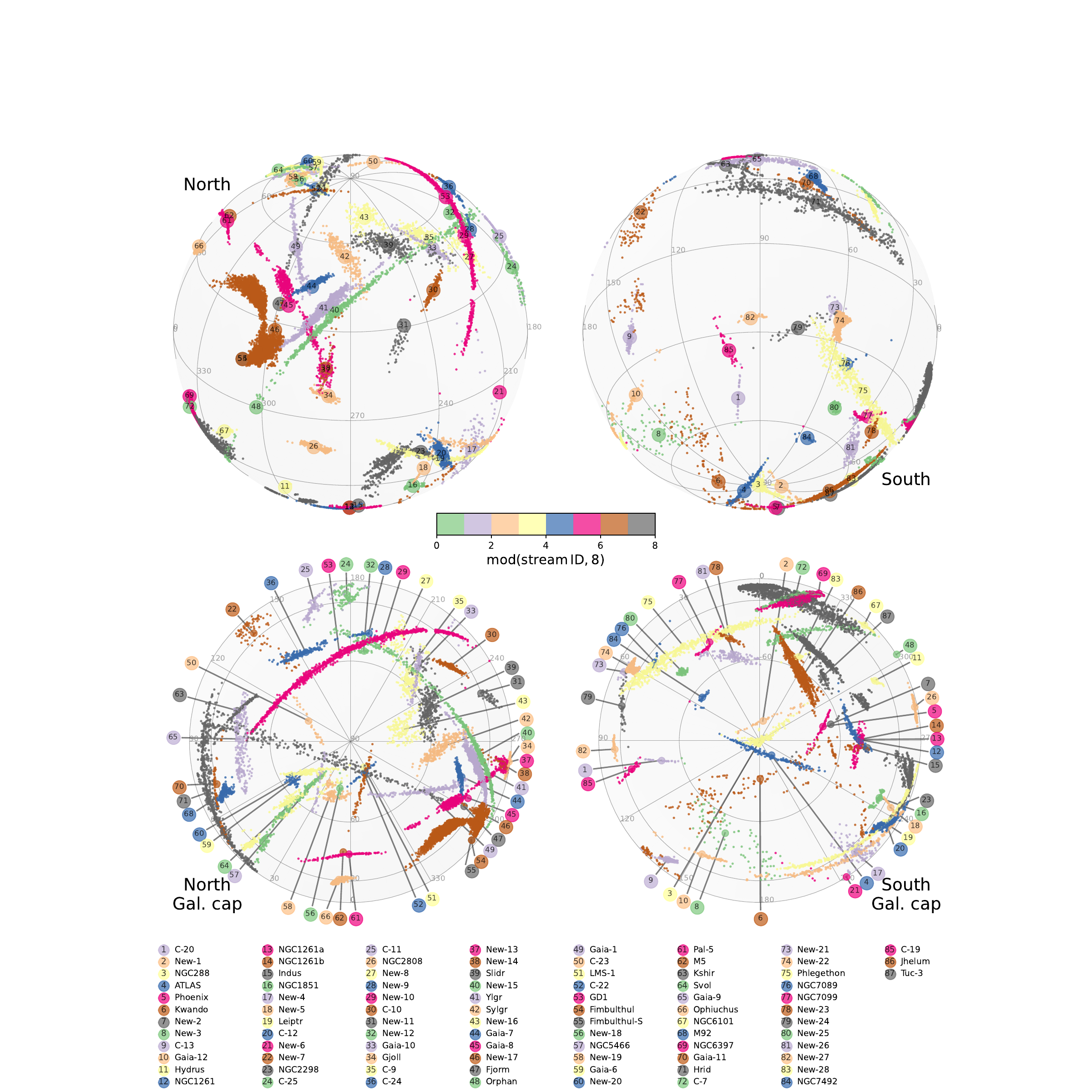}
\end{center}
\caption{Gallery of the stellar streams in the same projections as Figures~\ref{fig:Ball_mul} to \ref{fig:Ball_FeH}. The color of the points shows our chosen stream identification number in a modulo 8 representation.}
\label{fig:Ball_sID}
\end{figure*}

The bulk of the radial velocity data that we provide to the algorithm are the 33 million stars with radial velocity measurements published as part of \emph{Gaia} DR3. We  complemented these measurements with other large public spectroscopic surveys, cross-matching the full \emph{Gaia} DR3 survey with the APOGEE-2 survey \citep{2017AJ....154...94M}, the GALAH DR3 survey \citep{2018MNRAS.478.4513B}, the LAMOST DR7 survey \citep{2012RAA....12.1197C}, the Radial Velocity Experiment (RAVE DR5) \citep{2017AJ....153...75K}, the SDSS/Segue survey \citep{2009AJ....137.4377Y}, the \emph{Gaia}-ESO survey \citep{2022A&A...666A.121R}, and the S5 survey \citep{2019MNRAS.490.3508L}. We also included measurements of Palomar~5 stars by \citet{2009AJ....137.3378O} and \citet{2017ApJ...842..120I} as well as our previous velocity measurements of \texttt{STREAMFINDER} targets with CFHT/ESPaDOnS and NTT/EFOSC2 and VLT/UVES presented in Paper~II. For those sources where more than one measurement was available, we adopted the value with the lowest uncertainty.

A further difference with respect to Paper~II is that we now include stars from the \emph{Gaia} DR3 RR~Lyrae variable catalog \citep{2022arXiv220606278C}. Although we used the VIZIER database to recover any measured line of sight velocity of these stars (where available) we decided to ignore this information in the \texttt{STREAMFINDER} analysis, as RR~Lyrae are very strong radial velocity variables, and we could not reliably phase-correct the heterogenous archival velocity measurements, so for RR~Lyrae we set $\mathcal{P}_{v}=1$ in Eqn.~\ref{eqn:prob}. For these stars we use the \citet{2018MNRAS.481.1195M} metallicity G-band absolute magnitude relation to predict a distance given the SSP trial metallicity, and set $\mathcal{P}_{\rm LF}=1$ in Eqn.~\ref{eqn:prob}.

The algorithm proceeds star by star through the \emph{Gaia} survey, using each star as a reference to launch trial orbits within the adopted fiducial potential \citep{Dehnen:1998tk}. The potential is actually going to be finely constrained with the detected streams in the following sections, but see \citet{2018MNRAS.477.4063M} on the minor influence that a wrong potential in the first step has on the stream detection itself. The measured astrometry and photometry of the reference star are used to generate plausible trial orbits and the assumed stellar populations model provides the link between photometry and distance. When the line of sight velocity is available, it is used, otherwise we scan through the missing information as explained in Papers~I and II. For each trial orbit of the star we find (and record) all the neighbors within the ($10\deg$) search radius that can be associated within a chosen threshold of the stream model. We define $k$ to be the number of stars found in this way around a trial orbit centered on a particular star. 

In many fields of astronomy local measures of significance are used to define a detection, as the sources typically appear small (e.g. a star, or distant galaxy) compared to the field of view of the instrument or survey, and there is a well-defined ``background''. However, in the present situation we are interested in detecting structures that subtend very long arcs on the sky with a wide range of (typically) foreground contamination. For streams we therefore seek a {\it global} statistic to quantify detection. The log-likelihood of Eqn.~\ref{eqn:likelihood} can provide such a measure, but we would need to sum over the full \emph{Gaia} survey (with more than $10^9$ stars) in order to compare models of different structures. Note also that we would need to render the model more complex with each added stream; doing this in the Bayesian framework would require us to calculate the Bayesian evidence for each additional candidate structure to verify that it merits being added. This would be prohibitively computationally expensive. To simplify the calculation, we break up the log-likelihood into two terms ${\ln} \mathcal{L}  = {\ln} \mathcal{L}_1 + {\ln} \mathcal{L}_2$, where ${\ln} \mathcal{L}_1$ corresponds to the parameter space region $R_1$ containing $k$ stars (within say $3\sigma$ of the model in the orbital and stellar populations parameters) and ${\ln} \mathcal{L}_2$ is for the complement region $R_2$ with $n-k$ stars, where $n$ is the number of stars in $R_1+R_2$. Now for region $R_1$:
\begin{equation}
{\ln} \mathcal{L}_1 = \sum_{j=1}^k \ln \Big(\eta \mathcal{P}_{j, \rm model} + (1-\eta) \mathcal{P}_{j, \rm cont} \Big) - \ln \mathcal{P}_{j, \rm cont} \, ,
\label{eqn:likelihood_L1}
\end{equation}
where the final term provides a convenient zero-point so that ${\ln} \mathcal{L}_1=0$ if no stream is present. In region $R_2$ we make the simplification that the stream model probability is zero, so the corresponding log-likelihood is
\begin{align}
\begin{split}
{\ln} \mathcal{L}_2 \approx & \sum_{j=1}^{n-k} \ln \Big( (1-\eta) \mathcal{P}_{j, \rm cont} \Big) - \ln \mathcal{P}_{j, \rm cont} \\
= & (n-k)\ln(1-\eta) \, ,
\label{eqn:likelihood_L2}
\end{split}
\end{align}
where again we have included a zero-point choice so that ${\ln} \mathcal{L}_2=0$ when $\eta=0$. Adding Eqn.~\ref{eqn:likelihood_L1} and \ref{eqn:likelihood_L2} gives the total log-likelihood
\begin{equation}
{\ln} \mathcal{L} = (n-k) \ln(1-\eta) +\sum_{j=1}^k \ln \Big(1 - \eta + \eta \frac{\mathcal{P}_{j, \rm model}}{\mathcal{P}_{j, \rm cont}} \Big) \, ,
\label{eqn:likelihood_actual}
\end{equation}
which allows a straightforward statistical comparison of the detected features over the full sky, and has the convenient property of being zero when $\eta=0$.

Previously in Papers~I and II, we selected the stream solution $\eta$ with the highest likelihood according to Eqn.~\ref{eqn:likelihood_actual}, which is a choice that works well at high Galactic latitude, but becomes problematic at low latitude where the number of possible associations becomes very large. We remind the reader that our algorithm is not finding the best single model fit to the global \emph{Gaia} data, but rather our procedure checks whether a large-scale stream model centered at the phase space position of each of the billions of stars in the survey is statistically significant or not. Hence a given star may be considered as a member of many different proposed groups. This is of course very different to usual fitting problems in astronomy, where we fit the parameters of a model to a dataset, or in detection problems, where we say fit galaxy models to a large number of spatially independent regions of a pixellated image. For the present work we therefore decided to use the logarithm of the likelihood per star of the group as our selection statistic
\begin{equation}
{\ln} \mathcal{\hat{L}} = {\ln} \mathcal{L} - k \, ,
\label{eqn:new_likelihood}
\end{equation}
where $k$ is again the number of stars associated with the trial orbit\footnote{In the present context, the statistic in Eqn.\ref{eqn:new_likelihood} is reminiscent of the Akaike information criterion, ${\rm AIC} = 2 m - 2 {\ln} \mathcal{L}$, which is a statistical estimator designed to optimize the balance between goodness of fit and model complexity ($m$ is the number of parameters of the model). Because of the fact that stars can be re-used in different stream solutions, the higher $k$ is on average the more potential overlapping models there will be and so the higher the global model complexity will be.}. Optimizing this statistic forces the algorithm to find the orbit and associated stream fraction $\eta$ that maximises the number of high likelihood associations. By adjusting the acceptance threshold, stream maps built with this ${\ln} \mathcal{\hat{L}}$ statistic can be selected to be very similar at high Galactic latitude to those built previously using the ${\ln} \mathcal{L}$ statistic. However, close to the plane of the Galaxy the new maps do not show spurious solutions that possess a large number of low probability members (i.e. that are composed mostly of stars at a high number of standard deviations from the orbit and stellar populations model). 

\begin{figure*}
\begin{center}
\includegraphics[angle=0, viewport= 1 1 720 864, clip, width=\hsize]{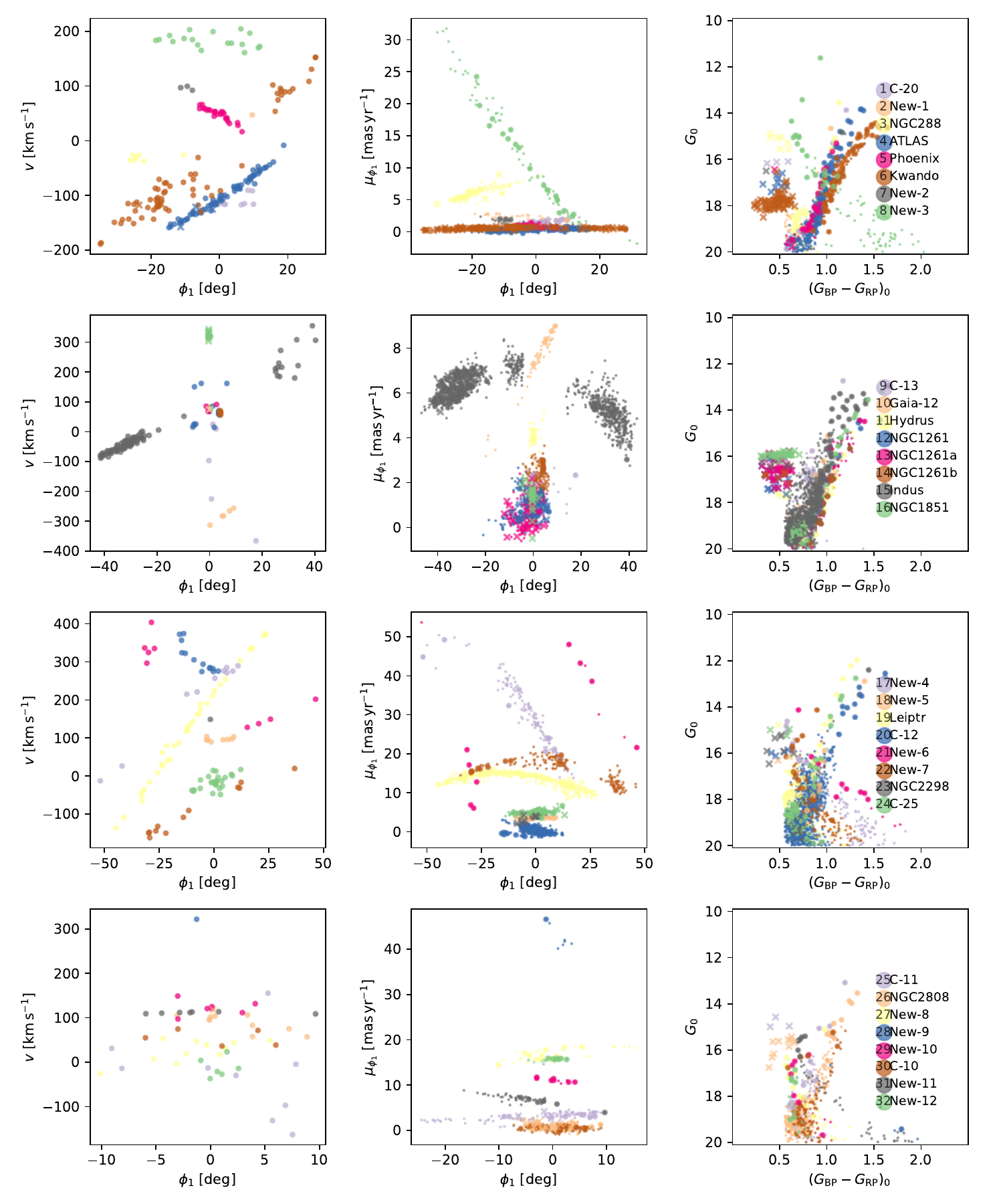}
\end{center}
\caption{Stream kinematic profiles and color-magnitude relations. The left column of panels shows the heliocentric line of sight velocity as a function of angle $\phi_1$ along the streams, while the middle column displays proper motion $\mu_{\phi_1}$ along the $\phi_1$ direction. The color-magnitude properties of the stellar populations can be seen in the right column of panels.}
\label{fig:params1}
\end{figure*}

\begin{figure*}
\begin{center}
\includegraphics[angle=0, viewport= 1 1 720 864, clip, width=\hsize]{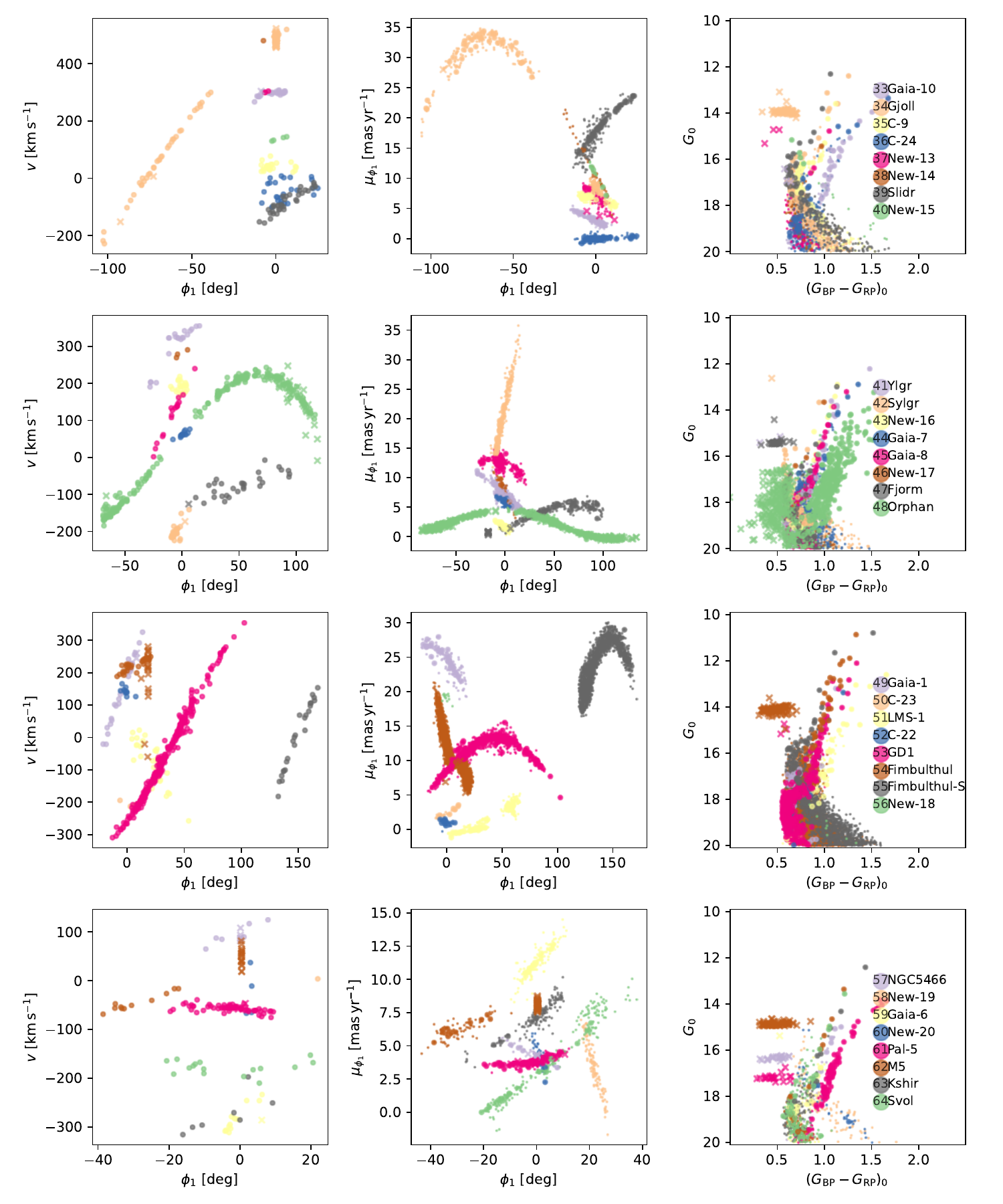}
\end{center}
\caption{Continuation of Figure~\ref{fig:params1}.}
\label{fig:params2}
\end{figure*}

\begin{figure*}
\begin{center}
\includegraphics[angle=0, viewport= 1 215 720 864, clip, width=\hsize]{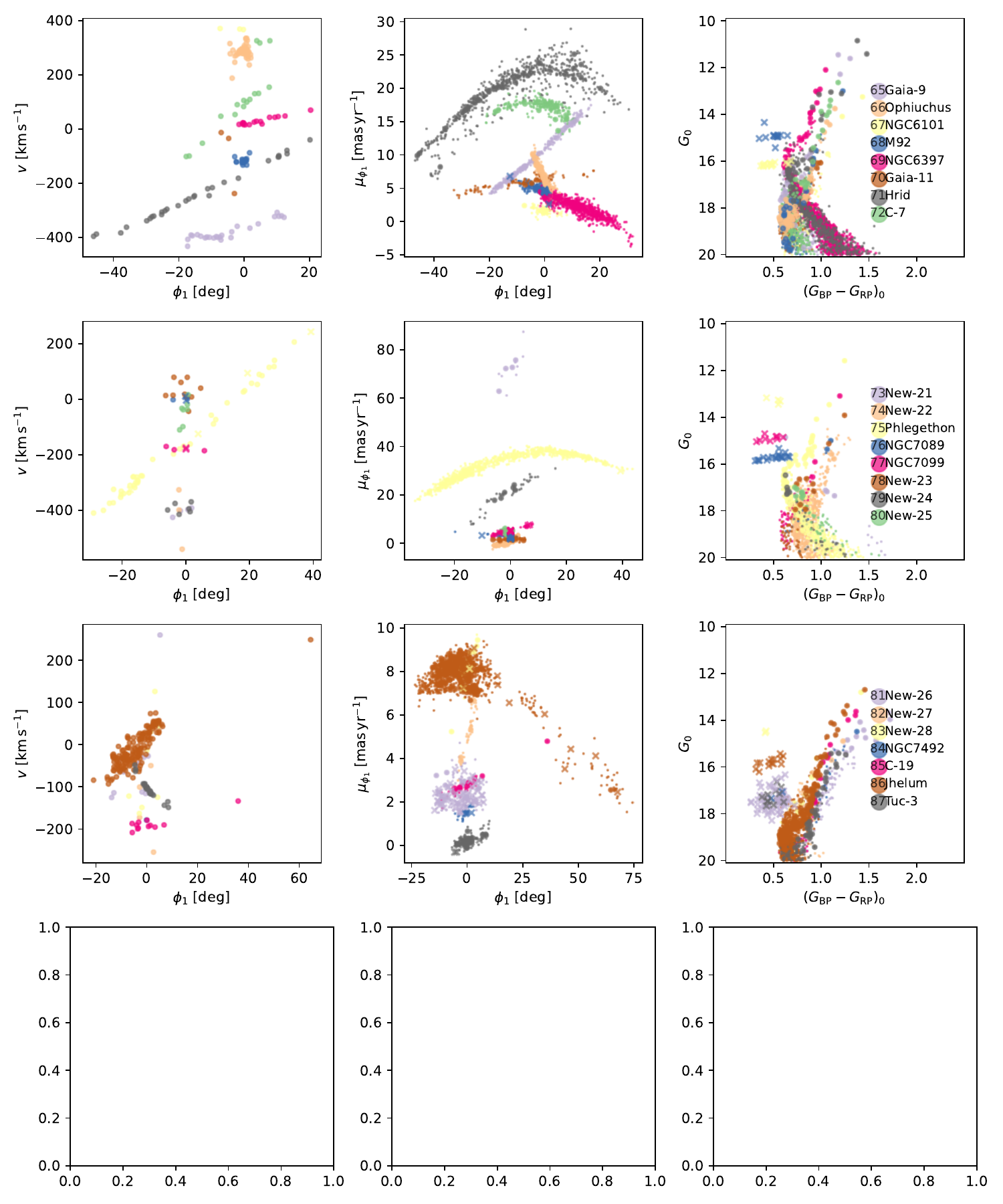}
\end{center}
\caption{Continuation of Figure~\ref{fig:params1}.}
\label{fig:params3}
\end{figure*}

\section{Spectroscopic Observations}
\label{sec:Observations}

We used the VLT/UVES spectrograph \citep{2000SPIE.4005..121D} to follow up selected \texttt{STREAMFINDER} sources detected in the \emph{Gaia} EDR3 and DR3 catalogs. These runs comprise runs 105.20AL.001 (2.5 nights in visitor mode), 110.246A.001 (40 hours service in service mode), and 111.2517.001 (3.6 nights in visitor mode). Our instrumental setup uses the DIC2 dichroic beamsplitter in the ``437+760'' setting, covering the wavelength ranges 3730--4990\AA\ and 5650--9460\AA. To reduce read noise, we binned the CCD in $2\times2$ pixel blocks, which in conjunction with a $1\scnd$0 wide slit yields a spectral resolution of approximately 40,000. Exposure times were selected on a star-by-star basis to reach $S/N\sim3$--5 for the fainter stars in the sample, so as to measure their radial velocities, but we set a minimum exposure time of $5$~min. For the brighter stars, this minimum exposure time also allowed some elemental abundances to be measured. The spectra were reduced with the ``esoreflex'' pipeline using daytime calibration arc lamps and flat-field images, resulting in extracted wavelength-calibrated one-dimensional spectra. 

We also secured observations with the IDS long-slit spectrograph on the 2.5m Isaac Newton Telescope in several runs over the course of 2022. Bright northern hemisphere stream stars were targeted with typically $\sim 1$~hour exposures at $G=16$~mag. The instrument was configured with the RED$+2$ detector, the R1200R grating with a central wavelength set to 8500\AA, a $1\scnd$ wide slit and the GG495 order sorting filter. 
 
The radial velocities of the target stars were measured with the IRAF \texttt{fxcor} algorithm, using the bright and relatively metal-poor star HD~182572 as a radial velocity standard. The UVES spectra were of sufficient quality and resolution to obtain excellent radial velocity measurements with $<1\kms$ uncertainty for stars to $G=18$~mag, while the IDS spectra resulted in velocity uncertainties of $\sim 10\kms$ at $G=16$~mag.

\section{\emph{Gaia} DR3 \texttt{STREAMFINDER} maps}
\label{sec:DR3_maps}

We ran the \texttt{STREAMFINDER} using 7 different SSP templates with metallicity ${\rm [Fe/H]= -2.2}$, $-1.9$, $-1.7$, $-1.5$, $-1.3$, $-1.1$, $-0.7$, and a common age $12.5\Gyr$, as we are primarily aiming to discover halo structures. We allowed the solutions to explore the distance range $[1,100]\kpc$. As in Paper~II, the \emph{Gaia} catalog was limited to sources with $G_0<20$~mag and V-band extinction $A_V<2$~mag, so as to mitigate against variations in extinction and survey depth due to the satellite's scanning law\footnote{To correct for interstellar extinction, we assume that all the dust in the foreground, and interpolate between the pixels of the \citet{1998ApJ...500..525S} maps, adopting the \citet{2011ApJ...737..103S} recalibration with $R_V=3.1$.}. The full sky was processed with the exception of circular regions around satellites; the radius of these masked regions was set to twice the tidal radius for globular clusters and 7 times the half-light radius for satellite galaxies, as detailed in Paper~II.

Figures~\ref{fig:Ball_mul} and \ref{fig:Ball_mub} show the resulting sky maps of the detections of 24,540 (non-RRLyrae) sources with $\ln{\mathcal{\hat{L}}}>19.8$ (note that with the standard log-likelihood the value $\ln{\mathcal{{L}}}=19.8$ corresponds to a $6\sigma$ threshold) colored according to proper motion along the Galactic longitude and latitude directions, respectively. The corresponding distance solutions are shown in Figure~\ref{fig:Ball_dis}, colored according to the distance modulus. Figure~\ref{fig:Ball_vel} displays the 2178 (non-RRLyrae) sources with measured line of sight velocities, while Figure~\ref{fig:Ball_FeH} shows the mean metallicity of the streams (listed in Table~\ref{tab:streams}) as calculated from the mean of the spectroscopic metallicity measurements (or photometric metallicities, for those structures without spectroscopic measurements). These maps display a rich tapestry of crisscrossing streams that possess coherent behaviour in proper motion, line of sight velocity and distance. In Figure~\ref{fig:Ball_sID}, we have selected a contrasting color scheme to allow easier visual disambiguation of the overlapping structures that can be discerned in  Figures~\ref{fig:Ball_mul} to \ref{fig:Ball_FeH}.

To facilitate subsequent analyses, we create a local coordinate system on a per-stream basis by fitting a great circle to each structure using a least squares criterion (we consider that the uncertainties in sky position are negligible). For the streams with known globular cluster progenitors, we require the fit to intersect the globular cluster center, and we also use the right ascension of the cluster to define the zero-point of the great circle coordinates $(\phi_1,\phi_2)$ along and perpendicular to the stream. The coordinate $\phi_1$ is oriented so that the stream's velocity is approximately parallel to it\footnote{The stream velocity vector need not be perfectly parallel to $\phi_1$ due to projection and reflex motion.}, while $\phi_2$ is in the direction perpendicular to the great circle. For stream  \#48 (Orphan-Chenab) we chose the zero-point proposed by \citet{2019MNRAS.485.4726K}, while for stream \#53 (GD-1) we fully adopted the $(\phi_1,\phi_2)$ coordinate system of \cite{2010ApJ...712..260K}. The poles and zero-point of the $(\phi_1,\phi_2)$ coordinates are provided in Table~\ref{tab:streams}. We also estimate the stellar mass of each stream, using the metallicity listed in Table~\ref{tab:streams} together with the fitted stream distance models (calculated as explained in Section~\ref{sec:Stream_Fitting_isolated}), to correct for the missing stars in the stellar population, assuming that the survey is complete to $G_0=20$~mag. These stellar masses are also listed in Table~\ref{tab:streams}, but we stress that they are rough approximations and involve many assumptions including that we have identified the full extent of each stream and that only a single stellar population is present.

Figures~\ref{fig:params1} to \ref{fig:params3} display the individual stream structures in an array of three columns. The first column shows the line of sight velocity as a function of the angle $\phi_1$ along each stream. The proper motion along the $\phi_1$ direction is displayed in the middle column, while the last column displays the color magnitude diagram of the population. The same color scheme as in Figure~\ref{fig:Ball_sID} is employed.

\begin{table*}
\caption{The first 10 rows of the {\tt STREAMFINDER} catalog of 24,540 stars detected in \emph{Gaia} DR3.}
\label{tab:data}
\hskip -3cm
\begin{tabular}{lccccccccccc}
\hline
\hline
DR3 ID & $\alpha$ & $\delta$ & $\varpi$ & $\mu_\alpha$ & $\mu_\delta$ & $G_0$ & $(G_{BP}-G_{RP})_0$ & $v_h$ & $\delta v_h$ & s & $s_{\rm ID}$ \\
 & $\deg$ & $\deg$ & $\mas$ & $\masyr$ & $\masyr$ & mag & mag & $\kms$ & $\kms$ &  &  \\
2746748683780680704 & 0.131771 & 8.022269 & -0.523 & -0.338 & 1.390 & 19.35 & 0.74 &  &  &  & 1 \\
2746696319539384320 & 0.328548 & 7.886247 & 0.045 & -0.683 & 1.425 & 19.48 & 0.59 &  &  &  & 1 \\
2745893989584959488 & 0.714213 & 7.207192 & 0.090 & -0.553 & 1.405 & 15.83 & 1.04 & -115.03 & 0.35 & 13 & 1 \\
2745895810651091456 & 0.714858 & 7.301568 & -0.001 & -0.439 & 1.322 & 18.75 & 0.83 &  &  &  & 1 \\
2745605054248759296 & 0.992495 & 6.646490 & -0.413 & -0.760 & 1.801 & 19.04 & 0.68 &  &  &  & 1 \\
2745591005412035584 & 1.135276 & 6.569216 & -0.055 & -0.251 & 1.656 & 18.00 & 0.89 &  &  &  & 1 \\
2742045419713596416 & 1.655054 & 5.419648 & 0.186 & -0.531 & 1.219 & 19.22 & 0.75 &  &  &  & 1 \\
2742028759537330176 & 1.740425 & 5.219596 & 0.047 & -0.572 & 1.522 & 19.59 & 0.71 &  &  &  & 1 \\
2740941544398892032 & 2.436206 & 4.142149 & 0.545 & -0.651 & 1.321 & 18.60 & 0.79 &  &  &  & 1 \\
2740973155358226432 & 2.751825 & 4.308954 & -0.239 & -0.391 & 1.247 & 19.20 & 0.68 &  &  &  & 1 \\
\hline
\hline
\end{tabular}
\tablecomments{Column 1 provides the \emph{Gaia} DR3 identification of the star, 2--6 list the EDR3 equatorial coordinates $\alpha$ and $\delta$, parallax $\varpi$ and proper motions $\mu_\alpha (* \cos(\delta))$, $\mu_\delta$. The extinction corrected magnitude $G_0$, and color $(G_{BP}-G_{RP})_0$ used in the {\tt STREAMFINDER} are listed in columns 7 and 8. Columns 9 and 10 list the best measured heliocentric line of sight velocity, as derived from the corresponding source ``s'' in column 11. The source identifications ``s'' are: 1=APOGEE, 2=GALAH, 3=\emph{Gaia} RVS, 4=LAMOST, 5=S5, 6=SDSS, 7=BOSS, 8=ESPaDOnS (this work), 9=AAOmega (from \citealt{2017ApJ...842..120I}), 10=FLAMES (from \citealt{2017ApJ...842..120I}), 11=UVES (from \citealt{2009AJ....137.3378O}), 12=EFOSC (this work), 13=UVES (this work), 14=INT (this work), 15=\citealt{2022MNRAS.514.1664Y}, 16=\citealt{2021ApJ...911..149L}, 17=\citealt{2020AJ....159..287C}, 18=\citealt{2018ApJ...866...22L}, 19=\citealt{2018MNRAS.479.5343K}, 20=\citealt{2018ApJ...869..122L}, 21=\citealt{2020ApJ...892..137S}, 22=\citealt{2015ApJ...808..108W}, 23=VIZIER, 24=GES, 25=DESI. Finally, column 12 provides a unique stream identification label.}
\end{table*}

The nature of the detections can be appreciated better by considering first their dynamical properties, so we will next use the streams to understand the global properties of the Milky Way, which will then give us access to their orbits, and then return to a discussion of individual objects in Section~\ref{sec:DR3_Detections}.

\section{Stream Fitting with corrected orbits in an Isolated Milky Way}
\label{sec:Stream_Fitting_isolated}

We begin our modelling efforts by considering the Milky Way to be an isolated axisymmetric galaxy, whose center defines the origin of an inertial frame around which its satellites will orbit. In this context we seek to find the most likely Galactic mass model that admits progenitor orbits that fit the stream data simultaneously. As we have mentioned before, stellar streams do not precisely delineate the orbital path of their progenitors. To overcome this complication we will proceed in an iterative manner, to find plausible functions $\Delta_\Theta(\phi_1)$ in the derived mass model that correct the offset between the stream and the progenitor orbit at $\phi_1$ for each observable phase-space coordinate $\Theta$.

While we could have tried to fit the whole dataset of 87 streams and $>24,500$ stars, to find the best mass model, we restrict ourselves in the present contribution to a conservative sample containing velocity-confirmed members and RR-Lyrae. We further impose that the individual streams should be long ($>10\deg$ in length) containing at least 5 line of sight velocity measurements with a clear linear trend. Since we most probably only detect a small stream segment of the closest streams, we also select those structures at heliocentric distance $>2.5\kpc$ (this limit is somewhat arbitrary and in future work we will re-examine in detail the orbits of the closer streams). We set aside the Orphan stream for the initial analysis, because although it has excellent data, it is probably affected by the Large Magellanic Cloud \citep{2019MNRAS.487.2685E}. We have previously shown \citep{2019NatAs...3..667I} that the ``Fimbulthul'' stream possesses complex structure due to the internal rotation of its progenitor, the massive cluster $\omega$~Cen, and so we decided not to model this structure either. The sample of 29 long and thin streams we fit here are flagged ``1'' in the column titled ``sample'' in Table~\ref{tab:streams}. The dataset comprises 1,397 data points: 1,098 stars with measured line of sight velocities, 5 globular clusters (with phase space parameters derived from the compendia in \citealt{2019MNRAS.484.2832V} and \citealt{2021MNRAS.505.5957B}), and 294 \emph{Gaia} RR~Lyrae (for which we ignore any measured line of sight velocities due to concerns over their variability).

We constructed a similar MCMC orbit fitting algorithm to that presented in \citet{2018ApJ...865...85I}, but altered to allow for simultaneous adjustment to many streams. The MCMC driver algorithm we use was originally developed by \citet{2011ApJ...738..186I}, based on the ``walkers'' scheme proposed by \citet{2010CAMCS...5...65G}. The ``walkers'' algorithm (that simulates a population of agents sampling the likelihood at different parameter space locations) is very efficient as it is affine-invariant and adapts automatically to the multi-dimensional covariance of the problem.

The Milky Way potential is generated via multipole expansion from analytical double exponential density models (to represent the disk components) or double power law density models (to represent the spheroidal components), following \citet{Dehnen:1998tk}. We allow each disk component to have up to three parameters: central surface density $\Sigma_d$, a scale length $h_R$ and scale height $h_z$ 
\begin{equation}
\rho_d(R,z) = \frac{\Sigma_d}{2 h_z} \exp \Bigg( - \frac{R}{h_R} - \frac{|z|}{h_z} \Bigg) \, .
\label{eqn:disk}
\end{equation}
In contrast we allow the spheroidal components to have 6 parameters: central density $\rho_0$, density flattening $q_m$, inner power-law slope $-\beta$, outer power-law slope $-\gamma$, scale radius $r_0$ and truncation radius $r_t$
\begin{equation}
\rho_s(s) = \rho_0 \Bigg( \frac{s}{r_0} \Bigg)^{-\gamma} \Bigg( 1 + \frac{s}{r_0} \Bigg)^{\gamma-\beta} e^{-s^2/r_t^2} \, ,
\label{eqn:spheroid}
\end{equation}
where $s = \sqrt{R^2+ z^2/q_m^2}$ is an ellipsoidal coordinate. Each orbit obviously requires 6 parameters, however for those streams with known progenitor we consider their center to be perfectly determined, leaving 4 free parameters, while for all others we anchor the orbit at $\phi_1=0\deg$ (except for GD-1, as discussed below), leaving 5 free parameters. 

Hence a full parameter exploration of the problem would have in excess of 150 parameters and would therefore be very challenging for an MCMC approach. However, the orbits can be fit independently of one another, which motivated our choice to implement a nested MCMC search. The outer MCMC loop simply proposes the Galaxy model parameters (and the components of the Solar peculiar motion vector, if so desired). The inner loop adopts those Galaxy model parameters and attempts to fit each stream independently. We implemented an option for two algorithm variants for the inner loop. The first option is another independent MCMC search with 500 iterations, where every 25 iterations of the inner loop the step size is adjusted to try to maintain a 25\% acceptance ratio (independently for each stream). The best solution and step size (for each stream) are recorded as if they were parameters of the outer loop and reused in the next run of the inner loop. The second option is to use a Nelder-Mead ``downhill simplex'' search \citep{1992nrfa.book.....P}. We adopted a log-likelihood tolerance of $10^{-4}$, which gave robust results in our tests (each such search requires $\approx 1,000$ likelihood evaluations). The Nelder-Mead method can be sensitive to the chosen starting point of the search, so we re-run it with different randomly-drawn positions until the best likelihood solution is found again to within the chosen tolerance. We found that both the MCMC and downhill simplex options for the inner loop give similar results, but the latter is computationally cheaper to reach a given accuracy, while the former allows one to estimate confidence intervals on the stream parameters. The results shown below were calculated using the downhill simplex as the optimization method of the inner loop\footnote{Although we first employ the MCMC method in the inner loop to obtain an estimate of the uncertainties of the stream model parameters, which are then reused in the generation of the (Gaussian) random starting points for the downhill simplex method.}.

The outer loop uses 96 MCMC ``walkers'', and we adjust the stretch factor of the proposed walker steps dynamically (see \citealt{2010CAMCS...5...65G}) every 25 iterations aiming to obtain a 25\% acceptance ratio. We run the algorithm for at least 200,000 ``burn-in'' iterations, discard that data and then continue for a further 500,000 iterations (i.e. for a total of 500,000$\times$1,000 evaluations of the combined likelihood of the 29 streams, with each orbit being sampled at a resolution of up to 10,000 points). For every step of the outer loop, the Galaxy parameters and the stream parameters of that step are recorded as the MCMC chain that is the desired final output of the algorithm\footnote{Due to the low cost of the individual orbit integrations, which nevertheless require a non-negligible amount of memory to store their paths, this nested procedure proved to be very difficult to parallelize efficiently with standard parallelization libraries such as OPENMP/MPI. Consequently, we developed a custom parallelization scheme using persistent jobs (eliminating overheads due to memory allocation and data ingestion), one per machine core, that wait to be handed the different input parameters farmed out by the master process that handles the MCMC exploration.}. We consider that the MCMC exploration converged, when the parameter statistics of the first and second half of the 500,000 iteration-long chains are statistically identical. One such optimization of the Galaxy and stream parameters takes typically $\approx$ 5,000 CPU hours.

The log-likelihood objective function we choose to fit to the sample attempts to model the expected combination of thin streams, each enveloped by a wider ``cocoon'' population and a small amount of contamination. The adopted objective function is therefore:
\begin{align}
\begin{split}
{\ln} \mathcal{L} = 
&\sum_{\rm{data}} {\ln} \big[ \, (1-\zeta-\chi) \mathcal{P}_{\rm stream}^{\rm fit}  + \zeta \mathcal{P}_{\rm cocoon} + \chi \mathcal{P}_{\rm contam} \big] \\
& + {\ln} \mathcal{L}_{\rm ancillary}
\label{eqn:likelihood_fit}
\end{split}
\end{align}
where $\zeta$ and $\chi$ are the fraction of stars in the cocoon and contamination, respectively. We conservatively set $\chi=0.01$. In our examination of the GD-1 stream in Section~\ref{sec:stream53} below we deduce $\zeta=0.27$ for that system; however to be conservative we set $\zeta=0.1$ for all streams in our MCMC analysis. The stream model
\begin{equation}
\mathcal{P}_{\rm stream}^{\rm fit}  = \mathcal{P}_{\rm width} \times \mathcal{P}_{\rm color} \times \mathcal{P}_\mu \times \mathcal{P}_{\rm \varpi} \times \mathcal{P}_{v}
\end{equation}
is exactly the same stream probability criterion as used in the \texttt{STREAMFINDER} (Eqn.~\ref{eqn:prob}) but ignores the luminosity function information (as we are assuming at this point that the stars have been appropriately selected). Also, in contrast to the \texttt{STREAMFINDER}, the orbital track is corrected by the function $\Delta_\Theta(\phi_1)$ introduced above, so as to better approximate the corresponding stream (these functions will be discussed in more detail in Section~\ref{sec:Orbit_Stream_Correction_functions} below). The cocoon model is assumed to be identical to the stream model, except that we widen the model dispersions by a factor of 5 in line of sight velocity (approximately the ratio inferred in Section~\ref{sec:stream53} for GD-1), proper motion, distance and width on the sky (i.e. equivalent to an intrinsic $25\kms$ velocity dispersion and an intrinsic $250\pc$ stream width). In the present context the contamination is clearly a very complex multi-dimensional function and further compounding the complexity of our model does not seem like a good idea. So instead we adopted a pragmatic contamination model, inspired by the ``good-and-bad data model'' of \citet{sivia2006data} (their Section 8.3.2). According to their model the ``good'' fraction have Gaussian residuals $R \equiv (F-D)/\sigma$ between the data $D$ and the model prediction $F$ given the expected dispersion $\sigma$ (due to the model spread and the measurement uncertainties), while for the ``bad'' fraction these residuals are simply scaled down by some factor $\tau$ (\citealt{sivia2006data} use the symbol $\gamma$). In this way we choose the contamination model to be exactly the same as the stream model (not the cocoon), but with $\tau=10$, effectively inflating the $\sigma$ of the  stream model by that factor (for bright stars with small measurement uncertainties this corresponds to a velocity dispersion of $50\kms$ and a spatial dispersion of $0.5\kpc$). This renders the objective function (Eqn.~\ref{eqn:likelihood_fit}) much less sensitive to the presence of outliers, even though the assumed contamination fraction (1\%) is quite small.

The ${\ln} \mathcal{L}_{\rm ancillary}$ term in Eqn.~\ref{eqn:likelihood_fit} is used to include constraints from ancillary data sets, all in the form of simple $\chi^2$-like terms. So for instance, when we force the rotation curve of the proposed potential models $v_{c}^{\rm model}$ to be consistent with Galactic rotation curve measurements $v_{c}^{\rm Eilers}$ and associated uncertainties $\delta v_{c}^{\rm Eilers}$ \citep{2019ApJ...871..120E}, we include
\begin{equation}
- \sum_{\rm{data}} \frac{1}{2} \Bigg (\frac{ v_{c}^{\rm Eilers} - v_{c}^{\rm model}}{\delta v_{c}^{\rm Eilers}}  \Bigg )^2
\end{equation}
in ${\ln} \mathcal{L}_{\rm ancillary}$. We limit the \citet{2019ApJ...871..120E} rotation curve data to $R<15\kpc$, as the outer stellar disk is both substantially warped and flared beyond that radius (see, e.g. \citealt{2006A&A...451..515M}).

Further constraints on the radial mass distribution are available thanks to studies of the terminal velocity of \ion{H}{1} in the inner Galaxy. The high-resolution profiles of \citet{2007ApJ...671..427M,2016ApJ...831..124M} are conveniently measured with respect to the Local Standard of Rest, so we can directly compare them to the predictions of the mass models 
\begin{equation}
v_{\rm term}^{\rm model}(\ell)={\rm sgn}(\sin(\ell)) v_c(R | sin(\ell)|)-v_c(R_\odot) \sin(\ell) \, .
\end{equation}
The \ion{H}{1} terminal velocity curves as a function of angle show substantial localised bumps on scales of a few degrees (see, e.g. Figure~1 of \citealt{2016ApJ...831..124M}), implying correlated motion that our large-scale mass model cannot possibly reproduce. To mitigate against this, we decided to average the profiles over $2\deg$ intervals. Also, following \citet{2017MNRAS.465...76M} we retain only those measurement with $|\sin(\ell)|>0.5$ to avoid regions that are heavily affected by the presence of the Galactic bar. We assume an uncertainty of $6.2\kms$ of the terminal velocity measurements, as derived from the scatter between the measurements and the linear model fit by \citet{2016ApJ...831..124M}.

We similarly add the datum of $71\pm 6 \msun {\rm pc}^{-2}$ for the vertical force at $z=1.1\kpc$ above the Galactic midplane measured by \citet{1991ApJ...367L...9K}.  We also include the stellar density measurements towards the north Galactic cap of \citet{2017ApJ...848..129I} decomposed into thin and thick disk components according to photometric metallicity; the sample at Galactic latitude $b>70\deg$ is used (shown in their Figure~12f) truncated to $z<5\kpc$ as their thin disk profile is noisy at larger $z$. As indicated in \citet{2017ApJ...848..129I} the stellar number density profiles are transformed into {\it relative} mass density profiles with simulations that use PARSEC isochrone models and take into account the selection function of their survey. They adopt an average age of $5\Gyr$ and $10\Gyr$ for the thin and thick disks, respectively, and note that the difference in the computed density corrections between these two ages is only 20\%, implying that their measured relative vertical density profiles are not very sensitive to those reasonable assumptions on population age. The \citet{2017ApJ...848..129I} study chose not to constrain the normalization of the density profiles, so in the present work the thin and thick disk components are fit to those vertical profiles data while allowing for a free multiplicative constant.

Neither the \citet{2019ApJ...871..120E} rotation curve measurements nor the long streams we have selected probe the inner regions of the Milky Way. We therefore chose for the present work to simply adopt the Galactic bulge component of the model previously fit by \citet{2017MNRAS.465...76M}. The truncation radius of the halo component is fixed at $r_t=1000\kpc$, far beyond the expected virial radius of the Milky Way. 

In our explorations we realized that the dataset provides constraints on the peculiar velocity of the Sun that seem plausible given published studies, but that these constraints are somewhat degenerate with the freedom we allow to models of the potential. We defer an investigation into this issue to a future contribution, and therefore fix the peculiar velocity of the Sun to the values found by \citet{2010MNRAS.403.1829S}. Likewise, we also fix the distance to the Galactic center to $R_\odot=8.178\kpc$ \citep{2019A&A...625L..10G}, and ignore the small uncertainties on this value.

We calculate the $M_{200}$ mass of the models, defined as the mass out to a radius of $r_{200}$ where the average interior density of the model is 200 times the critical density $\rho_c=3 H^2/8\pi G$ of the universe, and we assume a Hubble parameter $H=70 \kms \mpc^{-1}$. Clearly, the calculated $M_{200}$ value involves a long extrapolation from the region constrained by the chosen sample of 29 streams. Nevertheless, we consider $M_{200}$ to be a useful constraint on the models, and require the solutions to be in the range $[0,2]\times 10^{12}\msun$. The upper bound on $M_{200}$ of $2\times 10^{12}\msun$ corresponds to half of the Local Group mass of $4\times 10^{12}\msun$ (95\% upper limit) estimated by \citet{2014MNRAS.443.2204P}. The other bounds are that the masses of all components, as well as the halo density flattening parameter $q_{m, h}$, should be positive\footnote{These bounds are implemented in the MCMC procedure simply by subtracting a very large number from $\ln\mathcal{L}$ if the limits are crossed.}. The thin and thick disk scale lengths are required to be in the range $[1,10]\kpc$, we impose that the thin disk scale height should be in the range $[0.05,0.5]\kpc$ and the thick disk scale height should be in the range $[0.05,5]\kpc$. Finally, we require that the thick disk scale height should be larger than that of the thin disk.

In our first modelling efforts we considered what we thought was the simplest useful Galaxy model, containing only a fixed bulge, a halo, and a single disk component. However, we noticed that if we left out the \citet{2017ApJ...848..129I} vertical density constraint, the resulting best solutions required a disk component with an unrealistically large scale height ($\sim 1\kpc$). This result is perhaps not surprising in retrospect, since the selected streams mostly reside in halo regions at high extra-planar distance, where the effect of the thin disk is less pronounced. So, evidently, a minimal Galaxy model for our purposes also entails including a thick disk component. Because our data provide poor constraints on the potential at low Galactic latitude, we also adopt the two gas disk models proposed by \citet{2017MNRAS.465...76M}, which include an atomic gas disk of scale height $85\pc$ and mass $1.1\times10^{10}\msun$, as well as a molecular gas disk of scale height $45\pc$ and mass $1.1\times10^{9}\msun$.

Before showing our results with this model we will next present a refined --- but computationally much more costly --- method that uses test particles to follow tidal dissolution in a moving Local Group potential. This model is also used to calculate the correction functions $\Delta_\Theta$ employed in the orbit fitting described above.

\section{Stream Fitting with test particles in the Local Group}
\label{sec:Stream_Fitting_test_particles}

Contrary to the assumptions underlying the models presented in the previous section, the Milky Way is of course not an isolated galaxy, but a member of the Local Group, the other major components of this grouping being the Andromeda and Triangulum galaxies approximately $800\kpc$ away, as well as the Large Magellanic Cloud, the Small Magellanic Cloud and the Sagittarius dwarf galaxy that reside within the halo of our Galaxy. Predicting the paths of stellar streams in the Local Group requires some consideration of the combined gravitational effect of these six bodies, and perhaps additional mass concentrations depending on the particular path of the stream progenitor.

Yet, running full N-body simulations of the Local Group with a resolution sufficient to fit the observed stellar streams is far beyond current computational abilities, so we need to make some simplifying steps to tackle the problem. Here, we will assume that the major galaxies can each be modelled as moving potentials with a mass distribution that does not vary in time when viewed from their center of mass. Since this assumption becomes worse further back in time, we will only consider integrating backwards for at most $5\Gyr$. 

We integrate the path of the Local Group members using a symplectic leapfrog scheme with direct summation of the forces from all other bodies. The major galaxies (the Milky Way, Andromeda and the LMC) are assumed to be truncated at their (static) virial radius. This allows us to treat the bodies as point masses when considering their effect on the other Local Group galaxies that are beyond the virial radius. When including M31 or M33 (whose influence we neglect in the present work), we also take into account a Dark Energy term, using an additional acceleration term $\ddot{\bf{x}}=H_0^2 \Omega_\Lambda \bf{x}$, where $\bf{x}$ in this case is the radial vector from the center of mass of the Local Group.

The dynamics of the LMC and Sagittarius galaxies are particularly difficult because we expect their orbits to have been substantially affected by dynamical friction. Here we use the Chandrasekhar dynamical friction model to integrate the massive bodies backwards in time, and adopt the formula of \citet{2022MNRAS.511.2610C} to approximate the Coulomb Logarithm. We have confirmed with several spot-tests that the \citet{2022MNRAS.511.2610C} method predicts well the behavior in N-body simulations of the spiralling-in path of the LMC when the Milky Way is represented by a live NFW halo.

\begin{figure}
\begin{center}
\includegraphics[angle=0, viewport= 1 10 460 345, clip, width=\hsize]{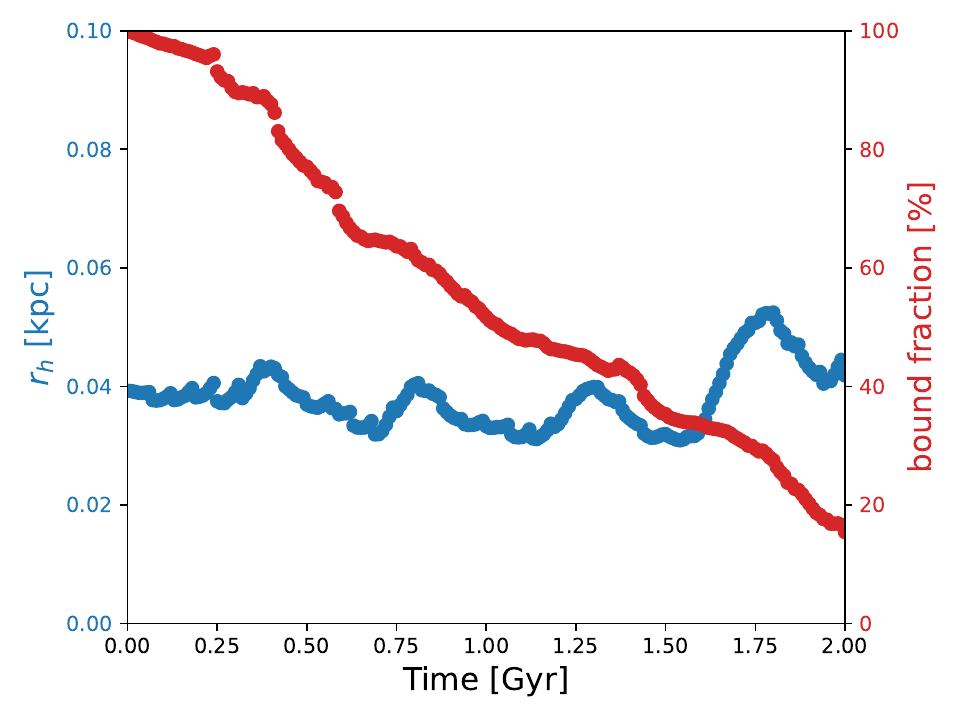}
\end{center}
\caption{Tidal dissolution of an N-body GD-1 progenitor model, showing the approximate constancy of the half mass radius as the system loses almost all its mass. For this experiment, a Plummer sphere of $10^5$ particles with a scale radius $a=50\pc$ and a mass of $3\times10^4\msun$ was evolved using the FalcON \citep{2002JCoPh.179...27D} N-body integrator. A softening length of $1\pc$ was adopted.}
\label{fig:rhalf}
\end{figure}

The orbits of the stream progenitors are integrated backwards in time for a time $T$ in this dynamic Local Group potential, starting from a trial present day phase space coordinate. At the initial time $-T$, we instantiate a set of $N=$10,000 mass-less test particles that follow a spherical Plummer model (parameters: mass, scale radius)\footnote{For streams that have dwarf galaxy progenitors an exponentially-truncated NFW model would be more appropriate \citep{2021MNRAS.505...18E}, and would be very simple to implement into this machinery.}. We then integrate this system forward in time until the present day. However, instead of following the full N-body evolution of the progenitor, we approximate tidal dissolution as follows. We use the analytic potential of the progenitor to approximate the evolving system, so that the forces on the test particles are the sum of the forces from the Local Group galaxies, plus the force due to the progenitor. If a particle moves beyond 10 scale radii from the progenitor, it is considered to be lost, and we decrease the mass of the progenitor by $1/N$. In contrast, we assume that the Plummer scale radius $a$ does not change during the tidal disruption. The advantage of this scheme is that it has saved us from having to follow the internal N-body dynamics of the progenitor, and it can also be parallelized trivially since the test particles are independent of each other ($\mathcal{O}(N)$ computational complexity). Our method is inspired by those of \citet{2012A&A...546L...7M,2023A&A...673A..44F}, but we implement mass loss in the progenitor, as well as massive perturbers. \citet{2012A&A...546L...7M} demonstrated that test-particle integration in non-varying host globular clusters can reproduce very similar stream extent and morphology as full N body simulations, but for the majority of the \texttt{STREAMFINDER} streams we need to be able to account for the full dissolution of the progenitor.

The validity of our simplifying approximation of fixing the Plummer scale radius can be judged from Figure~\ref{fig:rhalf}, where we show a separate N-body experiment of the evolution of a plausible GD-1 stream progenitor. For this test we use the Galactic potential model \#1 of \citet{Dehnen:1998tk} and integrate therein a Plummer model for $2\Gyr$. Over this period of time where the system loses almost all its mass (red profile) the half mass radius $r_h = (2^{2/3}-1)^{-1/2} a$ (blue profile) remains close to constant. An equivalent result is discussed in \citet{2023arXiv231110134E} for collision-less systems that follow an exponential density profile, suggesting that our simplifying assumption of $a = \mathrm{const}$ is not unique to our choice of progenitor density profile. Note, however, that our idealized setup does not capture the effects of potential internal collisional dynamics on the progenitor structure.

We use exactly the same optimization machinery and the same objective function for the stream fits as for the orbit fits in Section~\ref{sec:Stream_Fitting_isolated} (except of course that the correction functions $\Delta_\Theta$ are ignored). Unfortunately this procedure of integrating test-particles is substantially more computationally expensive to optimize: most importantly we have to follow $N$ times more orbits, and there are two more parameters per stream (the progenitor mass and scale radius of the Plummer model), and the global potential requires 7 additional parameters (mass and initial phase space position) for each Local Group galaxy that we aim to include the potential of. While it would be possible to reduce $N$, there needs to be a sufficient number of tidally disrupted particles to model a stream, and our experiments showed that below $N \approx 10^4$ the stream model could become too noisy to allow a reliable comparison to data. 

In the present contribution we will only attempt a limited exploration of the parameter space of this complex model, and defer a full stream-fitting analysis with this machinery to a future contribution. 

\begin{figure}
\begin{center}
\includegraphics[angle=0, viewport= 1 10 450 450, clip, width=\hsize]{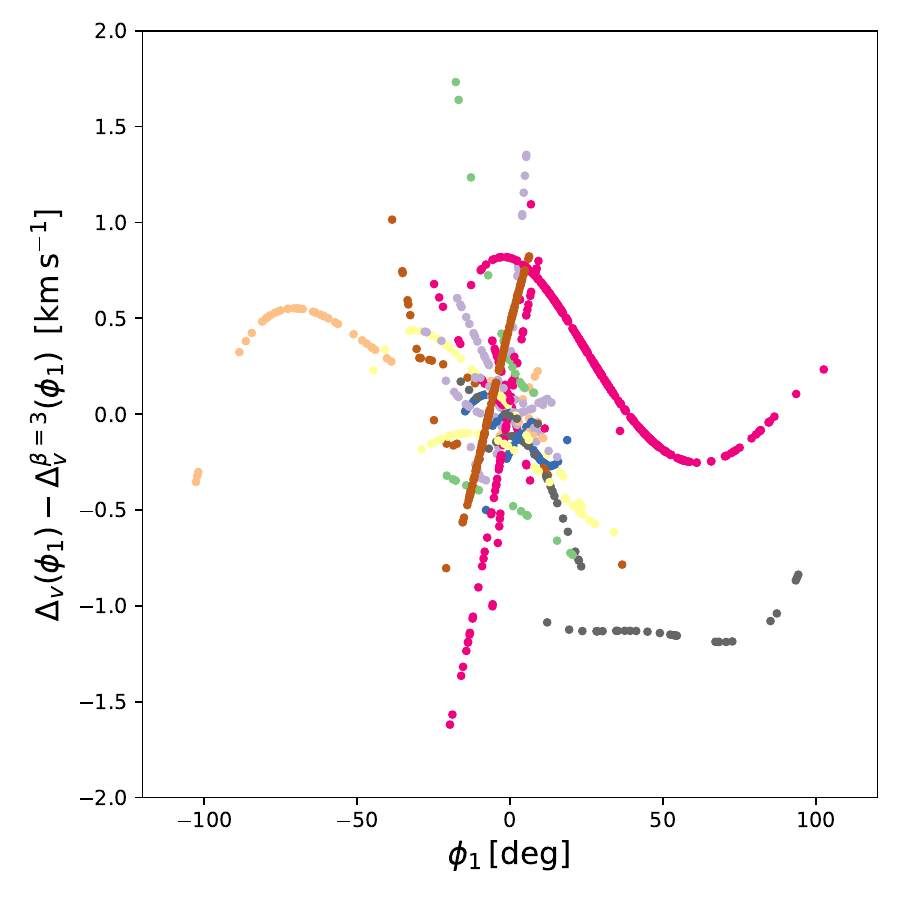}
\end{center}
\caption{Difference between the line of sight velocity correction functions $\Delta_v(\phi_1) = v_{\rm stream}-v_{\rm orbit}$ corresponding to the most likely Milky Way model and that resulting from imposing $\beta_h=3$. It can be seen that the correction functions differ by only $\approx 1\kms$. The points are plotted at the $\phi_1$ locations of the stars in the 29 selected streams, and we again use the color map of Figure~\ref{fig:Ball_sID}.}
\label{fig:Delta}
\end{figure}

\section{Orbit-Stream Correction functions}
\label{sec:Orbit_Stream_Correction_functions}

The test particle model introduced in the previous section is immediately useful to us as a means to calculate the $\Delta_\Theta(\phi_1)$ correction functions needed for the corrected-orbit fitting described in Section \ref{sec:Stream_Fitting_isolated}. We started by running the test particle model optimization procedure with the Milky Way mass parameters fixed at the values found by \citet{2017MNRAS.465...76M}, and with all other Local Group galaxies set to zero mass. For the five systems with known globular cluster progenitors (Gj\"oll/NGC~3201, Fj\"orm/M68, NGC~5466, Palomar~5, and M5, i.e. streams 34, 47, 57, 61 and 62) in our sample of 29 streams, we fixed the mass and half-mass parameters at the values measured by \citet{2018MNRAS.478.1520B}, and also fixed the line of sight velocities at the values given by \citet{2019MNRAS.484.2832V}; the distances and proper motions were allowed to vary within the uncertainties set by the measurements of \citet{2021MNRAS.505.5957B} and \citet{2019MNRAS.484.2832V}, respectively. For GD-1 we set the progenitor position to $\alpha=157\degg74$, as will be justified below in Section~\ref{sec:Stream_Fitting_results}. For the remaining 23 streams, we unfortunately have no constraint on the position of the progenitor, so it is set at the estimated mid-point of the stream ($\phi_1=0\deg$). The masses of the structures are set to the stellar mass estimates listed in Table~\ref{tab:streams}, but we conservatively set a floor of $2\times 10^4 \msun$ to these values (approximately the mass of our estimate of the  Palomar~5 progenitor). For these streams we also assume a progenitor Plummer scale length of $30\pc$. The correction function $\Delta_\Theta(\phi_1)$ is then calculated independently for each observable $\Theta$ (i.e. sky position, distance modulus, line of sight velocity and proper motions) as a fourth order polynomial fit to the best fit stream minus the corresponding progenitor orbit. This $\Delta_\Theta(\phi_1)$ is subsequently used in the corrected-orbit fitting algorithm to find an improved Milky Way mass model. We repeated this procedure a further two times, re-deriving $\Delta_\Theta(\phi_1)$ for the best mass model, and re-running the MCMC search of Section \ref{sec:Stream_Fitting_isolated}. We stopped at the third iteration as the derived Milky Way parameters were statistically identical to those in the second iteration. 

The procedure we have devised makes several simplifications to speed up the calculations, including the use of a single correction function $\Delta_\Theta(\phi_1)$ for each stream during an MCMC run. Ideally, a new correction function should be calculated for each stream for every set of Milky Way and stream parameters examined in the MCMC search\footnote{Of course this would be pointless in practise, as it would be more accurate and more computationally efficient to simply use the test-particle machinery developed in Section~\ref{sec:Stream_Fitting_test_particles} and forgo the corrected-orbit fitting approach altogether.}. In Figure \ref{fig:Delta} we show the difference between the correction functions in the best-fit model, and in the best model when imposing a halo outer power-law parameter of $\beta_h=3$ (so as to be consistent with an NFW profile). In these two quite different Milky Way models the correction functions are very similar, within $\approx 1\kms$ in line of sight velocity (which is the easiest observable to interpret when comparing streams, as there is no distance dependence).

\section{Stream fitting results}
\label{sec:Stream_Fitting_results}

\begin{figure}
\begin{center}
\includegraphics[angle=0, viewport= 1 10 576 864, clip, width=\hsize]{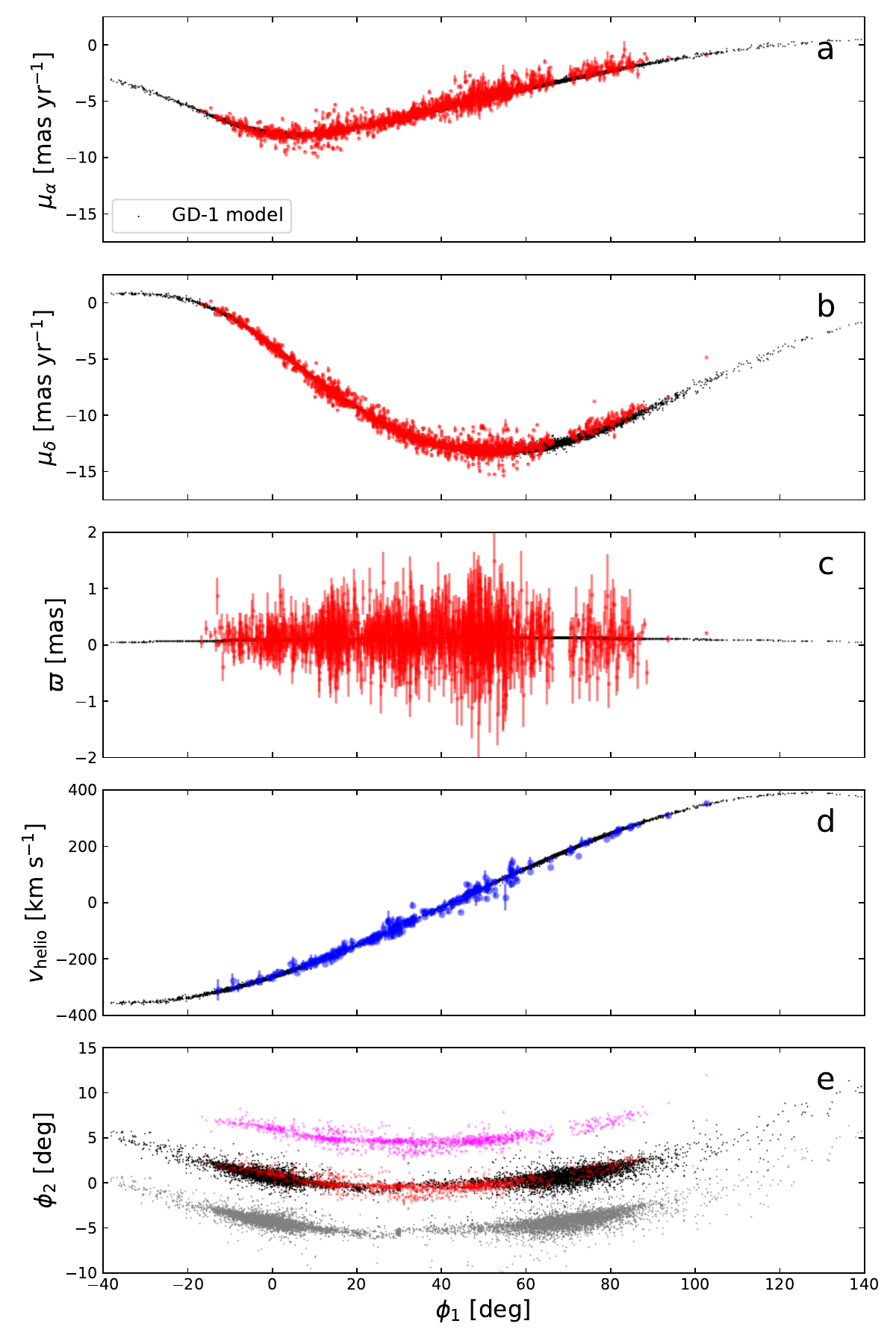}
\end{center}
\caption{Comparison of the GD-1 data (stream \#53, red points) with a best-fit model (black points) in a potential model that is constrained to have fixed halo power law parameters $\gamma_h=1$ and $\beta_h=3$, fixed scale heights of $0.3\kpc$ and $0.9\kpc$ respectively for the thin and thick disk components, together with a fixed thick disk scale length of $3.6\kpc$ (all these parameters are also fixed in the \citealt{2017MNRAS.465...76M} model). The model fits the halo central density, halo scale parameter, the disk scale length and the masses and relative fractions of the thin and thick disks. Although all GD-1 data are displayed, only those stars with measured radial velocity (blue points) were used in the fit. The GD-1 model has an initial mass of $3\times 10^4\msun$, a Plummer scale length of $50\pc$, and is integrated over $2.5\Gyr$. To aid visual comparison, the sky positions (e) include a version of the model shifted by $-5\deg$ (grey) as well as the data shifted by $+5\deg$ (magenta) in $\phi_2$. The gap in the data visible in all panels at $\phi_1\approx 68\deg$ is a consequence of our choice in running the \texttt{STREAMFINDER} to excise \emph{Gaia} data around dense sources (in this case M~67).}
\label{fig:panels_GD1}
\end{figure}

We begin by attempting to fit a static axisymmetric model to the Galaxy ignoring any stellar stream data; to this end we use the objective function in Eqn.~\ref{eqn:likelihood_fit}, with the first term on the right hand side set to zero (i.e. we only fit to the ancillary data). For this first experiment, we fix the halo power law parameters to $\gamma_h=1$ and $\beta_h=3$ so as to reproduce the expected behavior of dark matter halos \citep{1997ApJ...490..493N}, set the halo matter flattening to $q_{m, h}=1$, set the thin and thick disk scale heights to $0.3\kpc$ and $0.9\kpc$ respectively, and set the thick disk scale length to $3.6\kpc$; these are all choices that were adopted by \citet{2017MNRAS.465...76M}. Our main difference with respect to that study is that we are using the \citet{2019ApJ...871..120E} rotation curve measurements. We find a thin disk scale length of $h_{R, d}=2.22\pm0.07 (2.53\pm0.14) \kpc$, a halo scale radius of $r_{0, h}=12.4 \pm 2.0 (19.0\pm4.9) \kpc$, a total stellar mass of $M_{\star}=4.63\pm0.33 (5.43\pm 0.57) \times 10^{10}\msun$, $M_{200}=0.86\pm0.09 (1.3\pm0.3) \times 10^{12} \msun$ and a mass interior to $50\kpc$ of $M_{R<50}=0.41\pm0.02 \times 10^{12} \msun$, where the values in brackets are those reported by \citet{2017MNRAS.465...76M}, when available. Figure~6 of \citet{2017MNRAS.465...76M} shows their posterior PDF of $M_{R<50}$ peaking at a considerably higher mass of $M_{R<50}\approx 0.51 \times 10^{12} \msun$. Our aim with this test was to obtain a baseline for the following streams analysis, and to explore how the improved data changes these inferences.

\begin{figure*}
\begin{center}
\includegraphics[angle=0, viewport= 1 1 1750 1750, clip, width=\hsize]{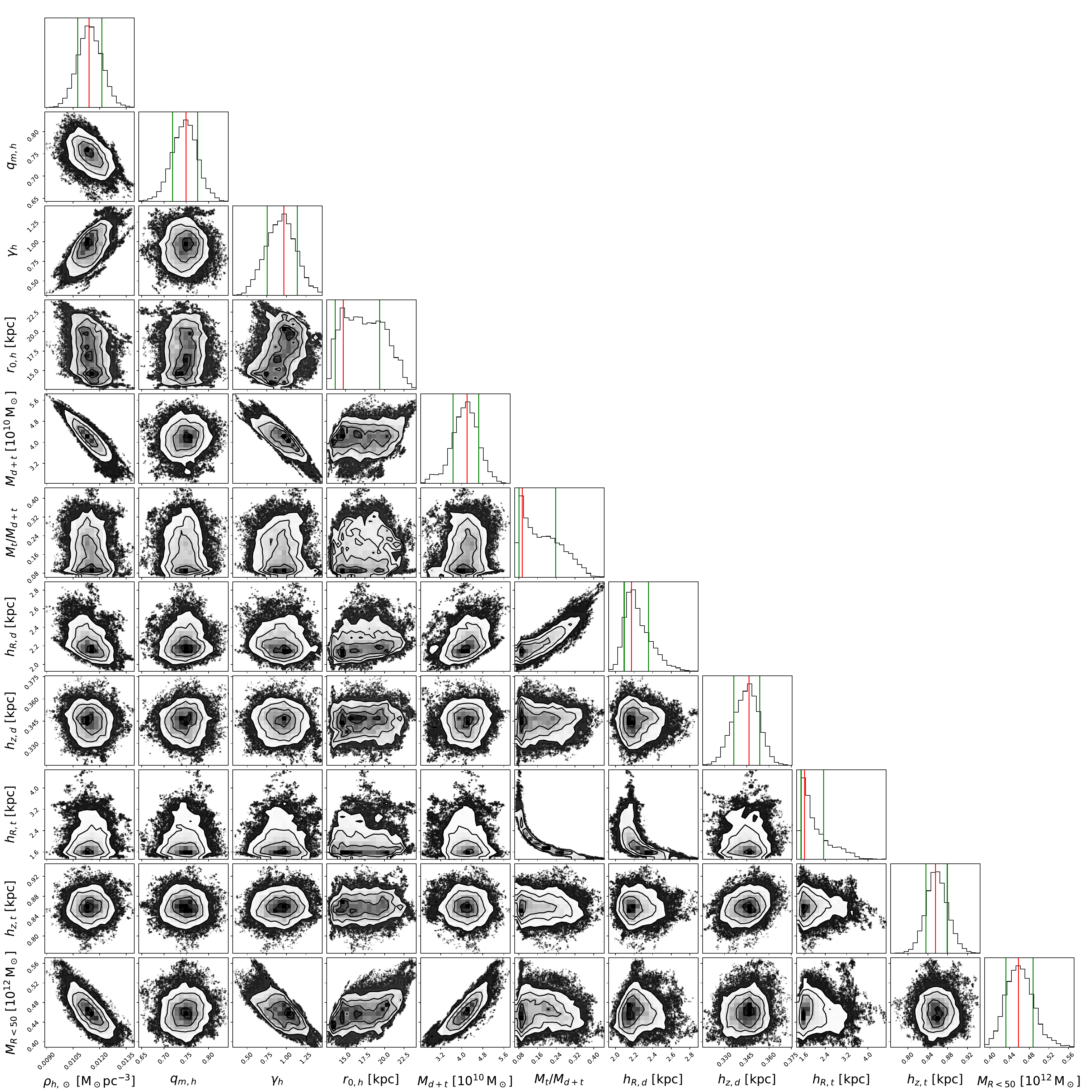}
\end{center}
\caption{Correlations between Milky Way parameters in a 6-component Galaxy model with fixed bulge and gaseous disks, but free halo (4 parameters, with fixed $\beta_h=3$), thin disk (3 parameters) and thick disk (3 parameters). The useful quantity $M_{R<50}$ is not a fitting parameter, but is derived from the mass model determined by the other parameters. This mass model determines the potential in which 29 stream structures are fit.}
\label{fig:corner_beta3}
\end{figure*}

\begin{figure*}
\begin{center}
\includegraphics[angle=0, viewport= 1 1 1900 1900, clip, width=\hsize]{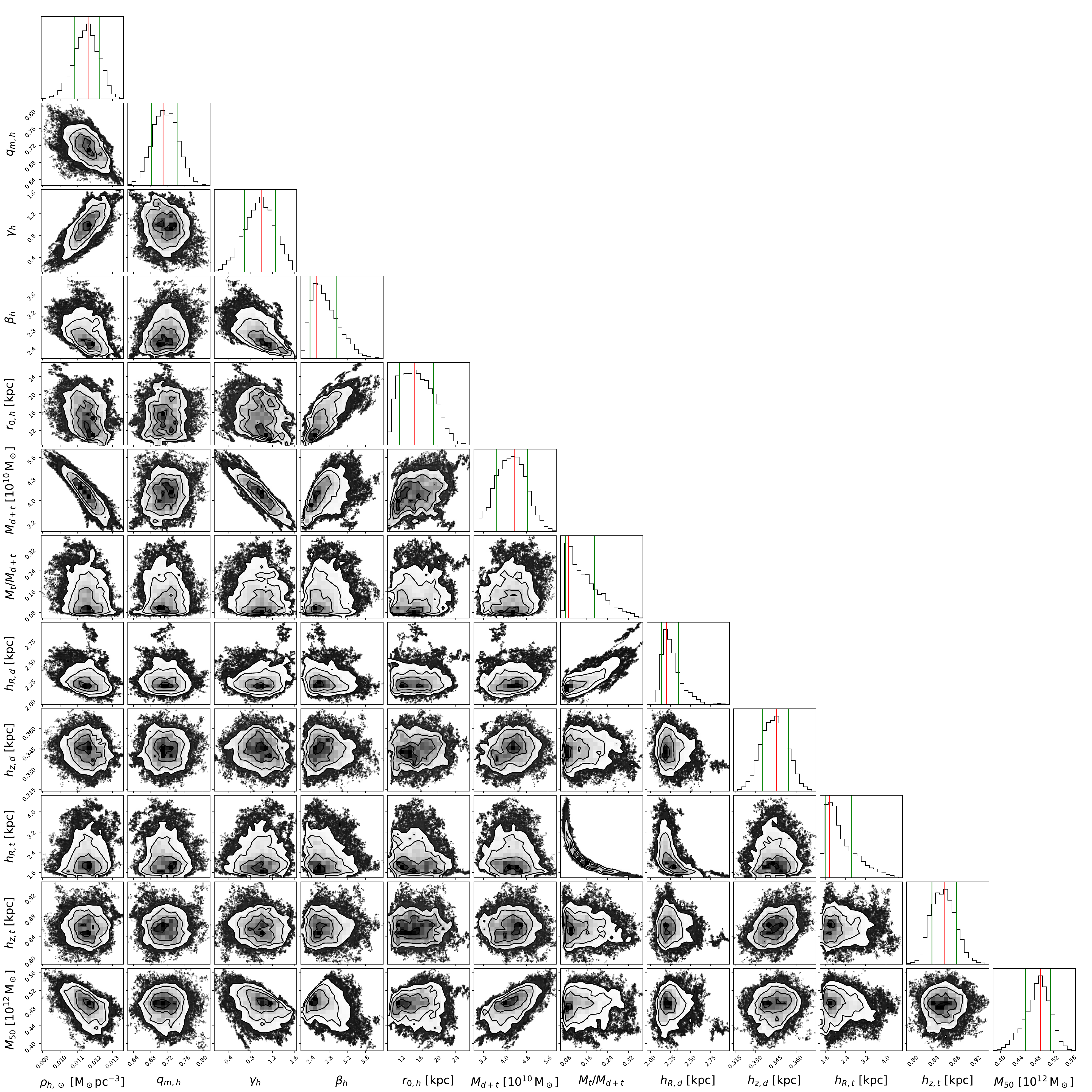}
\end{center}
\caption{As Figure~\ref{fig:corner_beta3}, but here the halo outer power law parameter $\beta_h$ is also allowed to vary, for a total of 11 free parameters of the Galactic mass model.}
\label{fig:corner_all_free}
\end{figure*}

We now repeat this experiment, but add in the GD1 stream, which is the structure for which we have the most numerous data (323 stars with line of sight velocity measurements and 3 RR~Lyrae). This time we allow the halo mass flattening parameter $q_{m, h}$ to also be free. With this dataset we find $h_{R, d}=2.16\pm0.08 \kpc$, $q_{m, h}=0.86\pm0.05$, $r_{0, h}=10.7 \pm 1.6 \kpc$, $M_{\star}=4.24\pm0.35 \times 10^{10}\msun$, $M_{200}=0.77\pm0.07 \times 10^{12} \msun$ and $M_{R<50}=0.39\pm0.01 \times 10^{12} \msun$, and the corresponding stream fit is shown in Figure~\ref{fig:panels_GD1}. While the stream model captures the large-scale behavior of the kinematics of GD-1, it possesses some small, yet systematic, deviations from the proper motion trend in the data, especially in the range $60\deg<\phi_1<90\deg$. The density distribution of the model particles (panel e) is also not a good match to the data, given that the model produces two high density peaks at $\phi_1 \approx 0\deg$ and $\phi_1 \approx 70\deg$, and a small remnant (at $\phi_1=29\degg7$) still remains, but is not present in reality. However, the large peaks in the data at $\phi_1 \approx 10\deg$ and $\phi_1 \approx 50\deg$, which we ascribed in an earlier contribution \citep{2020ApJ...891..161I} to epicyclic motion, can be seen to have counterparts in the model. Given this match, and given also that $\phi_1=32\degg4$ corresponds to the stellar mass-weighted midpoint of the stream\footnote{We calculate this midpoint by assuming that the survey is complete to $G_0=20$~mag, and correct for missing stars given the distance model and the metallicity listed in Table~\ref{tab:streams}. However, we did not correct for the masked region around M~67.}, we consider $\alpha=157\degg74$ to be currently the best estimate for the position of the center of the GD-1 progenitor. Henceforth we will fix the present day position of GD-1's progenitor to that location.

We continue the procedure of rendering the model more realistic, adding in the five streams in our sample that have known globular cluster progenitors (Gj\"oll/NGC~3201, Fj\"orm/M68, NGC~5466, Palomar~5, and M5). In addition to the previously fitted Galactic parameters, the scale heights of the thin and thick disk components, the scale length of the thick disk, and the halo inner power exponent are also allowed to vary. Compared to the least constrained model we present below, only the halo outer power law parameter remains fixed at $\beta_h=3$. The resulting fit has $h_{R, d}=2.31\pm0.14\kpc$, $h_{z, d}=345\pm9\pc$, $h_{R, t}=2.02\pm0.42\kpc$, $h_{z, t}=859\pm25\pc$, $\gamma_h=0.89\pm0.19$, $q_{m, h}=0.75\pm0.03$, $r_{0, h}=22.1 \pm 2.8 \kpc$, $M_{\star}=5.60\pm0.51 \times 10^{10}\msun$, $M_{200}=1.42\pm0.25 \times 10^{12} \msun$ and $M_{R<50}=0.50\pm0.03 \times 10^{12} \msun$. The most striking difference to the previous fit is that the stream kinematics require the thick disk scale length to be much shorter than the $3.6\kpc$ measured by \citet{2008ApJ...673..864J} and assumed by \citet{2017MNRAS.465...76M}; with a shorter thick disk scale length the halo scale radius $r_{0, h}$ becomes longer.

Exactly the same procedure is repeated, but this time using the full sample of 29 streams. As before, the Galactic potential model has a fixed bulge and gaseous disks, but a free halo, thin disk and thick disk. The results of the MCMC exploration with the halo outer power law parameter set to $\beta_h=3$ are shown in Figure~\ref{fig:corner_beta3}, while in Figure~\ref{fig:corner_all_free} $\beta_h$ is also allowed to vary. Table~\ref{tab:fit_results} provides the summary statistics of the two experiments. We note that some of the marginalized distributions are clearly highly skewed, so the mean and root mean square values can be misleading. So instead we report the distribution mode which we find as follows: using the binning scheme shown in Figures~\ref{fig:corner_beta3} and \ref{fig:corner_all_free} (30 bins between the minimum and maximum values), we identify the peak bin, and refine the position of the mode assuming that the distribution is locally quadratic over the region of the peak and the two adjacent bins. The marginalized distributions are then divided into two subsets, containing the data with lower and higher values than the mode, and in each subset, we respectively find the position of the datum that is the 68.27\% lowest and the 68.27\% highest, and consider these values as the $1\sigma$ confidence intervals.

\begin{table}
\caption{Statistics of the parameters of the MCMC chains shown in Figures~\ref{fig:corner_beta3} and \ref{fig:corner_all_free} derived from fitting a flexible Galactic model with the sample of 29 streams. The first two sets of values use a fixed value $\beta_h=3$ for the halo outer power law parameter, while the final two sets of columns are for the experiment where $\beta_h$ is allowed to be free. The derived model masses are listed on the final 4 rows of the table. The $\mathcal{L}_{\rm max}$ columns show the parameter values of the highest likelihood datum in the MCMC chains.}
\footnotesize
\begin{center}
\begin{tabular}{lcccc}
\hline
   quantity & $\beta_h=3$ & $\mathcal{L}_{\rm max}^{\beta_h=3}$ & $\beta_h={\rm free}$ & $\mathcal{L}_{\rm max}^{\beta_h={\rm free}}$ \\
\hline
$\rho_{h, \odot}[\frac{\msun \pc^{-3}}{1000}]$        & $11.4\pm 0.7$       & $11.4$    & $11.6^{+0.7}_{-0.8}$      & $11.2$    \\
$q_{m, h}$                                       & $0.749^{+0.026}_{-0.030}$ & $0.735$  & $0.709^{+0.033}_{-0.026}$ & $0.727$  \\
$\gamma_h$                                       & $0.97^{+0.17}_{-0.21}$    & $1.03$   & $0.99^{+0.26}_{-0.30}$    & $1.14$   \\
$\beta_h$                                        & $3$                       & $3$      & $2.53^{+0.42}_{-0.16}$    & $2.54$   \\
$r_{0, \, h} [\kpc]$                             & $14.7^{+4.7}_{-1.0}$      & $21.4$   & $14.7^{+4.4}_{-3.3}$      & $19.5$   \\
$M_{d+t} [10^{10} \msun]$                        & $4.20^{+0.44}_{-0.53}$    & $4.17$   & $4.34^{+0.50}_{-0.65}$    & $4.27$   \\
$M_t/M_{d+t}$                                    & $0.095^{+0.143}_{-0.014}$ & $0.201$  & $0.09^{+0.10}_{-0.01}$    & $0.14$   \\
$h_{R, \, d} [\kpc]$                             & $2.172^{+0.184}_{-0.079}$ & $2.241$  & $2.198^{+0.153}_{-0.062}$ & $2.251$  \\
$h_{z, \, d} [\kpc]$                             & $0.347^{+0.007}_{-0.010}$ & $0.348$  & $0.345^{+0.009}_{-0.010}$ & $0.349$  \\
$h_{R, \, t} [\kpc]$                             & $1.62^{+0.72}_{-0.13}$    & $1.74$   & $1.77^{+0.86}_{-0.17}$    & $2.07$   \\
$h_{z, \, t} [\kpc]$                             & $0.856^{+0.030}_{-0.020}$ & $0.858$  & $0.861^{+0.022}_{-0.025}$ & $0.860$  \\
\hline
$M_{R<50}[10^{12} \msun]$                        & $0.458^{+0.030}_{-0.026}$ & $0.477$  & $0.490^{+0.024}_{-0.033}$ & $0.518$  \\
$M_{200}[10^{12} \msun]$                         & $1.09^{+0.19}_{-0.14}$    & $1.24$   & $1.70^{+0.20}_{-0.49}$    & $2.00$   \\
$M_d[10^{10} \msun]$                             & $3.44^{+0.44}_{-0.54}$    & $3.34$   & $3.52^{+0.54}_{-0.51}$    & $3.65$   \\
$M_t[10^{10} \msun]$                             & $0.40^{+0.61}_{-0.09}$    & $0.84$   & $0.39^{+0.43}_{-0.07}$    & $0.62$   \\
\hline
\end{tabular}
\end{center}
\label{tab:fit_results}
\end{table}

The halo mass flattening turns out to be very well constrained with $q_{m, h}=0.749^{+0.026}_{-0.030} (0.709^{+0.033}_{-0.026})$, where the value in brackets corresponds to the fit in which $\beta_h$ is allowed to vary. Interestingly the (the negative of) the inner power law slope of the halo is close to the universal NFW profile found in cosmological dark matter simulations. Nominally, we find a shallower outer power-law profile $\beta_h=2.53^{+0.42}_{-0.16}$, but one is consistent with the NFW profile at $1\sigma$. The values of the halo scale radius $r_{0, \, h}=14.7^{+4.7}_{-1.0} (14.7^{+4.4}_{-3.3}) \kpc$ are closely consistent, although we note that the maximum likelihood solutions are larger: $r_{0, \, h}=21.4 (19.5) \kpc$.  We find that $r_{0,\, h}$ is strongly correlated with $\beta_h$, such that larger scale radii require steeper outer power law slopes.

\begin{figure}
\begin{center}
\includegraphics[angle=0, viewport= 1 10 400 400, clip, width=\hsize]{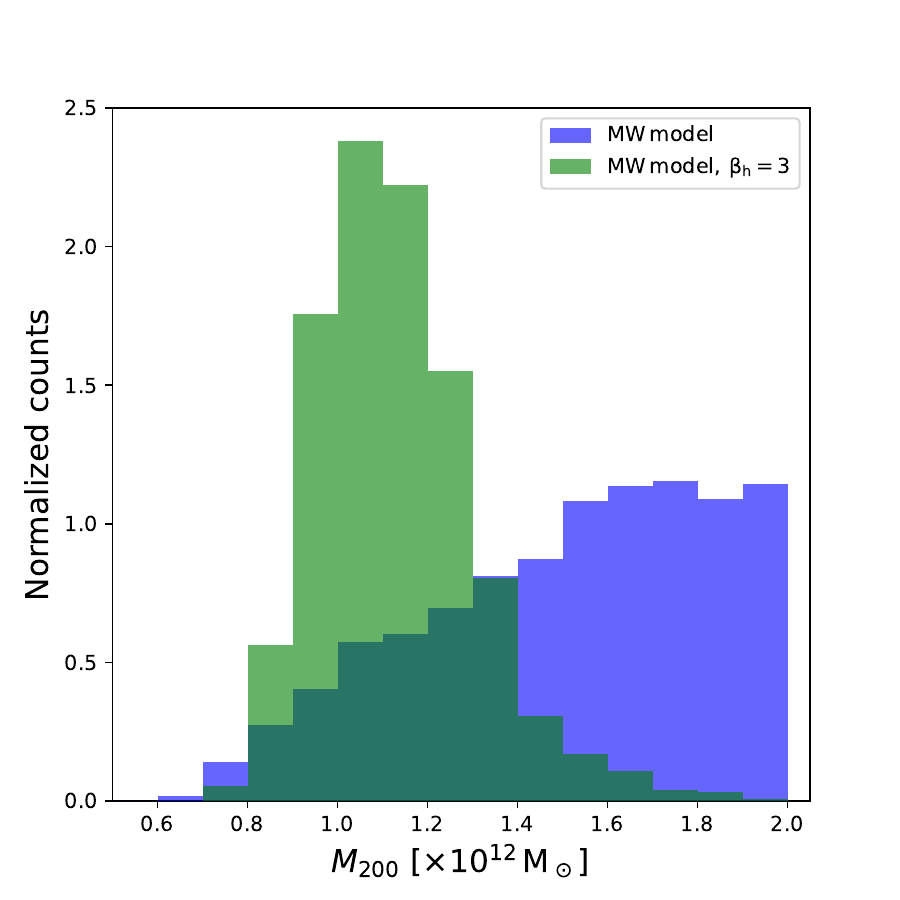}
\end{center}
\caption{The total $M_{200}$ masses of the Milky Way solutions shown in Figure~\ref{fig:corner_beta3} (green) and Figure~\ref{fig:corner_all_free} (blue).}
\label{fig:masses}
\end{figure}

The MCMC chain revealed strong correlations between the disk scale lengths and the central disk surface densities, and we also found that the combined mass of the disks $M_{d+t}=4.20^{+0.44}_{-0.53}(4.34^{+0.50}_{-0.65}) \times 10^{10} \msun$ is much better constrained than that of the individual components. Likewise, we found that the local halo density $\rho_{h, \odot}$ is much better constrained than the central halo density $\rho_{0, \, h}$. These realizations motivated our use of the combination of parameters shown in Figures~\ref{fig:corner_beta3} and \ref{fig:corner_all_free}. (The central surface density of the disk can be recovered simply using the fact that the total mass of an exponential disk is $M=2 \pi \Sigma h_R^2$, while the central density of the halo is related to the local density via Eqn.~\ref{eqn:spheroid}). For the thin and thick disk components, respectively, the scale heights are found to be $h_{z, d}=0.347^{+0.007}_{-0.010} (0.345^{+0.009}_{-0.010})\kpc$ and $h_{z, t}=0.856^{+0.030}_{-0.020}(0.861^{+0.022}_{-0.025})\kpc$ with scale lengths $h_{R, d}=2.172^{+0.184}_{-0.079}(2.198^{+0.153}_{-0.062})\kpc$ and $h_{R, t}=1.62^{+0.72}_{-0.13}(1.77^{+0.86}_{-0.17})\kpc$. The mass within $R=50\kpc$ of these two models is $M_{R<50} = 0.458^{+0.030}_{-0.026} (0.490^{+0.024}_{-0.033}) \times 10^{12} \msun$. Determining $M_{200}$ however, requires a substantial extrapolation out of the region where we have data. The distribution of $M_{200}$ masses of these Milky Way models as derived from the MCMC chains is shown in Figure~\ref{fig:masses}. The model with unconstrained $\beta_h$ is clearly affected by the imposed limit of $M_{200}<2\times 10^{12}\msun$. For this reason we prefer the $\beta_h=3$ model, although most of the other properties of the two models are consistent.

\begin{figure*}
\begin{center}
\includegraphics[angle=0, viewport= 1 10 600 860, clip, width=15.5cm]{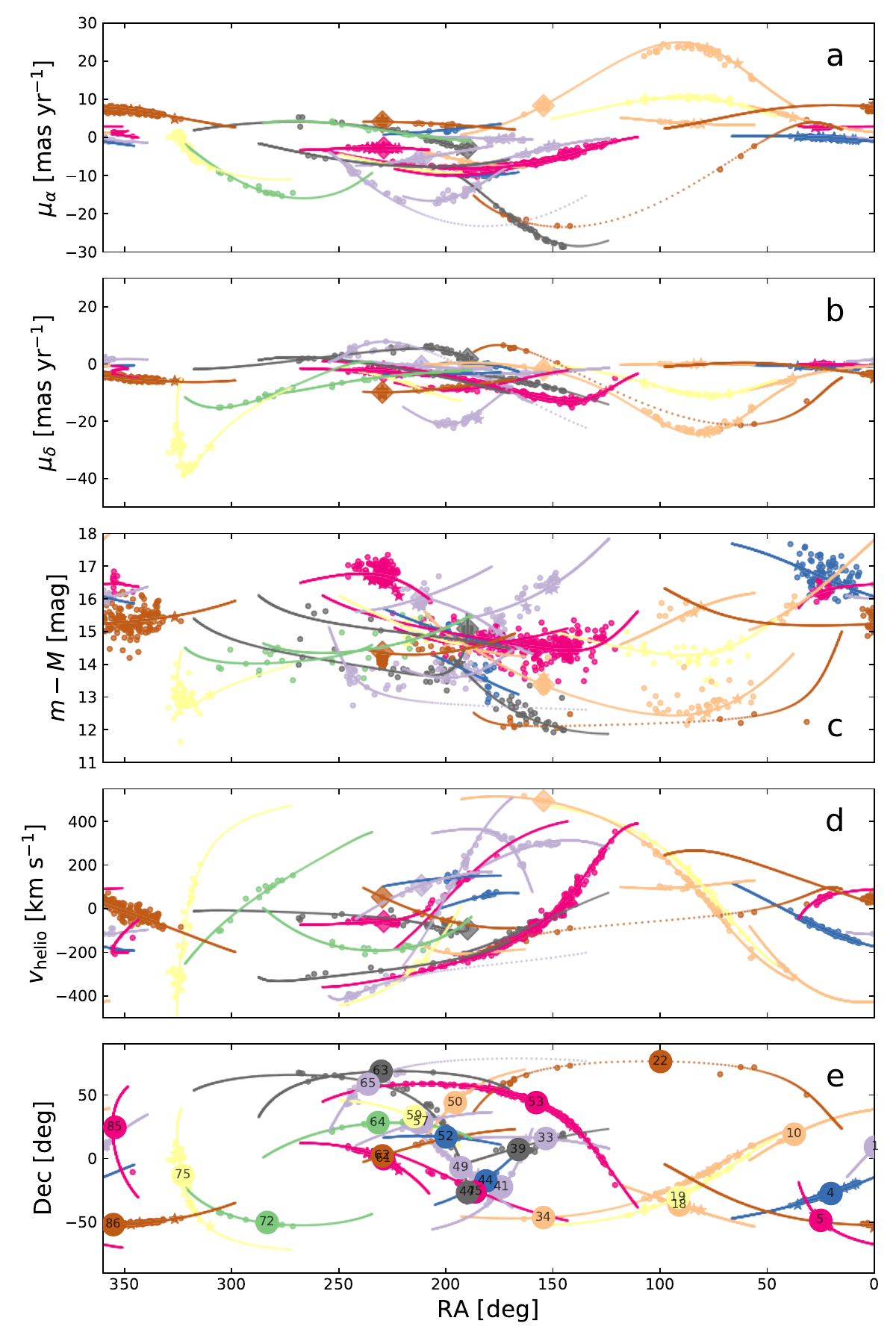}
\end{center}
\caption{Orbital fits to the sample of 29 long streams integrated in the best potential model with $\beta_h$ allowed to vary. The large and small four-cornered markers show, respectively, the positions of the center of globular clusters and RR~Lyrae. The filled circles are stars in the streams with measured line of sight velocities. For each parameter on the ordinate axes, we show the corresponding behavior of the best fit orbits. The color scheme of Figure~\ref{fig:Ball_sID} is re-used.}
\label{fig:fig_orbit_fits}
\end{figure*}

\begin{figure*}
\begin{center}
\includegraphics[angle=0, viewport= 1 10 600 860, clip, width=15.5cm]{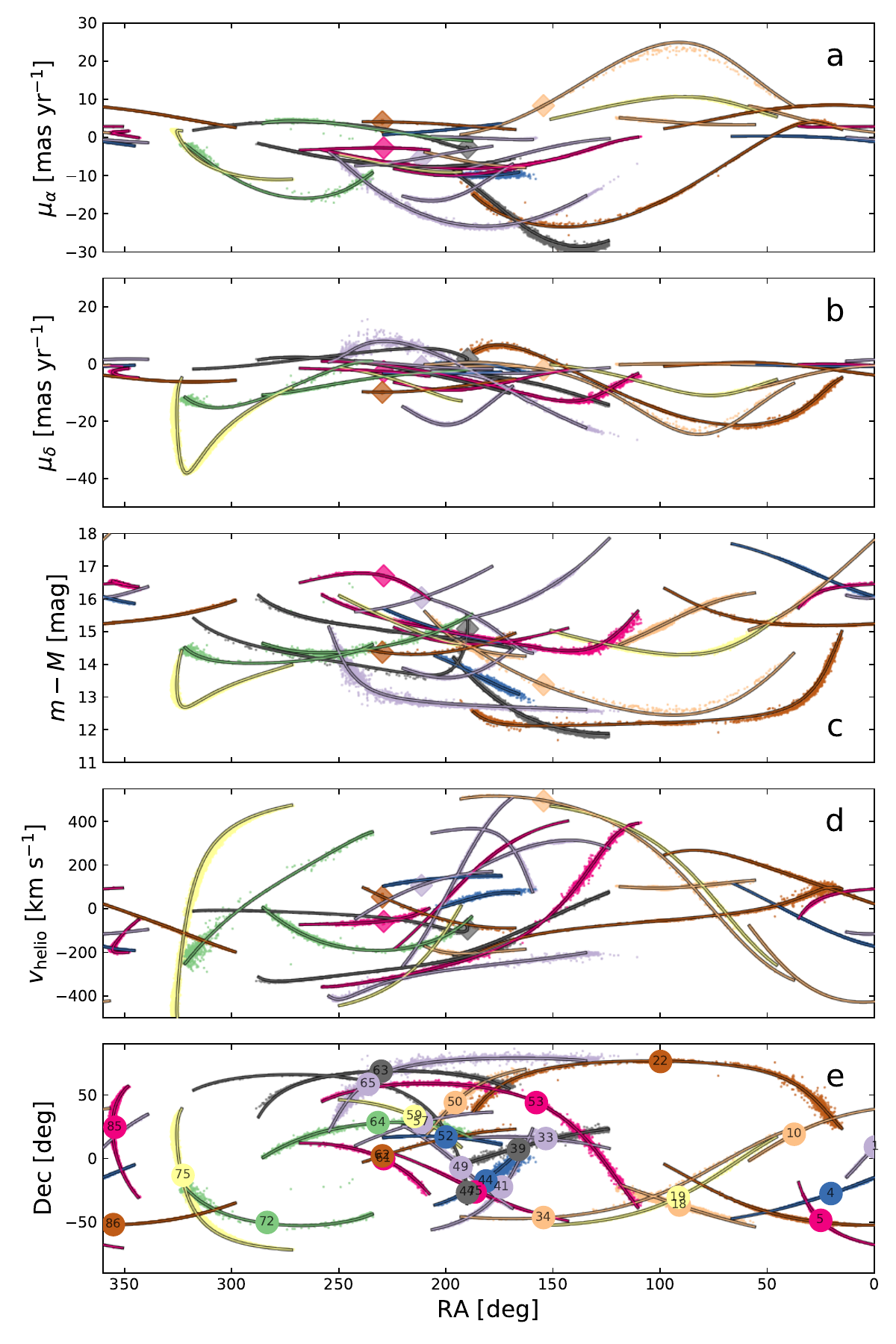}
\end{center}
\caption{As Figure~\ref{fig:fig_orbit_fits}, but comparing the progenitor orbits to the test particle stream simulations.}
\label{fig:fig_stream_fits}
\end{figure*}

The fitted orbits of the progenitors of the 29 streams integrated in the best-fit potential from Figure~\ref{fig:corner_all_free} are shown in Figure~\ref{fig:fig_orbit_fits}. These orbits can be seen to approximate quite closely the observational data, demonstrating that the $\Delta_\Theta$ correction functions are not required to be large for these systems. Similarly, in Figure~\ref{fig:fig_stream_fits}, we compare these orbits to the simulated streams resulting from the test-particle integrations in the same potential, which again shows a close correspondence. Our fitting procedure has clearly managed to encapsulate the large-scale properties of the kinematic and hence of the dynamical behavior of this sample of streams.

\begin{figure*}
\begin{center}
\includegraphics[angle=0, viewport= 60 15 800 390, clip, width=\hsize]{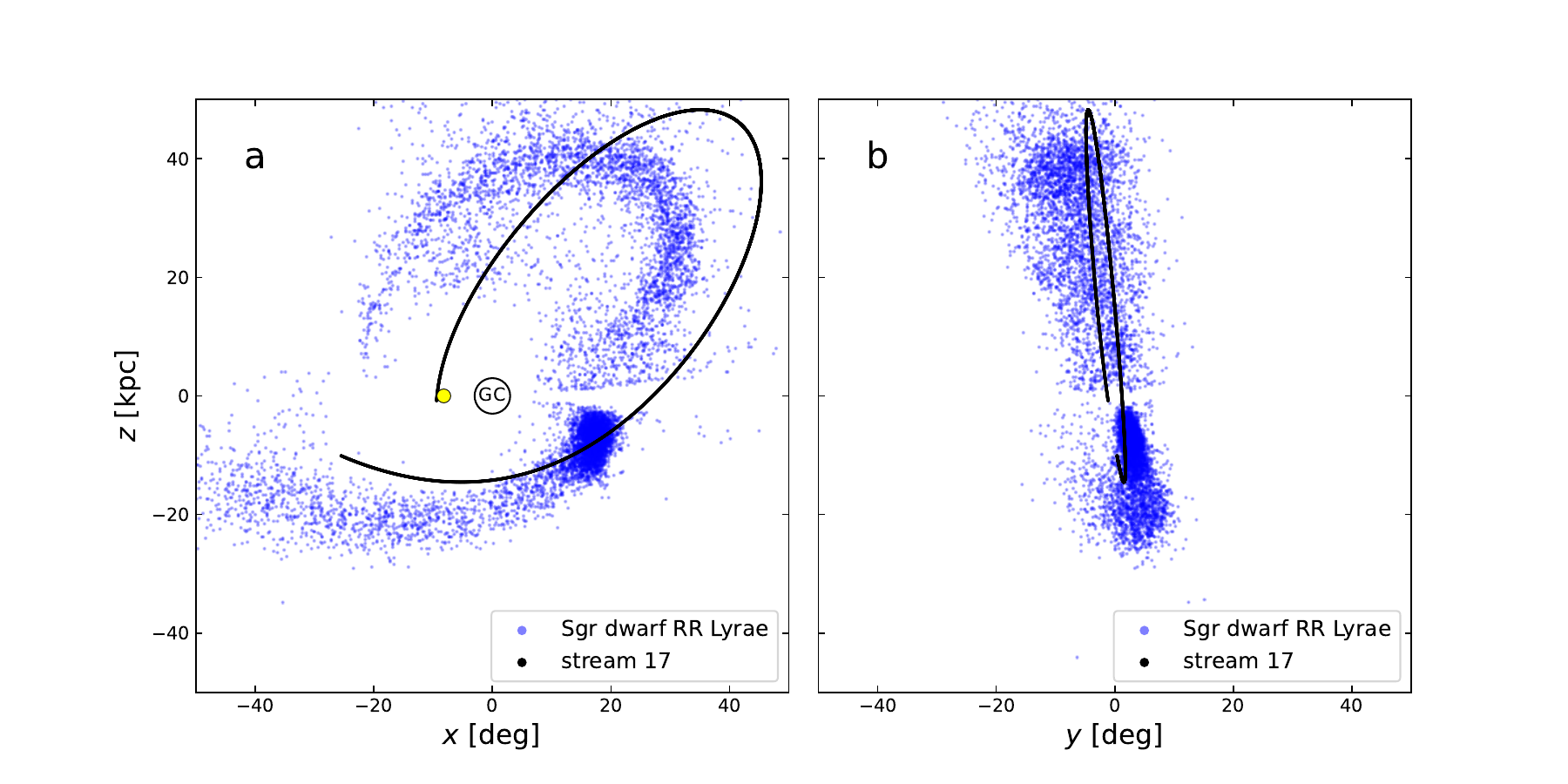}
\end{center}
\caption{Orbit of stream 17 compared to the present day structure of the Sagittarius dwarf galaxy, as seen from \emph{Gaia} DR3 RR~Lyrae extracted using the astrometric filter of \citet{2020ApJ...891L..19I}. The orbit is shown integrated backwards in time for $1\Gyr$, and can be seen to correspond very closely in position to this sample of RR~Lyrae stars. The position of the Sun is marked with a yellow dot.}
\label{fig:fig_Sgr}
\end{figure*}

\section{\emph{Gaia} DR3 Detections}
\label{sec:DR3_Detections}

Many of the structures displayed in Figures~\ref{fig:Ball_mul}--\ref{fig:params3} were presented previously in Papers~I and II, but 29  are new detections (the objects labelled ``New'' in Table~\ref{tab:streams} and stream \#11 emanating from the Hydrus dwarf galaxy, \citealt{2018MNRAS.479.5343K}). We have verified that they do not correspond to previously-reported detections in the compendium of \citet{2023MNRAS.520.5225M} or in the phase-space analysis of \citet{2023A&A...670L...2D}.

As may be expected, the stream-like structures discovered in each new \emph{Gaia} data release generally have a smaller number of member stars than those found in earlier releases, simply because more populated streams were easier to find first. And despite our efforts in trying to obtain follow-up spectroscopy, in 16 of the new streams we have secured fewer than 5 radial velocity measurements. However, some of the new features were harder to discover simply because they are more diffuse than those found earlier. For instance, streams 8, 17 and 22 are examples of very metal-poor structures (${\rm [Fe/H]}=-2.07$, $-2.71$, and $-2.18$, respectively) that subtend a long angle over the sky ($62\deg$, $69\deg$, and $80\deg$, respectively) are nearby ($\approx 1.2$, $\approx 1.5$, and $\approx 1.25\kpc$ distant, respectively) and display a well-defined velocity gradient. 

We will discuss the new findings in detail in a future contribution, but here we will nevertheless briefly mention some interesting highlights, either because the new stream is particularly interesting, or because the new data or analysis conveys some new insight on the previously known structure. We will discuss these highlighted streams in order of their ${\rm s_{ID}}$ index.

%\subsection{Stream \#10 (\emph{Gaia}-12)}
%\label{sec:stream10}

%Stream \#10 (\emph{Gaia}-12 in \citealt{2021ApJ...914..123I}) is an approximately $30\deg$ long structure $\approx 13.5\kpc$ away whose orbit extends out to a Galactocentric distance of $76\kpc$ (in the fitted potential model with $\beta_h=3$). It is one of the streams in the ``long-thin'' sample and its properties can be appreciated from an inspection of Figures~\ref{fig:fig_orbit_fits} and \ref{fig:fig_stream_fits}. Its metallicity is extremely low (${\rm [Fe/H]}=-3.28$) according to the \citet{2023arXiv230801344M} catalog based on \emph{Gaia} DR3 BP/RP spectro-photometry.

%During our spectroscopic follow-up campaign we secured UVES observations of 4 stars in the structure (\emph{Gaia} IDs: 81654921799327872, 33586201873793792, 32011598143700992, 29098953186515200). Since these stars are giants, and we have a fitted distance to the orbit, their metallicities can be straightforwardly measured from the \ion{Ca}{2} equivalent width, from which we find ${\rm \langle [Fe/H]} \rangle = -3.2$, using the empirical calibration of \citet{2010A&A...513A..34S}. Stream \#10 therefore appears to be almost as extreme as the C-19 stream,  which is the most metal-poor stellar system known \citep{2022Natur.601...45M}, and hence most likely a fossil of the early universe. 

\subsection{Stream \#17 (New 4)}
\label{sec:stream17}

This newly-discovered structure has 120 candidate members including 10 radial velocity measurements and 22 with excellent parallaxes ($\varpi/\delta\varpi>10$) which are all within $2\kpc$ of the Sun. A chemical abundance analysis with the MyGIsFOS code \citep{2014A&A...564A.109S} of the stars observed at high resolution with the UVES instrument will be presented elsewhere. However, we note here that the two well-measured stars (with \emph{Gaia} IDs: 3089847099636770560, 3074553030332697344) have ${\rm [Fe/H]}$ metallicities of $-2.75$, and $-2.67$, respectively, which places this system close to the most metal-poor known.

Another surprising property of this stream is its orbit. We use the machinery described in Section~\ref{sec:Stream_Fitting_isolated} to fit the phase space data, using again only the velocity-confirmed members. The resulting progenitor path is displayed in Figure~\ref{fig:fig_Sgr} in relation to the Sun (yellow circle), the Galactic Center ``GC'' and the positions of the RR~Lyrae members of the Sagittarius dwarf galaxy in \emph{Gaia} DR3, which we have selected using the parameter filter described in \citet{2020ApJ...891L..19I}. It therefore seems highly probable that the stream \#17 formed in the early universe as a globular cluster satellite of the Sagittarius dwarf, and was brought into the Galaxy with it. 

\begin{figure}
\begin{center}
\includegraphics[angle=0, viewport= 10 15 565 855, clip, width=\hsize]{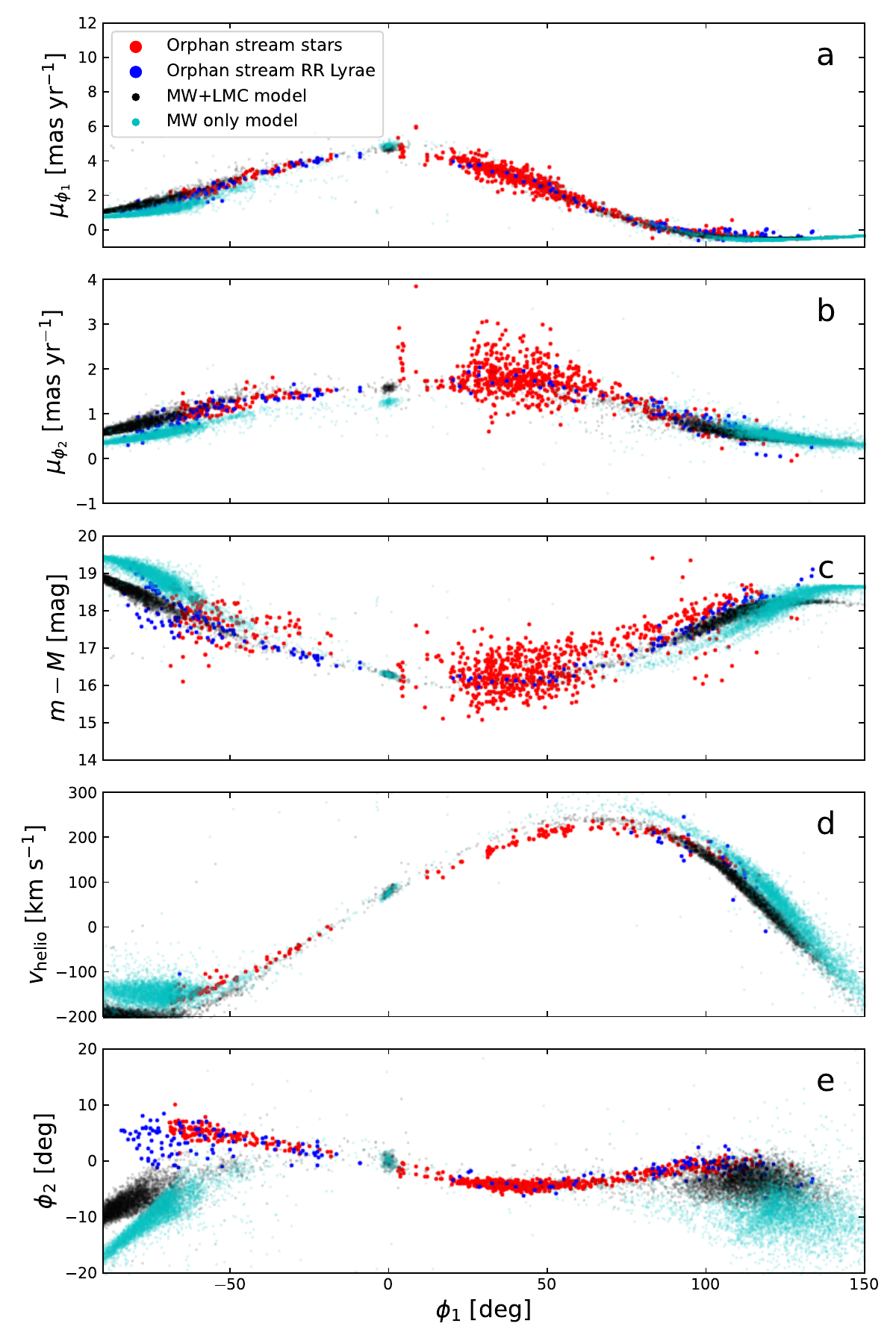}
\end{center}
\caption{Fit to stream 48 (Orphan-Chenab). Here, the potential includes a moving model for the Milky Way as well as a massive moving LMC, while the stream progenitor loses mass over a period of $5\Gyr$. Black dots show the best-fit stream model with both the Milky Way and the LMC, where the latter has a mass of $M_{\rm LMC}=1.5\times 10^{11}\msun$. In contrast, the green dots show the result of setting $M_{\rm LMC}=0$. The Milky Way host is represented by the model derived previously from orbit fitting (corresponding to the mass model with $\beta_h=3$ fit in Figure~\ref{fig:corner_beta3}).}
\label{fig:fig_Orphan}
\end{figure}

\subsection{Fitting stream \#48 (Orphan) with the LMC}
\label{sec:stream48}

The notable missing structure in the sample of 29 long thin streams which we set aside from our MCMC exploration above is stream \#48 (``Orphan-Chenab'', \citealt{2007ApJ...658..337B}), which is known to be strongly affected by the LMC \citep{2019MNRAS.487.2685E}. We will now attempt to fit this structure using our test-particle algorithm, which was built to handle precisely this sort of situation with the host galaxy and one or several massive perturbers moving under each-other's gravitational influence.

Our new contribution to the problem is that we can now impose what we expect is currently the most accurate Milky Way mass model, at least for the spatial region contained by the 29 fitted streams in Figure~\ref{fig:fig_orbit_fits} and \ref{fig:fig_stream_fits}. Onto our now moving Milky Way model, we add a moving NFW potential to model the LMC\footnote{To isolate the effect of the Milky Way and the LMC, all other Local Group galaxies are set to zero mass.}. We take the value of the center and the proper motions of the LMC from \citet{2018A&A...616A..12G}, we adopt the distance from \citet{2019Natur.567..200P}, and the radial velocity from \citet{2002AJ....124.2639V}. We further estimate from \citet{2014MNRAS.441..378L} that a plausible concentration for the NFW model of the LMC to be $c_{\rm LMC}=12.5$. Following \citet{2023MNRAS.521.4936K}, we model the progenitor with a mass of $2.67\times 10^7\msun$ and a Plummer scale radius of $1\kpc$.

In Figure~\ref{fig:fig_Orphan} we show the result of our stream modelling using just the preferred Milky Way model with $\beta_h=3$ found in Section~\ref{sec:Stream_Fitting_results} (green dots), while the maximum-likelihood fit upon adding the LMC has a mass of $M_{\rm LMC}=1.5 \times 10^{11}\msun$ (black dots). However, these models clearly fail to reproduce faithfully the large-scale behavior of the stream, and are much less convincing than the fits of \citet{2023MNRAS.521.4936K}. Further work is needed to ascertain whether axisymmetric double power-law halo models are capable of encapsulating the properties of the Milky Way sufficiently well all the way from the inner to the outer regions of the Galaxy.

\begin{figure}
\begin{center}
\includegraphics[angle=0, viewport= 15 15 570 870, clip, width=\hsize]{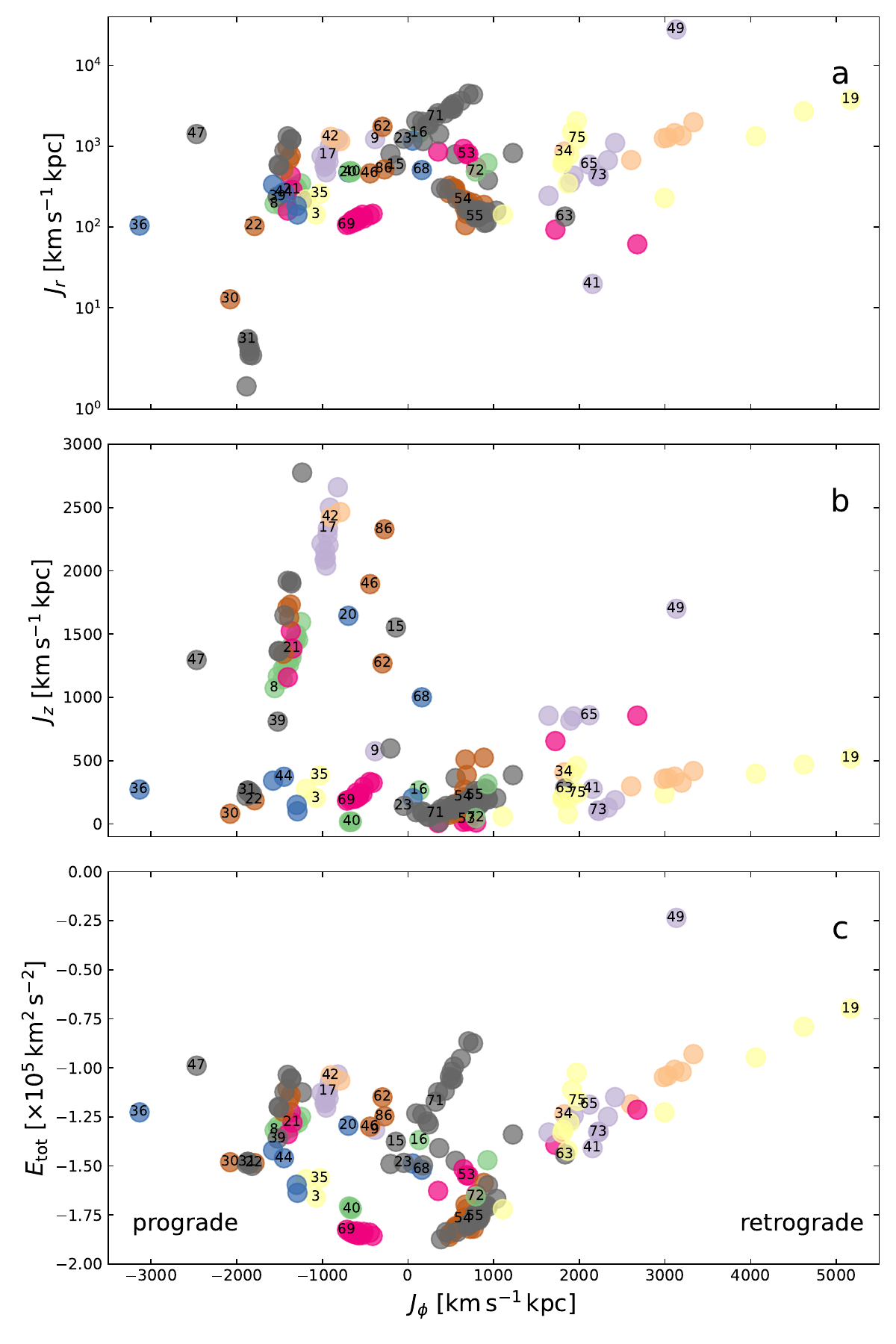}
\end{center}
\caption{Actions and energy of the 190 stars with measured radial velocities and excellent ($>7\sigma$) astrometric parallaxes, calculated in the \citet{2017MNRAS.465...76M} potential model (for easier comparison to earlier work). Color encodes the stream ${\rm s_{ID}}$ identification as in Figure~\ref{fig:Ball_sID}.}
\label{fig:fig_actions}
\end{figure}

\begin{figure}
\begin{center}
\includegraphics[angle=0, viewport= 10 10 400 400, clip, width=\hsize]{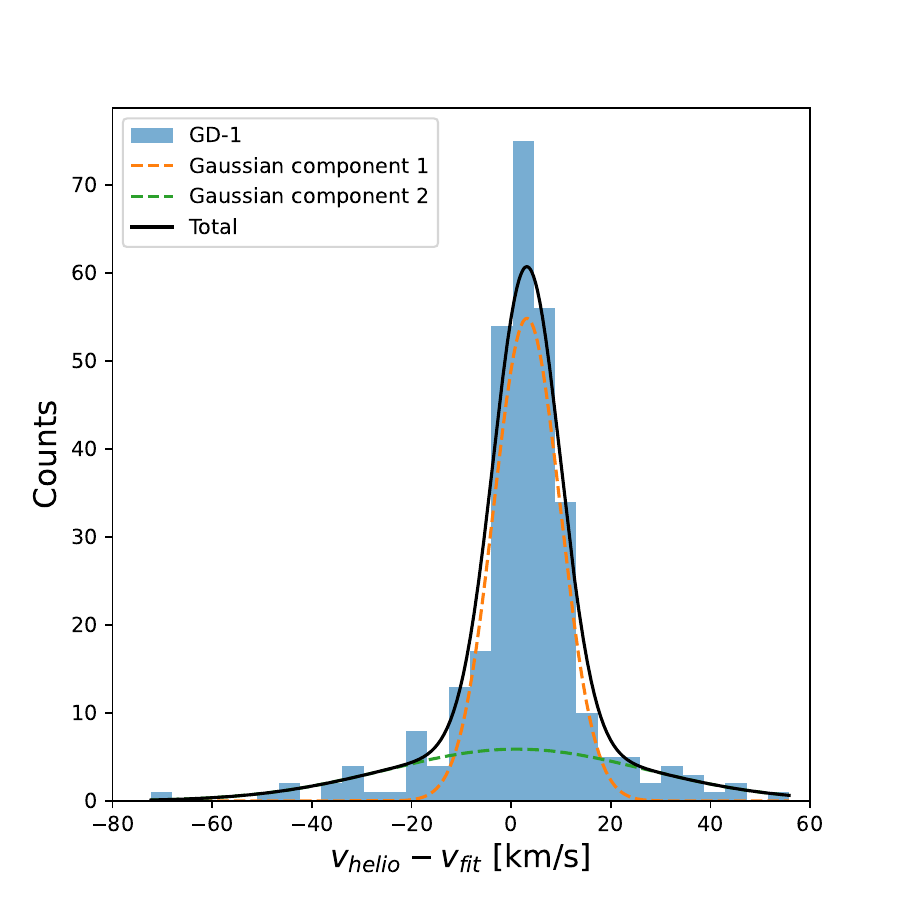}
\end{center}
\caption{Distribution of line of sight velocity offsets in the GD-1 stream from the profile previously fit by \citet{2020ApJ...891..161I}. The large wings to this distribution strongly suggest the presence of at least two Gaussian components.}
\label{fig:fig_GD1_Gaussians}
\end{figure}

\begin{figure}
\begin{center}
\includegraphics[angle=0, viewport= 10 10 430 430, clip, width=\hsize]{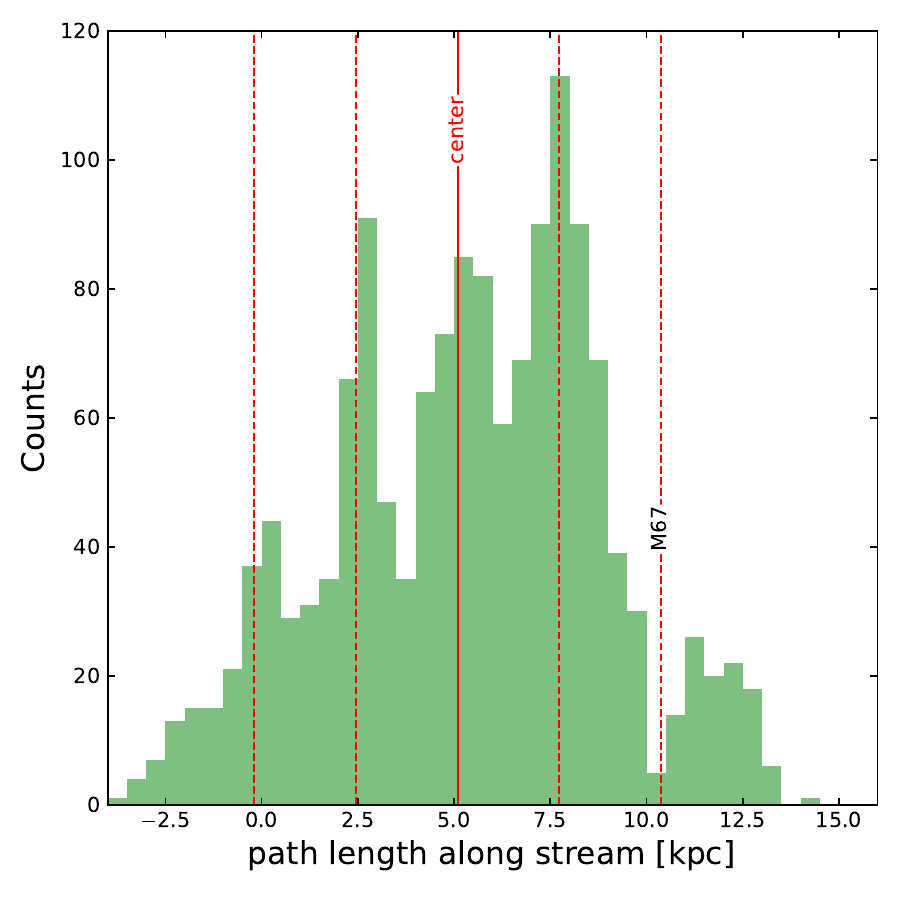}
\end{center}
\caption{Density as a function of physical path length along the GD-1 stream. The path-length zero-point is set to $\phi_1=0$. The probable current location of the progenitor is marked, and vertical dashed lines show the periodicity derived by \citet{2020ApJ...891..161I}. Our data masking at the position of the old open cluster M67 is responsible for the reduced counts at the marked location, which unfortunately coincides with a predicted peak.}
\label{fig:fig_GD1}
\end{figure}

\subsection{Stream \#49 (\emph{Gaia}-1)}
\label{sec:stream49}

Many of the streams possess stars with high-velocity along the line of sight (see Figures~\ref{fig:params1}-\ref{fig:params3}) and many possess high proper motion. It is interesting to consider those stars for which the \emph{Gaia} parallaxes $\varpi$ are sufficiently good to also identify stars with high total velocity. To this end we show in Figure~\ref{fig:fig_actions} the actions $(J_r,J_\phi,Jz)$ and orbital energy $E_{tot}$ of the objects in our \texttt{STREAMFINDER} sample that have $\varpi/\delta\varpi>7$ and measured radial velocity. (In this work we have chosen negative $J_\phi$ to correspond to prograde rotation). The outlier in this plane is stream \#49 (\emph{Gaia}-1, \citealt{2018MNRAS.481.3442M}) which thanks to one well-measured \emph{Gaia} star (ID 3578950965917470208, with $\varpi/\delta\varpi=9.5$) can be seen to have very high radial action $J_r$ and to have energy that puts it close to being unbound in the (relatively heavy) \citet{2017MNRAS.465...76M} potential. Indeed, this star is actually unbound in the lighter MW2014 potential model \citep{2015ApJS..216...29B}. At its present radial location of $r=8.95\pm0.36\kpc$, the 1\% lower limit to the mass of the Keplerian potential that would bind this star is $0.41\times 10^{12}\msun$. We note that the magnitude of the velocity of this star is $816\pm82\kms$, where the uncertainty is almost completely due to parallax uncertainty. The direction of motion of this stream is such that it is just now arriving into the inner Galaxy from the outer halo. So unless this system  is an inter-Galactic traveller, the requirement that it is bound to the Milky Way places a strong lower limit on the total mass of our Galaxy. Furthermore, its metallicity is ${\rm [Fe/H]=-1.8}$, which is not extremely metal-poor for a stream, so the possibility that this is a pristine system being accreted for the first time seems very implausible. 

As stream \#49 is in the selected sample of ``long thin'' streams, this object is already taken into account in our MCMC search. However, the constraint that the stars should be bound to the Milky Way was not used. This would be an easy upgrade to add into the algorithm in future work.

In any case, it is interesting that a single star, whose astrometry will be even more accurately measured in future \emph{Gaia} data releases, can be used to set a lower bound on the Galaxy's mass. The fact that this star is a member of a stream allows us to obtain a clearer view of its origin and possible history of previous interactions.

\subsection{Stream \#53 (GD-1)}
\label{sec:stream53}

We will briefly revisit the kinematics of stream \#53 (GD-1, \citealt{2006ApJ...643L..17G}) given the improved data on the system. Our very limited aim here is to motivate the likelihood model that was used as a template in the fitting software detailed above in Sections~\ref{sec:Stream_Fitting_isolated}-\ref{sec:Stream_Fitting_test_particles}.

The mean line of sight velocity profile of GD-1 as a function of $\phi_1$ was fit previously by \citet{2020ApJ...891..161I}. The distribution of residuals from that function is shown as a histogram in Figure~\ref{fig:fig_GD1_Gaussians}, and appears to show significant wings. To quantify this possibility, we again used an MCMC procedure to fit the dataset with a single Gaussian model, and reran the fit with a two-component Gaussian model, including of course the individual velocity measurement uncertainties when computing the likelihoods. The $p$-value for the measured difference in log-likelihood between the maximum likelihood fits of the two models is $\approx 1.2\times 10^{-7}$, indicating that the more complex model with two Gaussians provides a significantly better fit to the data compared to the simpler single Gaussian model. 

The two component fit has means $\mu_1=3.2\pm0.7\kms$, $\mu_2=0.8\pm3.9\kms$, dispersions $\sigma_1=7.4\pm1.1\kms$, $\sigma_2=29.1\pm6.1\kms$ and a fraction $0.73\pm0.09$ in component 1. Thus there is strong evidence that GD-1 has a cocoon-like envelope around it, as suggested by the simulations of \citet{2020ApJ...889..107C} and \citet{2023ApJ...953...99C}, which posit that stream progenitors were accreted inside dark matter subhalos.

Finally, given the new DR3 dataset, and given the center for GD-1 calculated from the mid-point of the mass-weighted stellar population along the stream presented in Section~\ref{sec:Stream_Fitting_results}, we believe it is useful to consider the counts along the stream as a function of physical path length, as shown in Figure~\ref{fig:fig_GD1}. To achieve this coordinate conversion we use the distance profile derived in \citet{2020ApJ...891..161I}. The epicyclic spikes with periodicity $2.64\kpc$ identified in that contribution are very clear in this new representation. While dark matter subhaloes are expected to cause gaps in streams \citep{2012ApJ...748...20C}, they are highly unlikely to produce such regular behavior.

\begin{figure*}
\begin{center}
\includegraphics[angle=0, viewport= 10 10 740 740, clip, width=\hsize]{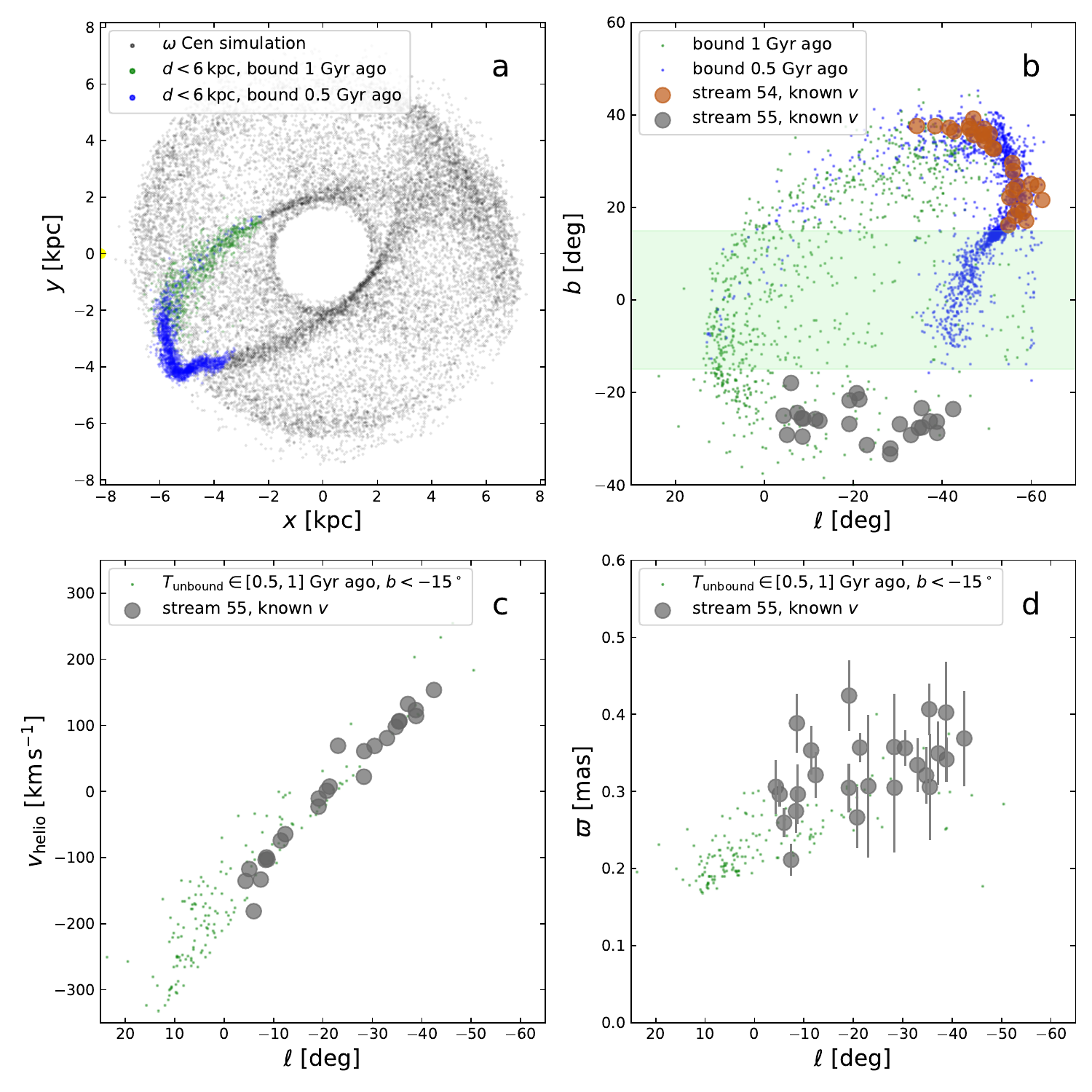}
\end{center}
\caption{Observations and model of streams \#54 and \#55. The present day snapshot of the \citet{2019NatAs...3..667I} $\omega$~Cen model is shown in the $x$-$y$ plane (a), oriented such that the location of the sun lies at $(-R_\odot,0)$. Particles that were still bound in the simulation $0.5\Gyr$ ago and currently lie at a heliocentric distance $d<6\kpc$ are colored blue, while those also at $d<6\kpc$ that were bound $1\Gyr$ ago are colored green. The same colored particles are shown again in panel (b), along with the velocity-confirmed members of stream \#54 (brown) and \#55 (grey, reusing the color scheme of Figure~\ref{fig:fig_actions}). The filled green area within $15\deg$ of the Galactic plane highlights the region where high source density and extinction render it very challenging to detect streams. The model particles of the older (green) part of the stream that lie below $b=-15\deg$ are reproduced in panels (c) and (d) which show, respectively, the line of sight velocity and parallax behavior of both the data and simulations.}
\label{fig:fig_wCen}
\end{figure*}

\subsection{Streams \#54 ($\omega$~Cen/Fimbulthul) \& \#55}
\label{sec:stream54}

One of the early successes of our \texttt{STREAMFINDER} program was the discovery of a tidal tail of $\omega$~Cen \citep{2019NatAs...3..667I}, the most massive globular cluster in the Milky Way, and long thought to be the central remnant of an accreted dwarf galaxy \citep{2000LIACo..35..619M, 2003MNRAS.346L..11B}. A total of 22 stars of this stream (\#54) currently have good parallaxes ($\varpi/\delta\varpi>7$) and velocities, and consequently appeared in the action and energy diagram displayed in Figure~\ref{fig:fig_actions} (which we discussed previously in relation to stream \#49). Interestingly, a further 18 stars (with the same quality cut) belong to another structure identified by the \texttt{STREAMFINDER} (stream \#55) that is almost coincident in the action-energy diagram. The mean values and rms scatter of these integrals of motion are: $(J_r,J_\phi,J_z)=(207\pm60,663\pm110,198\pm120) \, \kms \, \kpc$ and $E_{tot}=-1.78\pm0.06\times10^5 \, {\rm km^2 \, s^{-2}}$ for stream \#54 and $(J_r,J_\phi,J_z)=(217\pm160,797\pm190,200\pm60) \, \kms \, \kpc$ and $E_{tot}=-1.73\pm0.11\times10^5 \, {\rm km^2 \, s^{-2}}$ for stream \#55. This coincidence strongly suggests a common origin to the two groups.

The association between the two streams is explored further in Figure~\ref{fig:fig_wCen}. Panel (a) shows the end result of the N-body simulation of $\omega$~Cen presented by \citet{2019NatAs...3..667I}. A peculiarity of that model is that it incorporated an initial rotation of the cluster model in order to reproduce the observed present day rotation of $\omega$~Cen \citep{2013ApJ...772...67B}, which was also found to be necessary to reproduce the peculiar knee-shaped structure of the stream as seen in projection on the sky. In panel (a) particles at a heliocentric distance of $d<6\kpc$ that were still bound to the cluster $1\Gyr$ ago are displayed in color, with the particles that were no longer bound to the progenitor $0.5\Gyr$ ago shown in green. The corresponding distribution on the sky is displayed in Galactic coordinates in (b), along with the locations of streams \#54 and \#55. Although the modelled stream produces a young spatially thin (blue) component, the progenitor is sufficiently massive (with a present day model mass of $4.2\times 10^6\msun$) to scatter particles into an older broader (green) feature as well. The older part of the stream is also visible in panel (b) in the form of a population of particles below $b=-15\deg$, approximately coincident on the sky with the location of stream \#55. Panels (c) and (d) reproduce again the older (green) model stream selection from panel (b), but further trimmed to show only those particles with $b<-15\deg$. The line of sight velocity (c) and parallax (d) trends of this sub-sample of the model approximate the corresponding observed properties of stream \#55. Similar phase-space structure is present in the simulations of \citet{2023A&A...673A..44F}.

Hence this structure, like stream \#54 (``Fimbulthul''), is also trailing tidal arm debris from the disruption of $\omega$~Cen. The bulk of the sample is located $\approx 3\kpc$ away from the sun and covers much of the southern bulge region.

\section{Discussion and Conclusions}
\label{sec:Conclusions}

\subsection{Overview}

In this contribution we have presented a large catalog of stellar streams, extracted in a homogenous manner from the \emph{Gaia} DR3 dataset. We hunted for these structures using an improved version of the \texttt{STREAMFINDER} algorithm \citep{2018MNRAS.477.4063M}, employing a stream template with a spatial width of $100\pc$ so as to find structures resulting from the expected dissolution of globular clusters or very small dwarf satellite galaxies. This work is a continuation of our search efforts with the \emph{Gaia} DR2 \citep{2018MNRAS.481.3442M,2019ApJ...872..152I} and EDR3 \citep{2021ApJ...914..123I,2022ApJ...926..107M} catalogs. Despite the radial velocity information being very sparse, the new velocity constraints, along with other enhancements in the algorithm detailed in Section~\ref{sec:Alterations} have allowed us to improve stream statistics and detect more subtle, diffuse structures.

We also embarked on a large follow-up campaign of the stream sources with the VLT/UVES high resolution spectrograph (together with earlier observations in 2018--2020 with CFHT/ESPaDOnS in the northern hemisphere). Some low resolution spectroscopic observations with INT/IDS were also obtained. Combining these data with public spectroscopic surveys has resulted in a catalog containing $\approx 24,500$ thin stream stars including 2,178 line of sight velocities. This represents a significant expansion and refinement over previous catalogues. We expect the dataset to be of considerable astrophysical interest for charting the Galactic mass distribution, probing dark substructures and investigating alternative ideas for the workings of gravity.

\subsection{Metallicity distribution of stellar streams}

\begin{figure}
\begin{center}
\includegraphics[angle=0, viewport= 10 10 400 400, clip, width=\hsize]{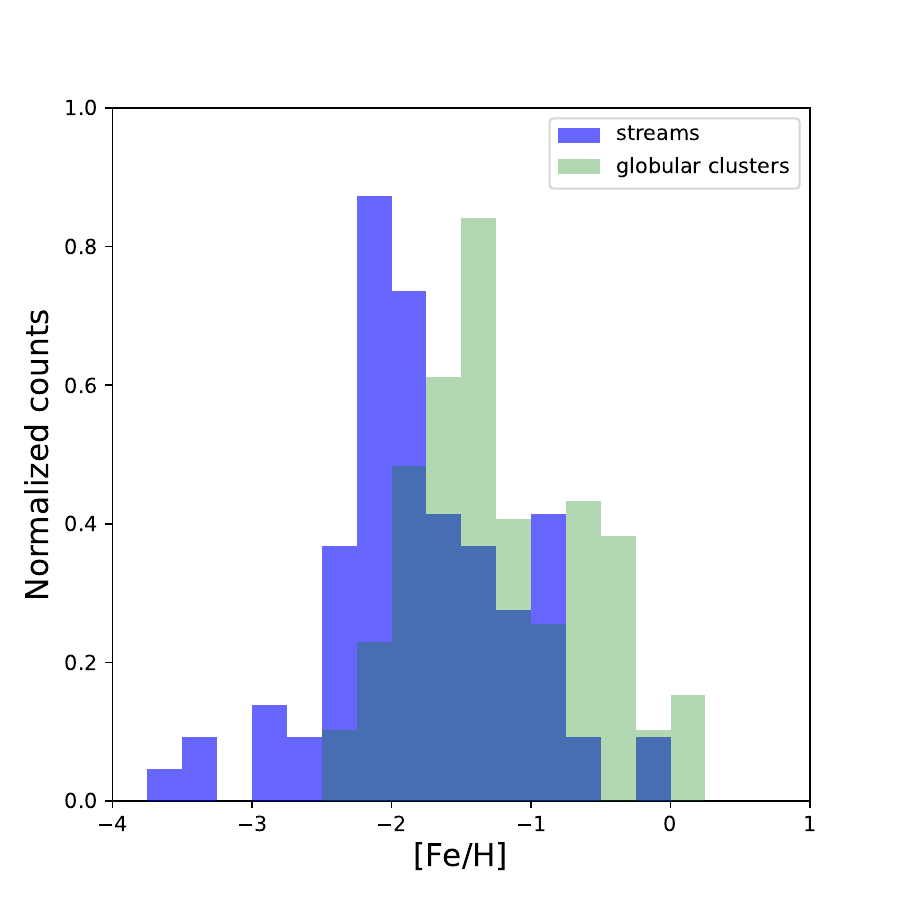}
\end{center}
\caption{Comparison of the metallicity distribution of the stellar streams detected in \emph{Gaia} DR3 with the \texttt{STREAMFINDER} with the full sample of Galactic globular clusters \citep{2010arXiv1012.3224H}.}
\label{fig:fig_FeH}
\end{figure}

We collated the metallicities of the detected streams from the literature, from individual metallicity measurements of member stars extracted from public spectroscopic and photometric surveys, and from our own follow-up spectroscopic observations. Updating the sample previously analysed in \citet{2022MNRAS.516.5331M}, in Figure~\ref{fig:fig_FeH} we compare the metallicity distribution function (MDF, using the values listed in Table~\ref{tab:streams}) with that of the globular clusters (GCs) listed in the compilation by \citet{2010arXiv1012.3224H}. Because of the adopted stream width template of $100\pc$, we expect the streams presented in this contribution to be mainly the remnants of GCs, although we note that the class of small and extremely low-luminosity systems that have recently been discovered \citep{2020ApJ...890..136M,2023ApJ...953....1C,2023arXiv231110147S}, and which may be very low mass dwarf galaxies, are also possible progenitor candidates. The peak of the MDF of the streams can be seen to be $\approx 0.75$~dex more metal poor than that of the surviving clusters.

Cosmological simulations show that metal-poor GCs experienced (on average) weaker tidal fields than metal rich GCs \citep{2017MNRAS.465.3622R}. Thus the observed difference in the MDF between streams and surviving GCs suggests that metal-poor GCs are more susceptible to disruption compared to their metal-rich counterparts, which could happen if  metal-poor GCs originally formed with lower densities or masses. Supporting this perspective, \citet{2017MNRAS.469.4359B} propose that metal-rich GCs may have formed in a denser initial state, characterized by shorter relaxation times. Their analysis seeks to explain the observed trend in dynamical mass-to-light ($M/L$) ratios with metallicity: metal-rich GCs, at least in M31, exhibit lower $M/L$ ratios compared to their metal-poor counterparts (\citealt{2011AJ....142....8S}, although we note that the formation history of M31 is likely to be strikingly different from our own). A denser formation with rapid relaxation times could accelerate the evolution of metal-rich GCs towards lower $M/L$ ratios, due to the preferential ejection of low-mass stars and stellar remnants. Additionally, a higher initial density in metal-rich GCs would confer a greater resistance to tidal disruption.

Alternatively, the MDF difference may be a manifestation of the fact that the most metal-rich GCs in the Galaxy tend to have a higher fraction of disk orbits or orbits confined to the innermost regions of the Galaxy. Both types of orbits produce streams that would be harder to detect with the \texttt{STREAMFINDER}, as the algorithm performs less well for very diffuse streams, streams in very high density regions, as in the bulge, or streams confined to the disk. \citet{2023A&A...673A..44F} have shown that orbits with low apocenters produce streams on the sky that are very different from ``classical'' streams such as Palomar~5, and thus they would be harder to detect by hunting along orbits. The differences between the stream and the GCs MDF may thus be a selection effect. Quantifying this requires computationally expensive completeness tests, and will be explored in a future contribution.

\subsection{MCMC exploration of streams and Milky Way }

Given this unprecedented dataset, we decided to also invest substantial effort to build analysis tools that would allow us to build an updated global model of the Milky Way. The computational task is daunting however, as reasonably realistic axisymmetric Galactic mass models have in excess of 10 parameters, while 6 parameters are needed to define the orbit of each stream progenitor, and at least a mass and a scale radius is required to define the simplest (Plummer) mass model for each of the latter. The approach we took was to devise two algorithms. The first algorithm is a simpler orbit-fitting analysis, which uses, we believe for the first time, a correction function for each of the observable parameters so as to better approximate stream tracks. The Galactic mass model parameters and the stream kinematics are optimised through a Markov Chain Monte Carlo (MCMC) search. The second algorithm integrates test particles around dissolving progenitors within a dynamic Local Group environment, where the massive host and perturbing galaxies are allowed to move each under the influence of the gravity of the other massive bodies, but with the massless test particles having no effect on other bodies. While the second algorithm is capable of a full MCMC exploration, in the current contribution we used it primarily to derive the orbit-to-stream corrections needed for the first approach. In future work we intend to use this algorithm to explore stream and satellite dynamics throughout the Local Group.

\subsection{Halo constraints}

The analysis presented in Section~\ref{sec:Stream_Fitting_results} started from a simple fit to extant circular velocity and terminal velocity data (together with some additional kinematic constraints) and progressively added in the information from GD-1, which is the best-measured stream, then from streams with known globular clusters, and finally from the selected full sample of 29 long and thin streams. Our hope had been that a model of the complexity of that presented here could be fully constrained by the present dataset, with no strong priors needed. However, it transpires that in allowing the halo outer power-law parameter $\beta_h$ to vary, we obtain solutions to the virial mass that reach up to our imposed limit of $M_{200}<2\times 10^{12}\msun$ (which corresponds to half of the 95\% upper limit to the mass of the Local Group, as determined by \citealt{2014MNRAS.443.2204P}). With this constraint, the most likely value of the outer slope parameter is $\beta_h\approx2.5$. Cosmological simulations of galaxy formation predict that dark matter halos have profiles have on average an asymptotic $r^{-3}$ slope in their outer regions \citep{1997ApJ...490..493N}. Yet there is also an environmental dependence \citep{1999MNRAS.310..527A}, with steeper slopes in galaxy clusters, and shallower slopes ($\beta\approx 2.3$--$2.7$) in groups and galaxy systems. Thus the PDF of $\beta_h$ found in Figure~\ref{fig:corner_all_free} is therefore not unexpected for the case of the Milky Way. Nevertheless, because the MCMC exploration does reach the $M_{200}$ limit when $\beta_h$ is free, we select the fits with the outer halo power-law parameter fixed to $\beta_h=3$ as our preferred model\footnote{Obviously, we could have incorporated the results of the \citet{2014MNRAS.443.2204P} analysis into our MCMC search as a Bayesian prior, but we deem that the present approach presents a clearer view of the constraints provided by the \texttt{STREAMFINDER} dataset.}. 

The fitted value of the halo inner power law parameter $\gamma_h=0.97^{+0.17}_{-0.21}$ is fully consistent with the NFW profile, and the scale radius $r_{0, \, h}=14.7^{+4.7}_{-1.0}\kpc$ is also consistent at $\approx 1\sigma$ with expectations for the Milky Way from $\Lambda$CDM based models \citep{2002ApJ...573..597K}, as is the inferred virial mass of $M_{200}=1.09^{+0.19}_{-0.14}\times 10^{12}\msun$. \citet{1999MNRAS.310..645W} introduced the mass interior to $50\kpc$ as a useful measure of the Galaxy, finding $M_{R<50}=0.54^{+0.02}_{-0.36}\times 10^{12}\msun$. Our fitted value $M_{R<50}=0.46\pm0.03\times 10^{12}\msun$ is fully consistent with this, and with the posterior distribution shown by \citet{2017MNRAS.465...76M} (their Figure~6). However, our fits appear to be in tension with the recent detection of a Keplerian decline in the Galactic rotation curve \citep{2023A&A...678A.208J}, which implies a total Milky Way mass of $0.206^{+0.024}_{-0.013}\times 10^{12}\msun$; in future work it will be interesting to explore how the two analyses can be tallied. 

We note however that our results should not be taken as inconsistent with those of \citet{2017MNRAS.465.1621P} and \citet{2017MNRAS.465..798C}, who found a cored halo with an inner slope shallower than 0.6 in the central $2\kpc$, since the present study is not well suited to constrain the mass profile in that region, given the 29 streams considered here. A detailed study of the dark matter distribution in this central region would need a realistic model of the bar (e.g., \citealt{2017MNRAS.465.1621P, 2019A&A...626A..41M}).

\begin{figure}
\begin{center}
\includegraphics[angle=0, viewport= 10 10 400 400, clip, width=\hsize]{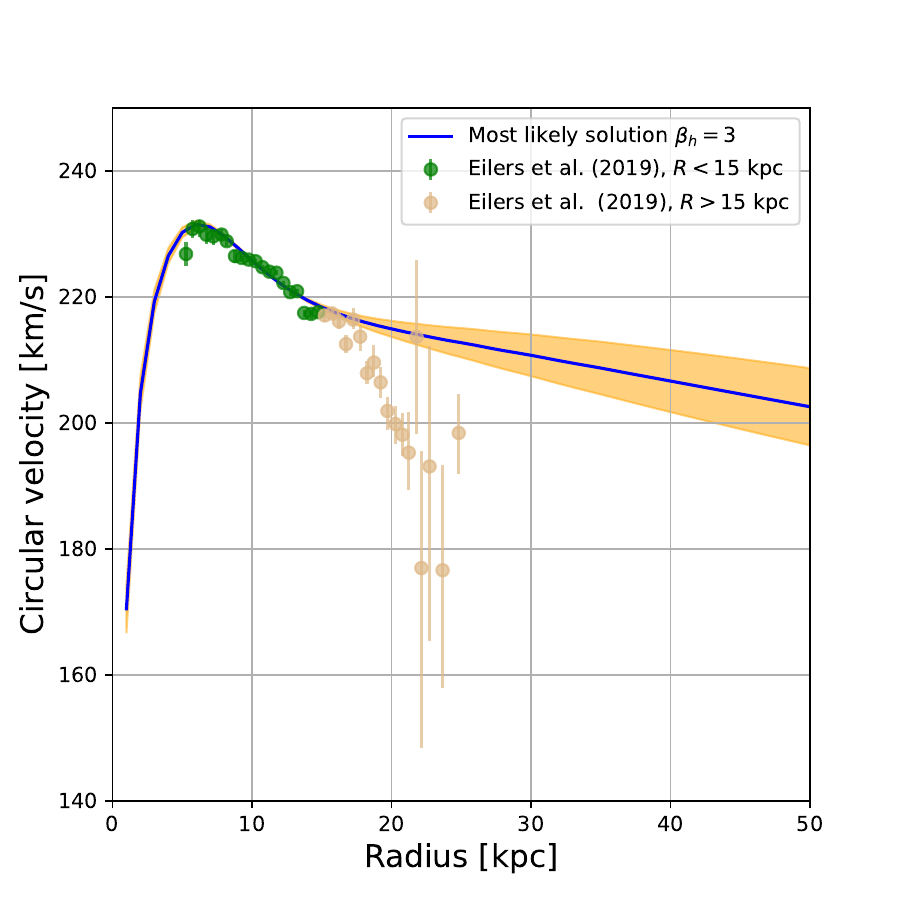}
\end{center}
\caption{Circular velocity of the most likely Galaxy model with $\beta_h=3$ (purple line). The shaded (yellow) region shows the $1\sigma$ spread derived from the models explored in the MCMC chain. We show for comparison the \citet{2019ApJ...871..120E} dataset; the selected subset at $R<15\kpc$ where the effects of the Galactic warp and flaring are less likely to be problematic is highlighted in green.}
\label{fig:Vc}
\end{figure}

\begin{figure}
\begin{center}
\includegraphics[angle=0, viewport= 1 10 400 400, clip, width=\hsize]{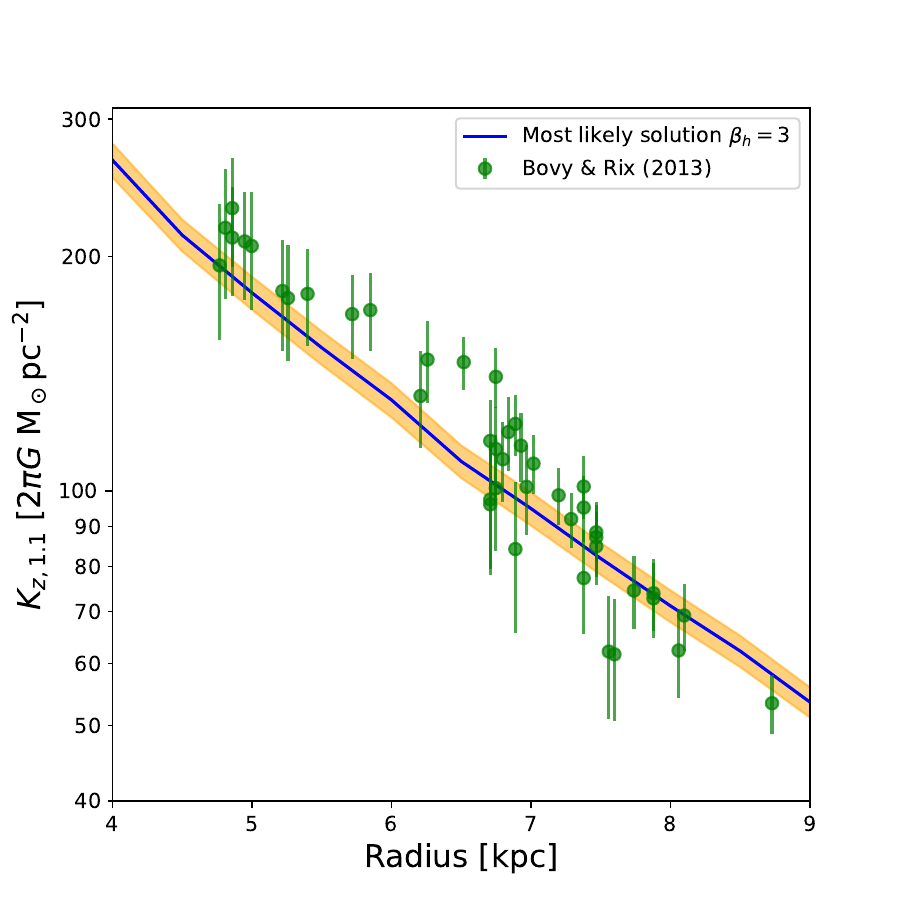}
\end{center}
\caption{Profile of the $K_{z, 1.1}$ vertical acceleration $z=1.1\kpc$ above the Galactic plane. The best Milky Way model (purple) together with its ($1\sigma$) uncertainties derived from the MCMC chain are compared to the analysis of \citet{2013ApJ...779..115B}. Note that these data were not used to derive the mass model.}
\label{fig:Kz}
\end{figure}

The local dark matter density, assuming that it is given solely by the density of the halo component, is tightly constrained at $\rho_{h, \odot}=0.0114\pm0.0007 \msun \pc^{-3}$. This is consistent at the $2\sigma$ level with the best-fit value of $\rho_{h, \odot}=0.0101\msun \pc^{-3}$ found by \citet{2017MNRAS.465...76M}. 

\subsection{Disk constraints}

Our fits put strong constraints on the large-scale mass and structure of the disk components of the Milky Way. The thin disk is found to have a scale length of $h_{R, \, d}=2.17^{+0.18}_{-0.08}\kpc$ and scale height of $h_{z, \, d}=0.347^{+0.007}_{-0.010}\kpc$ while the same properties of the thick disk are, respectively, $h_{R, \, t}=1.62^{+0.72}_{-0.13}\kpc$ and $h_{z, \, t}=0.86^{+0.03}_{-0.02}\kpc$. The combined mass of the two stellar disks is $M_{d+t}=4.20^{+0.44}_{-0.53} \times 10^{10} \msun$,  of which $9.5^{+14.3}_{-1.4}$\% resides in the thick disk. In terms of the surface density at the Solar radius, we find that the local thick disk constitutes $12.4\pm0.7$\% of the combined disk surface density, while the local thick disk density fraction is $5.4\pm0.4$\%. The disk structural parameters we measure are contained within the range reported in the literature review and meta-analysis by \citet{2016ARA&A..54..529B}. To help visualize the behavior of our best-fit model, we show its rotation curve in Figure~\ref{fig:Vc}.

As a sanity check of (primarily) the stellar disk components of our model, we compare in Figure~\ref{fig:Kz} the profile of the $K_{z, 1.1}$ vertical acceleration at a position of $z=1.1\kpc$ above the Galactic plane to the values of this parameter derived from the analysis of \citet{2013ApJ...779..115B}. We did not use their dataset in our model fitting because the uncertainties on their datapoints are correlated, yet one can see that the model follows the $K_{z, 1.1}$ trend reasonably well. Like the \citet{2017MNRAS.465...76M} model, ours slightly under-predicts the \citet{2013ApJ...779..115B} $K_{z, 1.1}$ data at small radius, but our model follows the profile closely at $R\simgt 6.5\kpc$.

Our model is fitted primarily to dynamical constraints, and so the properties of the disk models we find should reflect a decomposition of their mass distribution given the imposed double-exponential analytic form. In contrast, when the thick disk is defined using chemical abundances, the resulting thin and thick disk masses have been found to be very similar \citep{2014ApJ...781L..31S}. This difference may be the result of using mass models that are too simple for these components in the dynamical analysis; broken exponentials may be more appropriate \citep{2022MNRAS.513.4130L}.

\subsection{Mass to light ratio}

It is interesting to consider the mass to light ratio of the luminous components of the Galaxy given that our model is fit to dynamical constraints and \emph{relative} star counts. The extinction-corrected $K$-band absolute magnitude was measured from \emph{COBE} data \citep{2001ApJ...556..181D} to be $M_K=-24.02$, and \citet{2007ApJ...662..322H} estimate a 0.1~mag uncertainty on this value from a comparison to earlier measurements. Adopting $M_{K\odot}=3.3$ \citep{2003ApJS..149..289B}, we find a $K$-band mass to light ratio of the stellar components of the Milky Way of $M_{\star, \rm MW}/L=0.61\pm 0.08\msun/\lsun$, which is substantially lower than the $K$-band $M/L=0.95\pm0.03\msun/\lsun$ derived by \citet{2003ApJS..149..289B} from a stellar-populations analysis of 2MASS and SDSS galaxies. However, those authors use a universal initial mass function (IMF) tailored to produce maximum-disk solutions, and they comment that values $\approx 30$\% lower are obtained with a \citet{1993MNRAS.262..545K} IMF. Hence with plausible assumptions about the IMF our Milky Way stellar mass estimate based on stream dynamics is fully consistent with the stellar mass based on stellar evolution models and the integrated luminosity of the Milky Way.

\subsection{Toward a non-parametric model of the Galaxy}

The very tight constraints we have derived on the local dark matter and on the structure of the Galactic components are clearly in part an artefact of the rigidity of the analytic functions (Eqns.\ref{eqn:disk} and \ref{eqn:spheroid}) used to describe them. In future work we hope to overcome this limitation; indeed, our motivation in developing the orbit to stream correction functions $\Delta_\Theta(\phi_1)$ was to be able to reuse them to derive a non-parametric Galactic model using the machinery developed in \citet{2021ApJ...915....5I}. That method uses an unsupervised machine learning procedure to derive the transformation from positions and velocities along orbits to canonical actions and angles, and thereby effectively builds a data-driven Hamiltonian of the Galaxy.

\subsection{Highlighted streams}

While a detailed exposition of the properties of the new streams is beyond the scope of the present paper, and will be presented in a future contribution, in Section~\ref{sec:DR3_Detections} we nevertheless briefly discussed some highlights and insights from the new catalog. 
\begin{itemize}
\item Stream \#17 is a very metal poor stream, with a metallicity of ${\rm [Fe/H]=-2.7}$ that is below the expected floor for cluster formation \citep{2019MNRAS.487.1986B}. It is currently passing through the Solar neighborhood, yet it appears to have been a satellite of the Sagittarius dwarf galaxy. It will be fascinating to model the dynamics of this system further to ascertain when and under what conditions its backward path can be made to coincide precisely with that of the Sagittarius dwarf. We suspect that this will require the perturbing effect of the LMC to also be taken into account, for which the stream-fitting machinery presented in Section~\ref{sec:Stream_Fitting_test_particles}, with its capability of implementing the forces due to several perturbers at once, should be ideal. The extreme proximity of this extragalactic system may also be useful for stellar population and chemical analyses.
\item For stream \#48 (Orphan-Chenab), we briefly investigate with our dynamical stream fitting software whether our newly-fitted Milky Way mass model affects the conclusion that a massive LMC is needed to reproduce the observed phase space structure of this stream. While the best-fit solution has $M_{\rm LMC}=1.5 \times 10^{11}\msun$ (consistent with \citealt{2019MNRAS.487.2685E}), we find that our models do not yet give a satisfactory representation of the stream's large-scale morphology, which we suspect is due to an insufficient flexibility of the adopted Milky Way halo density model. 
\item Stream \#49 is remarkable due to its very large radial action: it is presently arriving from the outer halo and so its high velocity members place useful constraints on the total mass of the Galaxy. Indeed, the astrometry of one of its bright member stars was very well measured by \emph{Gaia} (ID 3578950965917470208), and the derived total energy of this star is in tension with the popular MW2014 Galactic potential model \citep{2015ApJS..216...29B}, unless the system is not bound to the Milky Way, which seems highly unlikely.
\item We revisit stream \#53 (GD-1), and show that the line of sight velocity distribution is best modelled as a combination of two components, a kinematically cold part with dispersion $7.4\pm1.1\kms$ and a hot component with dispersion $29.1\pm6.1\kms$. Such kinematic structure is consistent with simulations where globular clusters form at random locations within dark matter subhalos and are subsequently accreted onto large galaxies \citep{2020ApJ...889..107C,2023ApJ...953...99C}. We also used the updated catalog to clearly demonstrate the periodicity of the spikes in the number counts along this stream when viewed as a function of physical distance along the structure: such periodicity is highly unlikely to be due to random encounters with dark matter subhalos.
\item Finally, we argue that stream \#55 is a tidal feature of $\omega$~Cen, the most massive of the Milky Way's globular clusters, and is a broader and significantly closer counterpart to the previously discovered stream \#54 (``Fimbulthul''). This trailing arm structure skirts the southern bulge at a heliocentric distance of $\approx 3\kpc$. It appears to be constituted of stars that were lost to $\omega$~Cen between 0.5 to $1\Gyr$ ago, whereas the stream \#54 stars were lost more recently than $0.5\Gyr$ ago. We suspect that the observed relative densities of streams \#54 and \#55 could be used in future work to refine the disruption model and so better understand the dynamical history of this unique and important cluster.
\end{itemize}

\subsection{Limitations}

While we have tried to render the analysis as realistic as possible, there are several limitations to our approach. First of all, our Milky Way model is axisymmetric, which is clearly unrealistic especially in the inner Galaxy where dynamics is dominated by the bar. Indeed, some of the streams we removed from the sample were rejected precisely because we feared that their small pericenters made them highly susceptible to the bar, and hence that they would be poorly reproduced with our axisymmetric model. The recent work of \citet{2023A&A...678A.180T} shows that it is now possible to overcome such limitations, and following their lead we hope to be able to include non-axisymmetric components in our MCMC exploration in the near future. Similarly, it may be useful to include spiral arms in the modelling, though we suspect their influence will be relatively minor for the halo streams we have considered.

Using a double power-law mass model with a single shape parameter may also be too limiting, as hinted by the analysis in Section~\ref{sec:stream48}. It may be necessary to consider a basis function approach (e.g. \citealt{2022MNRAS.514.1266P}) to model the mass distribution in a more realistic manner. 

In our future work we intend to use the algorithms developed here to explore the combined effect of the LMC, of the Sagittarius dwarf, and of other potential perturbers on the full population of stellar streams.

\subsection{Concluding remarks}

The upcoming huge WEAVE and 4MOST spectroscopic campaigns that will soon be launched promise an exciting future for this field of study. The present work was based on a relatively small number of velocity-confirmed stream members and our statistics could easily be improved by an order of magnitude with observations from those spectroscopic surveys. Furthermore, the next \emph{Gaia} DR4 release will improve substantially the proper motion and parallax accuracy, most probably allowing yet more systems to be detected with our algorithm, and tightening the constraints on the known structures.

However, the most exciting use of this data now is to use it to try to shed light on the dark matter, and ascertain whether the observed stream morphology, structure, sub-structure and dynamics allow us to prefer one dark matter model over others.

\vfill\eject
 
\acknowledgments

RI, NM, DA, BF, GM, RE and AS acknowledge funding from the Agence Nationale de la Recherche (ANR project ANR-18-CE31-0006, ANR-18-CE31-0017 and ANR-19-CE31-0017), from CNRS/INSU through the Programme National Galaxies et Cosmologie, and from the European Research Council (ERC) under the European Unions Horizon 2020 research and innovation programme (grant agreement No. 834148). FR acknowledges support provided by the University of Strasbourg Institute for Advanced Study (USIAS), within the French national programme Investment for the Future (Excellence Initiative) IdEx-Unistra. MB acknowledges the financial support to this research by INAF, through the 
Mini Grant assigned to the project ``Chemo-dynamics of the Accreted Halo of the Milky Way (CHAM)'' (P.I. M. Bellazzini), and by the Italian {\em Ministero dell'Universit\`a e della Ricerca} through the PRIN 2022 Grant 2022LLP8TK\_001 ``LEGO – Reconstructing the building blocks of the Galaxy by chemical tagging'' (P.I. A, Mucciarelli). GT acknowledges support from Agencia Estatal de Investigaci\'on del Ministerio de Ciencia en Innovaci\'on (AEI-MICIN) under grant number PID2020-118778GB-I00/10.13039/501100011033, the MICIN under grant number FJC2018-037323-I, and the AEI under grant number CEX2019-000920-S. AAA acknowledges support from the Herchel Smith Fellowship at the University of Cambridge and a Fitzwilliam College research fellowship supported by the Isaac Newton Trust. ES acknowledges funding through VIDI grant ``Pushing Galactic Archaeology to its limits'' (with project number VI.Vidi.193.093) which is funded by the Dutch Research Council (NWO). This research has been partially funded from a Spinoza award by NWO (SPI 78-411). This research was supported by the International Space Science Institute (ISSI) in Bern, through ISSI International Team project 540 (The Early Milky Way).

We gratefully acknowledge the High Performance Computing centre of the Universit\'e de Strasbourg for a very generous time allocation and for their support over the development of this project.

We wish to thank Pablo Gal\'an, Rosa Hoogenboom, Sara Vitali, Cl\'ar-Br\'id Tohill, Sara Vitali and Paige Yarker for their invaluable help in supporting the INT observations reported here.

This work has made use of data from the European Space Agency (ESA) mission \emph{Gaia}(\url{https://www.cosmos.esa.int/gaia}), processed by the {\it Gaia} Data Processing and Analysis Consortium (DPAC, \url{https://www.cosmos.esa.int/web/gaia/dpac/consortium}). Funding for the DPAC has been provided by national institutions, in particular the institutions participating in the \emph{Gaia} Multilateral Agreement. 

Based on observations collected at the European Southern Observatory under ESO programmes 105.20AL.001, 110.246A.001, and 111.2517.001.

Funding for SDSS-III has been provided by the Alfred P. Sloan Foundation, the Participating Institutions, the National Science Foundation, and the U.S. Department of Energy Office of Science. The SDSS-III web site is http://www.sdss3.org/.

SDSS-III is managed by the Astrophysical Research Consortium for the Participating Institutions of the SDSS-III Collaboration including the University of Arizona, the Brazilian Participation Group, Brookhaven National Laboratory, Carnegie Mellon University, University of Florida, the French Participation Group, the German Participation Group, Harvard University, the Instituto de Astrofisica de Canarias, the Michigan State/Notre Dame/JINA Participation Group, Johns Hopkins University, Lawrence Berkeley National Laboratory, Max Planck Institute for Astrophysics, Max Planck Institute for Extraterrestrial Physics, New Mexico State University, New York University, Ohio State University, Pennsylvania State University, University of Portsmouth, Princeton University, the Spanish Participation Group, University of Tokyo, University of Utah, Vanderbilt University, University of Virginia, University of Washington, and Yale University.

Guoshoujing Telescope (the Large Sky Area Multi-Object Fiber Spectroscopic Telescope LAMOST) is a National Major Scientific Project built by the Chinese Academy of Sciences. Funding for the project has been provided by the National Development and Reform Commission. LAMOST is operated and managed by the National Astronomical Observatories, Chinese Academy of Sciences.

\software{STREAMFINDER \citep{2018MNRAS.477.4063M}, Armadillo \citep{Sanderson2016}, IRAF \citep{1986SPIE..627..733T,1993ASPC...52..173T}, GalPot \citep{Dehnen:1998tk}}

\begin{table*}
\caption{Streams detected in {\it Gaia} DR3.}
\label{tab:streams}
\footnotesize
\begin{center}
\begin{tabular}{rlcrrrrrrrll}
\hline
   s$_{\rm ID}$ & name &  sample      & $\alpha_0$ & $\alpha_{\rm pole}$ & $\delta_{\rm pole}$ & $n$ & $n_v$ & $M_\star \, [\msun]$ & $\langle {\rm [Fe/H]}\rangle$ & Ref & Comment \\
\hline
 1 & C-20         & 1 &   0.561 &   275.038 &    -30.9   &   29 &    8 &   1200 &  -2.93 & 1    &                                             \\
 2 & New-1        & 0 &   6.705 &   273.23  &     -5.703 &   36 &    1 &    530 &  -2.22 & 2    &                                             \\
 3 & NGC288       & 0 &  13.188 &   101.35  &      1.686 &  173 &    7 &   2200 &  -1.32 & 3    &                                             \\
 4 & ATLAS        & 1 &  20.185 &    75.623 &     47.653 &  208 &   80 &  19000 &  -2.2  & 4    &                                             \\
 5 & Phoenix      & 1 &  24.341 &   309.372 &     12.542 &   58 &   26 &   3000 &  -2.7  & 5    &                                             \\
 6 & Kwando    & 0 &  24.930  &   112.058 &      9.835 &  138 &   62 &  40000 &  -2.29 & 6    & Very wide, rough ${\rm[Fe/H]}$ estimate     \\
 7 & New-2        & 0 &  27.124 &   296.02  &     -1.042 &   11 &    3 &    420 &  -1.85 & 2    &                                             \\
 8 & New-3        & 0 &  34.127 &    53.727 &    -85.092 &   90 &   18 &     73 &  -2.07 & 7    &                                             \\
 9 & C-13         & 0 &  35.678 &   311.207 &     -6.984 &  100 &    5 &    750 &  -0.25 & 8    & Velocities unclear, strange CMD             \\
10 & Gaia-12      & 1 &  37.357 &   147.363 &     43.851 &   46 &    5 &   1200 &  -3.28 & 2    &                                             \\
11 & Hydrus       & 0 &  37.389 &    63.965 &      9.619 &   39 &   14 &   7700 &  -3.48 & 2    & New stream                               \\
12 & NGC1261      & 0 &  48.068 &    18.701 &     31.13  &  173 &   23 &   8100 &  -1.27 & 3    & Velocities messy, orbit apocenter           \\
13 & NGC1261a     & 0 &  48.068 &   153.876 &    -10.473 &  136 &    4 &   9700 &  -1.27 & 3    & Related to N1261?                           \\
14 & NGC1261b     & 0 &  48.068 &    84.128 &     32.219 &   70 &   12 &   5100 &  -1.27 & 3    & Related to N1261?                           \\
15 & Indus        & 0 &  56.206 &    38.115 &     19.901 & 1021 &   96 &  24000 &  -2.09 & 9    &                                             \\
16 & NGC1851      & 0 &  78.528 &   151.161 &     19.287 &  105 &   11 &   3400 &  -1.18 & 3    &                                             \\
17 & New-4        & 0 &  82.663 &   254.437 &    -72.963 &  120 &   10 &    120 &  -2.71 & 10   & Two clumps?                                 \\
18 & New-5        & 1 &  90.988 &    51.229 &     45.719 &   29 &    8 &    730 &  -1.76 & 2    &                                             \\
19 & Leiptr       & 1 &  91.505 &   152.738 &     39.307 &  412 &   37 &   3000 &  -2.17 & 2    &                                             \\
20 & C-12         & 0 &  94.401 &   289.858 &    -59.422 &  929 &   13 &  14000 &  -1.34 & 10   & Two overlapping structures?                 \\
21 & New-6        & 0 &  98.455 &   249.927 &    -87.804 &   13 &    9 &     16 &  -1.02 & 2    & Two structures?                             \\
22 & New-7        & 1 &  99.776 &   106.87  &    -14.425 &  146 &   12 &    340 &  -2.18 & 7    &                                             \\
23 & NGC2298      & 0 & 102.248 &   137.971 &     48.3   &  125 &    1 &   2000 &  -1.92 & 3    &                                             \\
24 & C-25         & 0 & 112.659 &   324.848 &     44.684 &  121 &   25 &   1300 &  -2.3  & 1    & Related to C-11?                            \\
25 & C-11         & 0 & 117.062 &   349.03  &     24.975 &  112 &    9 &   1100 &  -2.91 & 1    & Velocities unclear                          \\
26 & NGC2808      & 0 & 138.013 &   171.607 &     21.534 &  292 &   10 &   7300 &  -1.14 & 3    &                                             \\
27 & New-8        & 0 & 139.625 &   318.111 &     66.353 &   37 &   11 &    110 &  -2.23 & 7    & Velocities unclear                          \\
28 & New-9        & 0 & 142.991 &   237.5   &      6.097 &    7 &    1 &      3 &  -1.99 & 7    &                                             \\
29 & New-10       & 0 & 144.479 &   318.071 &     58.997 &   10 &    7 &     38 &  -1.95 & 7    &                                             \\
30 & C-10         & 0 & 145.970  &    55.813 &      5.518 &  158 &    5 &   1800 &  -0.91 & 8    & Thick RGB                                   \\
31 & New-11       & 0 & 147.682 &    61.538 &     17.134 &   47 &    7 &    100 &  -0.99 & 10   &                                             \\
32 & New-12       & 0 & 153.299 &   234.748 &    -11.648 &   18 &    6 &    110 &  -2.03 & 7    & Velocities unclear                          \\
33 & Gaia-10      & 1 & 153.312 &   156.208 &    -74.386 &  141 &   26 &   7600 &  -1.4  & 1    &                                             \\
34 & Gjoll        & 1 & 154.403 &   158.539 &     43.545 &  607 &   40 &   2100 &  -1.63 & 3    &                                             \\
35 & C-9          & 0 & 155.889 &   115.351 &    -66.055 &  183 &   35 &    630 &  -0.72 & 1    & Several clumps?                             \\
36 & C-24         & 0 & 158.186 &   228.19  &    -12.131 &  244 &   27 &   2700 &  -0.93 & 1    & Velocities unclear                          \\
37 & New-13       & 0 & 161.861 &    84.236 &     14.306 &   89 &    3 &    580 &  -1.85 & 2    &                                             \\
38 & New-14       & 0 & 162.209 &   264.091 &    -13.901 &   47 &    1 &    250 &  -1.19 & 2    & High velocity                               \\
39 & Slidr        & 1 & 166.076 &    92.169 &    -64.021 &  330 &   34 &    840 &  -1.7  & 1    &                                             \\
40 & New-15       & 0 & 170.248 &    69.207 &    -26.38  &   18 &    4 &     28 &  -0.95 & 2    & Compact                                     \\
\end{tabular}
\end{center}
\tablecomments{Columns 1 and 2 list the identification number s$_{\rm ID}$ and name of the stream; a value of ``1'' in column 3 marks whether the stream is part of the chosen sample of 29 streams that were used to constrain the Galactic potential. Columns 4--6 list, respectively, the zero-point in right ascension and position of the pole of the coordinate system used to derive the $\phi_1$ and $\phi_2$ stream coordinates. Columns 7 and 8 report the number of stars found in the structure, and the number with measured radial velocity, respectively. Column 9 reports the stellar mass estimate, based on star counts, while columns 10 and 11 list the metallicity of the structure and the source of the metallicity information, where 1=\citet{2022MNRAS.516.5331M}, 2=\citet{2023arXiv230801344M}, 3=\citet{2010arXiv1012.3224H}, 4=\citet{2021ApJ...911..149L}, 5=\citet{2020Natur.583..768W}, 6=\citet{2017ApJ...847..119G}, 7=SDSS, 8=LAMOST, 9=\citet{2020AJ....160..181J}, 10=This work, 11=\citet{2019ApJ...883...84R}, 12=\citet{2023MNRAS.521.4936K}, 13=\citet{2020ApJ...898L..37Y}, 14=\citet{2020AJ....159..287C}, 15=Gaia BP/RP, 16=\citet{2023ApJS..267....8A}, 17=\citet{2019MNRAS.490.3508L}. Some brief comments on the nature of the structure are given in column 12. We note that stream \#6 (``Kwando'', \citealt{2017ApJ...847..119G}) appears to be the same structure as the ``Cetus/Palca'' stream \citep{2022ApJ...930..103Y}. Note also that one star (\emph{Gaia} ID 3267948604442696448) in stream \#8 (``New-3'') is a member of group \#4 of \citet{2023A&A...670L...2D}, but closer inspection shows that these two structures are different, with stream \#8 showing narrow spatial and kinematic coherence.}
\end{table*}

\begin{table*}
\footnotesize
\begin{center}
\begin{tabular}{rlcrrrrrrrll}
41 & Ylgr         & 1 & 174.004 &   256.584 &     17.379 &  919 &   20 &  11000 &  -2.09 & 7    &                                             \\
42 & Sylgr        & 0 & 176.261 &   273.255 &    -60.454 &  256 &   33 &    700 &  -2.92 & 11   & High velocity scatter                       \\
43 & New-16       & 0 & 178.134 &    66.201 &     58.999 &  152 &   23 &    960 &  -2.01 & 7    &                                             \\
44 & Gaia-7       & 1 & 181.295 &    78.993 &    -35.727 &  282 &   15 &   1300 &  -0.88 & 10   &                                             \\
45 & Gaia-8       & 1 & 186.342 &   305.604 &    -48.129 &  453 &   18 &   3600 &  -1.77 & 10   &                                             \\
46 & New-17       & 0 & 188.285 &   102.418 &      5.746 &   44 &    3 &    170 &  -1.62 & 2    &                                             \\
47 & Fjorm        & 1 & 189.867 &   103.086 &     19.455 &  297 &   29 &   1700 &  -2.23 & 3    &                                             \\
48 & Orphan       & 0 & 191.105 &    68.052 &    -15.708 &  867 &  247 & 130000 &  -1.9  & 12   & Massive system                              \\
49 & Gaia-1       & 1 & 192.56  &   287.833 &    -35.545 &  200 &   28 &   1100 &  -1.8  & 7    &                                             \\
50 & C-23         & 1 & 195.564 &   122.347 &    -16.78  &   29 &    6 &    310 &  -2.36 & 1    &                                             \\
51 & LMS-1        & 0 & 198.23  &    43.633 &     62.255 &  358 &   29 &  15000 &  -2.09 & 13   & Massive and wide system                     \\
52 & C-22         & 1 & 199.904 &    56.016 &     68.974 &   39 &   11 &    470 &  -2.05 & 8    &                                             \\
53 & GD1          & 1 & 200     &   214.928 &    -29.881 & 1468 &  323 &  14000 &  -2.49 & 1    &                                             \\
54 & Fimbulthul   & 0 & 201.697 &   304.173 &    -22.859 & 3724 &   29 &  16000 &  -1.53 & 3    & Massive system                              \\
55 & Fimbulthul-S & 0 & 201.697 &    16.148 &      1.199 & 1734 &   25 &   4400 &  -1.53 & 10   & Massive system                              \\
56 & New-18       & 0 & 210.104 &   147.375 &    -50.685 &   13 &    0 &     29 &  -2.35 & 2    &                                             \\
57 & NGC5466      & 1 & 211.364 &   161.594 &    -50.069 &   43 &    6 &   1900 &  -1.98 & 3    &                                             \\
58 & New-19       & 0 & 214.041 &   302.311 &     -1.888 &  107 &    1 &    280 &  -0.94 & 2    &                                             \\
59 & Gaia-6       & 1 & 214.600   &   268.951 &    -41.694 &  145 &   14 &   1800 &  -1.53 & 1    &                                             \\
60 & New-20       & 0 & 221.839 &   294.781 &    -22.535 &   40 &    4 &    100 &  -0.79 & 2    & Compact, velocities unclear                 \\
61 & Pal-5        & 1 & 229.022 &   320.298 &    -54.042 &  129 &   69 &  17000 &  -1.41 & 3    &                                             \\
62 & M5           & 1 & 229.638 &   325.223 &     63.075 &   91 &   11 &    710 &  -1.29 & 3    &                                             \\
63 & Kshir        & 1 & 230.196 &   229.22  &    -21.937 &  141 &    7 &   2200 &  -1.83 & 7    &                                             \\
64 & Svol         & 1 & 231.711 &    28.918 &     60.957 &  234 &   16 &   2000 &  -1.98 & 1    &                                             \\
65 & Gaia-9       & 1 & 236.238 &   169.091 &    -13.272 &  233 &   25 &    950 &  -2.21 & 1    &                                             \\
66 & Ophiuchus    & 0 & 242.576 &    18.564 &    -79.685 &  391 &   72 &   4800 &  -1.98 & 14,3 &                                             \\
67 & NGC6101      & 0 & 246.451 &   227.979 &     16.954 &   94 &    3 &   4800 &  -1.98 & 3    &                                             \\
68 & M92          & 0 & 259.281 &   232.737 &    -43.4   &  202 &   23 &   2100 &  -2.31 & 3    &                                             \\
69 & NGC6397      & 0 & 265.175 &   268.821 &     35.811 & 1207 &   17 &   2500 &  -2.02 & 3    &                                             \\
70 & Gaia-11      & 0 & 267.612 &   195.778 &    -15.134 &   84 &    3 &   1100 &  -1.19 & 2    & Strange CMD, velocities unclear             \\
71 & Hrid         & 0 & 280.228 &   187.604 &      2.034 &  666 &   29 &   2000 &  -1.13 & 7    &                                             \\
72 & C-7          & 1 & 283.419 &    76.064 &    -35.517 &  239 &   16 &   1500 &  -1.52 & 10   & Two superposed streams                      \\
73 & New-21       & 0 & 314.963 &    50.439 &     18.37  &   12 &    4 &      8 &  -1.76 & 15   &                                             \\
74 & New-22       & 0 & 316.569 &   243.966 &    -56.299 &  762 &    3 &   7000 &  -0.87 & 2    & High velocity                               \\
75 & Phlegethon   & 1 & 322.720  &    55.023 &     -8.319 &  632 &   41 &   1700 &  -2.19 & 8    &                                             \\
76 & NGC7089      & 0 & 323.363 &   235.973 &     72.865 &   15 &    2 &    300 &  -1.65 & 3    &                                             \\
77 & NGC7099      & 0 & 325.092 &    79.501 &    -44.028 &   54 &    3 &    680 &  -2.27 & 3    &                                             \\
78 & New-23       & 0 & 325.847 &   211.35  &    -34.781 &   55 &   10 &    580 &  -0.98 & 2    & Velocities unclear                          \\
79 & New-24       & 0 & 329.825 &   188.254 &     66.646 &   44 &    6 &    120 &  -2.25 & 8    &                                             \\
80 & New-25       & 0 & 334.183 &    59.385 &     20.639 &   56 &    7 &    170 &  -0.64 & 16   & Compact, velocities unclear                 \\
81 & New-26       & 0 & 335.959 &    50.194 &     23.447 &  121 &   12 &  18000 &  -1.68 & 2    & Velocities unclear, superposed streams?                          \\
82 & New-27       & 0 & 341.875 &    23.599 &    -51.695 &   43 &    4 &    300 &  -0.17 & 2    & Velocities unclear                          \\
83 & New-28       & 0 & 342.224 &    80.278 &     -5.92  &   37 &    8 &    410 &  -2.08 & 2    & Velocities unclear                          \\
84 & NGC7492      & 0 & 347.111 &    89.005 &    -34.892 &   26 &    1 &   3300 &  -1.78 & 3    &                                             \\
85 & C-19         & 1 & 354.356 &    81.45  &     -6.346 &   46 &   12 &   3200 &  -3.58 & 1    &                                             \\
86 & Jhelum       & 1 & 355.231 &     5.299 &     37.216 &  986 &  160 &  17000 &  -2.12 & 9    &                                             \\
87 & Tuc-3        & 0 & 359.150  &   176.704 &    -30.396 &  102 &   34 &  12000 &  -2.49 & 17   &                                             \\
\hline
\end{tabular}
\end{center}
\end{table*}

\bibliography{Gaia_DR3Streams}
\bibliographystyle{aasjournal}

\end{document}